\documentclass[twocolumn]{aastex62}

\usepackage{amsmath}
\usepackage{graphicx}

\newcommand{\WRnum}{19}
\newcommand{\dustem}{\texttt{DustEM}}

\graphicspath{{./}{}}

\accepted{June 12, 2020}
\submitjournal{ApJ}

\shorttitle{Revisiting the Impact of Dust Production from Carbon-Rich Wolf-Rayet Binaries}
\shortauthors{Lau et al.}

\begin{document}

\title{Revisiting the Impact of Dust Production from Carbon-Rich Wolf-Rayet Binaries}

\correspondingauthor{Ryan Lau}
\email{ryanlau@ir.isas.jaxa.jp}

\author{Ryan M. Lau}
\affil{Institute of Space \& Astronautical Science, Japan Aerospace Exploration Agency, 3-1-1 Yoshinodai, Chuo-ku, Sagamihara, Kanagawa 252-5210, Japan}
\author{J.J. Eldridge}
\affil{Department of Physics, University of Auckland, Private Bag 92019, Auckland 1010, New Zealand}
\author{Matthew J. Hankins}
\affil{Division of Physics, Mathematics, and Astronomy, California Institute of Technology, Pasadena, CA 91125, USA}
\author{Astrid Lamberts}
\affil{Universit\'e C\^ote d'Azur, Observatoire de la C\^ote d'Azur, CNRS, Laboratoire Lagrange, Laboratoire ARTEMIS, France}
\author{Itsuki Sakon}
\affil{Department of Astronomy, School of Science, University of Tokyo, 7-3-1 Hongo, Bunkyo-ku, Tokyo 113-0033, Japan}
\author{Peredur M. Williams}
\affil{Institute for Astronomy, University of Edinburgh, Royal Observatory, Edinburgh EH9 3HJ}





\begin{abstract}

We present a dust spectral energy distribution (SED) and binary stellar population analysis revisiting the dust production rates (DPRs) in the winds of carbon-rich Wolf-Rayet (WC) binaries and their impact on galactic dust budgets.
\dustem~SED models of \WRnum~Galactic WC ``dustars" reveal DPRs of $\dot{M}_d\sim10^{-10}-10^{-6}$ M$_\odot$ yr$^{-1}$
and carbon dust condensation fractions, $\chi_C$, between $0.002 - 40\%$.  A large ($0.1 - 1.0$ $\mu$m) dust grain size composition is favored for efficient dustars where $\chi_C\gtrsim1\%$. Results for dustars with known orbital periods verify a power-law relation between $\chi_C$, orbital period, WC mass-loss rate, and wind velocity consistent with predictions from theoretical models of dust formation in colliding-wind binaries.
We incorporated dust production into Binary Population and Spectral Synthesis (BPASS) models to analyze dust production rates from WC dustars, asymptotic giant branch stars (AGBs), red supergiants (RSGs), and core-collapse supernovae (SNe).
BPASS models assuming constant star formation (SF) and a co-eval $10^6$ M$_\odot$ stellar population were performed at low, Large Magellanic Cloud (LMC)-like, and solar metallicities (Z = 0.001, 0.008, and 0.020). 
Both constant SF and co-eval models show that SNe are net dust destroyers at all metallicities.
Constant SF models at LMC-like metallicities show that AGB stars slightly outproduce WC binaries and RSGs by factors of $2-3$, whereas at solar metallicites WC binaries are the dominant source of dust for $\sim60$ Myr until the onset of AGBs, which match the dust input of WC binaries.
Co-eval population models show that for ``bursty" SF, AGB stars dominate dust production at late times ($t\gtrsim 70$ Myr).

\end{abstract}



\section{Introduction} \label{sec:intro}

Dust formation can surprisingly occur in the hostile circumstellar environment of Wolf-Rayet (WR) stars, which are descendants of massive O-stars characterized by fast winds ($\gtrsim1000$ km s$^{-1}$), hot photospheres ($T_*\gtrsim 40000$ K), and high luminosities ($L_*\sim10^5$ L$_\odot$; \citealt{Gehrz1974,Williams1987,Crowther2007}). All of the known dust-forming WR stars, hereafter referred to as ``dustars" \citep{Marchenko2007}, are of the carbon-rich (WC) sub-type. These WC dustars typically exhibit late spectral sub-types (WC7 - WC9), which are characterized by relatively cool WR photospheres ($T_*\sim40000 - 70000$ K; \citealt{Sander2019}). 

In order to explain the observed dust formation in the hostile environment of WC dustars, the binary nature is believed to play a key role: strong winds from the WC star collide with weaker winds from an OB-star companion and create dense regions in the wake of the companion's orbit that allow for dust to condense \citep{Williams1990,Usov1991, Cherchneff2015}. 

The density enhancements in the wind collision region facilitates rapid radiative cooling that lead to dust formation in addition to shielding newly formed dust grains from UV photons emitted by the central binary. Unambiguous evidence of dust formation regulated by binary influence was presented in high spatial resolution near-IR images of the WC9 binary WR104 by \citet{Tuthill1999} that revealed a remarkable dust ``pinwheel."

The binary orbital parameters such as the semi-major axis/orbital period and eccentricity modulate the dust formation in WC dustars. For example, systems with low eccentricity and short orbital periods ($\sim$ yr) like WR104 exhibit persistent dust formation in continuous pinwheel plumes resembling an Archimedean spiral that change in position angle as the WC star and its binary companion move in their orbit \citep{Tuthill2008}. Alternatively, WC dustars with high orbital eccentricity and longer orbital periods ($\gtrsim5$ yr) like WR140 exhibit periodic dust formation, where the onset of dust formation corresponds to periapse when the WC star and its companion are near their closest orbital separation \citep{Williams2009a}.

Despite the fact that WC dustars can produce copious amounts of dust ($\dot{M}_d\sim10^{-8} - 10^{-6}$ $\mathrm{M}_\odot\,\mathrm{yr}^{-1}$; \citealt{Williams1987,Hankins2016,Hendrix2016}) they have been commonly overlooked as significant sources of dust in the ISM of galaxies in the local and early Universe. The input from WC dustars compared to leading dust producers like asymptotic giant branch (AGB; \citealt{Boyer2012}) stars and core-collapse supernovae (SNe; \citealt{Dwek2011}) is thought to be low given their relative rarity as products of only the most massive O-stars. Additionally, at low metallicities it is difficult to form a WR star in the context of single star evolution \citep{Conti1975} due to a lack of metals that drive mass-loss and the expulsion of the hydrogen envelope. 

However, the influence of their binary nature on the formation of WR stars and their dust formation is not well-understood, especially given observations that reveal a majority ($\gtrsim70\%$) of massive stars are in binaries that will eventually interact with their companion over their lifetimes \citep{Sana2012}. Such binary interaction enables the formation of WR stars through mechanisms such as envelope stripping via Roche-Lobe overflow (\citealt{Mauerhan2015, DMI2017}). This provides formation channels of WC binaries even at low metallicities. Revisiting the impact of WC dustars is motivated by the need for additional dust input sources since SNe may in fact be net dust destroyers due to the shocks they drive into the surrounding ISM \citep{Temim2015}.

There are, however, major challenges in addressing the dust contribution from WC dustars in the ISM of galaxies across cosmic time. The measured dust production rates of WC dustars show a wide range of conflicting values in literature throughout the past several decades. 
This is due in part to uncertainties in distance estimates and the different techniques used to model dust emission, which is also difficult given the complex dust morphology observed around WC dustars. Another challenge is that binary stellar evolution tracks are complex and much more difficult to model than single star evolution.

In this paper we study the dust production from WC dustars by conducting a consistent dust spectral energy distribution (SED) analysis utilizing archival IR photometry and spectroscopy of a sample of \WRnum~Galactic dustars with distance estimates from Gaia DR2 \citep{BJ2018,RC2020}. Based on the results of our SED analysis, we incorporate dust production in Binary Population and Spectral Synthesis (BPASS; \citealt{BPASS,S2018}) models to investigate the dust input from WC dustars relative to AGB stars, red supergiants (RSGs), and core-collapse SNe at difference metallicities representative of different epochs in cosmic time. We also model the dust input in the well-studied Large Magellanic Cloud (LMC), where AGB stars are claimed to be a significant source of dust and SN are likely net dust destroyers \citep{Riebel2012,Temim2015}. This work presents the first implementation of BPASS to investigate dust production from binary stellar populations.

The paper is organized as follows. First, we describe the sample selection and the IR photometric and spectroscopic data sets and extinction correction utilized for the SED modelling (Sec.~\ref{sec:2}). In Sec.~\ref{sec:3}, we detail our dust SED modeling approach with \dustem, present the results of our analysis with descriptions of selected WC dustars, and compare our result with previous studies and the theoretical dust formation model by \citet{Usov1991}. In Sec.~\ref{sec:4}, we describe the BPASS dust models and present the results of constant star formation and co-eval population models in environments corresponding to low (Z = 0.001), LMC-like (Z = 0.008), and solar (Z = 0.020) metallicites. We also provide a comparison between BPASS model results and the LMC that highlight the importance of star formation history in modeling the dust input from massive stars. Lastly, we discuss the astrophysical implications of our study on the dust production from WC dustars (Sec.~\ref{sec:5}).

\section{WC Dustar Sample and Observations}
\label{sec:2}

\subsection{Dustar Sample Selection}

The goal of this analysis is to fit dust spectral energy distribution (SED) models to the mid-IR emission from WC dustars incorporating distance information from Gaia DR2 \citep{RC2020} to derive the dust ``shell" radii, mass, production rates, and condensation fraction. Dust SED models require spectroscopy or well-sampled photometry between $\sim2-30$ $\mu$m where the thermal emission from circumstellar dust dominates over the photosphere and free-free emission from ionized winds in the WC system (e.g.~\citealt{Williams1987}). Beyond $\sim30$ $\mu$m, background emission from cooler ISM dust may start to contaminate the IR SED. Dustar distances are adopted from the recent study by \citet{RC2020}, who perform a Bayesian analysis on Gaia DR2 parallaxes to 383 Galactic WR stars with priors based on HII regions and dust extinction.

Since a non-negligible fraction of known WC dustars exhibit variability on $\lesssim1$ yr timescales \citep{Williams2019}, the known variables in this sample are restricted to those that have contemporaneous mid-IR observations or mid-IR spectroscopy out to at least $\sim20$ $\mu$m taken in a single epoch. The dustars with no observed variability in this sample have $\sim10-20$ $\mu$m spectroscopic or photometric coverage with sufficiently high spatial resolution at longer wavelengths (FWHM$\lesssim 6$'') to distinguish possible contamination from surrounding IR sources or ISM emission. 

Out of the 88 currently known WC dustars in V 1.23 (July 2019) of the Galactic Wolf Rayet Catalogue \citep{RC2015}\footnote{\url{http://pacrowther.staff.shef.ac.uk/WRcat/}}, our sample consists of \WRnum~dustars shown in Table~\ref{tab:DustarTab} that meet the above criteria.

\subsection{Observations and Archival Data}

\subsubsection{Mid-IR Imaging of WR48a with VLT/VISIR}

Mid-IR imaging observations of WR48a (PID: 097.D-0707(A); PI - Lau) were performed  on the Very Large Telescope (VLT) at the ESO Paranal observatory using the VLT spectrometer and imager for the mid-infrared (VISIR, \citealt{Lagage2004}) at the Cassegrain focus of UT3 on 2016 June 15. Images of WR48a presented in this work were taken with the NeII$\_2$ filter ($\lambda_c$=13.04 $\mu$m, $\Delta \lambda=0.22$) and obtained using chopping and nodding to remove the sky and telescope thermal background emission. 

Given the size of the field of view with VISIR of $38\times38''$ in comparison to the extent of the dust emission from WR48a ($\sim5''$), an on-detector chop-nod configuration was used with 20'' chop and nod amplitudes. The total integration time on WR48a was 15 minutes. Raw images were accessed and downloaded from the ESO Science Archive Facility and processed using the Modest Image Analysis and Reduction (MIRA) software written by Terry Herter (See \citealt{Herter2013}).

The NeII$\_2$ imaging observations of WR48a achieved near diffraction-limited imaging with a measured Gaussian full-width at half maximum (FWHM) of 0.45''.

\subsubsection{Space-Based Mid-IR Spectroscopy and Photometry from ISO, Spitzer, and WISE}
\textbf{ISO/SWS.} Five WC dustars in this sample have archival 2.2 - 40 $\mu$m medium resolution ($R\approx 250 - 600$) spectra taken by the Short Wavelength Spectrometer (SWS; \citealt{deGraauw1996}) on the Infrared Space Observatory (ISO; \citealt{Kessler1996}) that were obtained in the WRSTARS program (PI - van der Hucht; \citealt{vdh1996}). Archival ISO/SWS spectra of these five dustars (WR48a, WR70, WR98a, WR104, and WR118) were downloaded from the database of SWS spectra processed and hosted by \citet{Sloan2003}\footnote{\url{https://users.physics.unc.edu/~gcsloan/library/swsatlas/atlas.html}}. The ``sws" files, which are the final versions of the SWS spectra corrected for segment discontinuities and overlaps, were used for the SED analysis.

In order to verify the quality of the SWS data and identify possible normalization issues between SWS bands, the ``pws" files that show the spectra prior segment-to-segment normalization were inspected. All five dustars with ISO/SWS data did not exhibit segment-to-segment discontinuities larger than $\sim10\%$ below 27.5 $\mu$m, which is the wavelength where the 3E band of the long wavelength section starts. Due to the flux discontinuities and larger flux uncertainties beyond 27.5 $\mu$m , only the 2.2 - 27.5 $\mu$m data were used for the SED fitting. The spectra of WR48a, WR98a, WR104, and WR118 were smoothed by a median filter with a 51-element kernel, and the spectrum of WR70, which had an ``sws" file with wavelengths sampled a factor of 4 higher times higher than the other 4 sources, was smoothed with a 201-element kernel. 

\textbf{Spitzer/IRS.} Six WC dustars have archival mid-IR spectra taken by the Infrared Spectrograph (IRS; \citealt{Houck2004}) on the Spitzer Space Telescope (\citealt{Werner2004}) across multiple programs throughout the \textit{Spitzer} ``cold mission" (2004 - 2009). Hi-Res (R$\sim600$) \textit{Spitzer}/IRS spectra of WR140 (PID - 124; PI - Gehrz), WR19, WR103, and WR53 (all 3 PID 199; PI - Houck) were downloaded from the Combined Atlas of Sources with Spitzer IRS Spectra (CASSIS; \citealt{Lebouteiller2015})\footnote{\url{https://cassis.sirtf.com/}}. The spectrum of WR19 was published in the WC dust chemistry study by \citet{Marchenko2017}, and the spectrum of WR103 was published by \citet{Crowther2006} in their UV to mid-IR spectral analysis comparing WC9 and [WC9] stars. \citet{Ardila2010} also included the spectra of WR53 and WR103 in their atlas of \textit{Spitzer} stellar spectra. Broad emission line features from WR103, WR140, and WR19 were masked, and the lower signal-to-noise ratio spectra of WR140 and WR19 were smoothed by a median filter with an 11-element kernel. 

In order to verify the photometric accuracy of the long wavelength ($>20$ $\mu$m) IRS spectra of WR140, the flux was checked against 24 $\mu$m photometry from the Multi-band Imaging Photometer for Spitzer (MIPS; \citealt{Rieke2004}) taken in the same program and within $\sim1$ month of the IRS observations (PID - 124; PI - Gehrz).
The 24 $\mu$m \textit{Spitzer}/MIPS photometry of WR140 extracted with a 35''-radius aperture and a 40 - 50'' background annulus is $F_\mathrm{24,MIPS}=1.07\pm0.05$\footnote{An aperture correction factor of 1.08 was included in this flux calculation in accordance with Table 4.13 of the MIPS Instrument Handbook for the aperture and background annulus radii used.} This is consistent with the IRS spectrum at 24 $\mu$m ($F_\mathrm{24,IRS}\approx0.8\pm0.4$ Jy).

The dustars WR48a and WR98a had coverage from both \textit{Spitzer}/IRS (PID - 40285, PI - Waters) and ISO/SWS, but the ISO spectra were adopted in this work for SED modeling due to its extended wavelength coverage and the sufficiently high signal-to-noise detection.

\textbf{Spitzer/IRAC and MIPS.} For the non-variable dustars, mid-IR \textit{Spitzer} photometry were adopted from the Galactic Legacy Infrared Mid-Plane Survey Extraordinaire (GLIMPSE; \citealt{Churchwell2009}) taken by the Infrared Array Camera (IRAC; \citealt{Fazio2004}) at 3.6, 4.5, 5.8, and 8.0 $\mu$m in addition to 24 $\mu$m photometry from the MIPS Galactic Plane Survey (MIPSGAL; \citealt{Carey2009}).

\textbf{WISE.} For dustars without any significant contamination or confusion from background emission or nearby sources with $\lesssim10''$, four band mid-IR photometry (3.4, 4.6, 12.1, and 22.0 $\mu$m) from the Wide-field Infrared Survey Explorer (WISE; \citealt{Wright2010}) in the ALLWISE program \citep{Cutri2013} were adopted. WISE W3 (12.1 $\mu$m) photometry is notably valuable since it bridges the wavelength gap between the \textit{Spitzer}/IRAC Ch4 (8.0 $\mu$m) and MIPS 24 $\mu$m measurements.

The \textit{Spitzer}, WISE, and ground-based 2MASS photometry provided in the MIPSGAL 24 $\mu$m point source catalog \citep{Gutermuth2015} were primarily utilized for the non-variable dustars without ISO spectroscopy.
\subsubsection{Ground-based IR Photometry}

Given their brightness in the mid-IR ($\gtrsim$ Jy), WC dustars have been targeted by ground-based observatories in the IR for many decades (e.g. \citealt{Williams1987}). In the near-IR, JHK fluxes from 2MASS \citep{Skrutskie2006} were incorporated in the SED modeling for the following dustars without ISO/SWS coverage: WR48b, WR53, WR59, WR95, WR96, WR103, WR106, WR119, WR121, and WR124-22. Mid-IR L-,M-,N-, and Q-band photometry published by \citet{Williams1987} taken by the ESO 3.6-m and the United Kingdom Infra-Red Telescope (UKIRT) was used for WR103, WR106, and WR119.

For WR137, one of the known periodic dust-makers, the SED is composed of contemporaneous J, H, K, L, M, N, and Q photometry taken in 1985.47 from UKIRT and published by \citet{Williams2001}. Near-contemporaneous JHKLMNQ photometry taken by UKIRT in 1993.5 was also adopted for the SED of the periodic dust-maker WR125 \citep{Williams1994}.

In the unique case of the periodic dust-maker WR140, ground-based mid-IR observations with UKIRT by \citet{Williams2009a} were linked with the space-based spectroscopic observations by \textit{Spitzer}/IRS. Although near-IR photometry of WR140 was not obtained around the same time as the mid-IR UKIRT and IRS observations, JHK measurements were taken during the previous dust formation epoch at an identical orbital phase ($\phi=0.43$; \citealt{Williams2009a}). Importantly, the epoch-to-epoch stability of the near-IR light curve is consistent to within $0.1$ mag.  The semi-contemporaneous and phase-consistent coverage of WR140 from 1.2 - 35 $\mu$m therefore enabled the dust SED analysis of this system.

Table ~\ref{tab:ModelDetails} provides a summary of the archival dataset used for each of the \WRnum~dustars.

\subsubsection{IR Extinction Correction}
Many of the WC dustars in this sample are heavily extinguished ($A_V\gtrsim5$) along the line of sight by the ISM, which is apparent from deep silicate absorption in their IR SEDs at 9.7 $\mu$m (e.g. \citealt{Chiar2001}). This silicate-dominated extinction is interstellar in nature since the circumstellar dust formed by the WC dustars is carbon-rich \citep{Roche1984}. In order to correct for interstellar extinction, the dustar SEDs were dereddened with the ``Local ISM" extinction curve from \citet{Chiar2006} and adopting the visual extinction, $A_v$ ($=1.1 A_\mathrm{V}$), derived by \citet{RC2020}\footnote{$A_v$ is referred to as $A_v^{WR}$ in \citet{RC2020}} except for WR98a due to its large distance and extinction uncertainties. The visual extinction towards WR98a is instead based on the value from \citet{vdh2001}. The $v$ filter ($\lambda_\mathrm{C}=5160\,\AA$) is on the narrow-band system introduced by \citet{Smith1968} specifically to study WR stars. The adopted $A_v$ for each dustar is shown in Tab.~\ref{tab:Adopted}. The selection of the local ISM extinction curve by \citet{Chiar2006} was motivated by their use of WC dustar spectra to define the shape of the 1.25 - 25 $\mu$m extinction. Since several of the dustar spectra extended beyond 25 $\mu$m, a consistent $\lambda^{-2}$ power-law was used in this work to extrapolate the extinction curve.

\section{WC Dustar SED Modeling Results and Analysis}
\label{sec:3}
\subsection{WC Dustar SED Modeling Overview}

The complex ``pinwheel" morphology of the spatially resolved circumstellar dust observed around several well-studied dustars such as WR104 \citep{Tuthill1999,Soulain2018} and WR98a \citep{Monnier1999} presents a complicated scenario for radiative transfer modeling. For example, \citet{Hendrix2016} performed radiative transfer dust models on 3D hydrodynamic simulations fitting the observed emission and morphology of the dust plume around WR98a. Alternatively, \citet{Williams1987} had fit simple analytic dust shell models to dustar SEDs assuming their dust emission is optically thin. \citet{Zubko1998} provided an updated SED analysis using a detailed theoretical model of dust shells that account for grain formation physics and dynamics on a sample of Galactic WC dustars with photometry from \citet{Williams1987}. Although the \citet{Zubko1998} SED models were performed before the circumstellar emission from WC dustars were spatially resolved, the derived dust production rates and carbon condensation fraction provide valuable benchmarks for comparison.

We performed dust SED fits to the \WRnum~WC dustars in our sample with the tool \dustem~\citep{Compiegne2011}
with single or double dust component models. \dustem~is a numerical tool that computes the dust emission in the optically thin limit heated by an input radiation field with no radiative transfer. Single dust component models were attempted for all dustars initially, and double components models were used for SEDs with unsatisfactory single component fits. The two different SED modeling methods are described as follows:

\begin{itemize}
\item Single component dustar models are assumed to have a single geometrically thin dust shell centered on the WC+OB binary.

\item Double component dustars are modeled with two geometrically thin dust ring components with different radii from the central binary utilizing the technique described below in Sec.~\ref{sec:2comp} in greater detail. 
\end{itemize}

\noindent
Note that in the single component model, the morphology of the dust shell (e.g. torus vs. spherical shell) does not affect the dust emission spectrum. 

In addition to circumstellar dust, the IR excess from WC dustars can originate from free-free emission in their ionized winds \citep{Cohen1975}. The IR excess due to free-free emission can be characterized by a power-law $F_{ff}\propto\lambda^{-0.96}$ \citep{Morris1993} and may therefore dominate at shorter IR wavelengths such as the J-band ($\lambda = 1.25$ $\mu$m). In order to remove contamination from free-free emission in the SEDs, we subtract this free-free emission power-law model normalized at the J-band flux for dustars where the J-band flux is $\gtrsim10\%$ of the mid-IR flux peak associated with circumstellar dust. In the unique case of WR140 where the emission from the central system has been spatially resolved from extended circumstellar dust, an interpolation is performed between the IR photometry of the stellar wind continuum measured by \citet{Williams2009a} to characterize and subtract this component. 

\subsection{Two-component Dust Model}
\label{sec:2comp}

\begin{figure}[t!]
    \centerline{\includegraphics[width=1\linewidth]{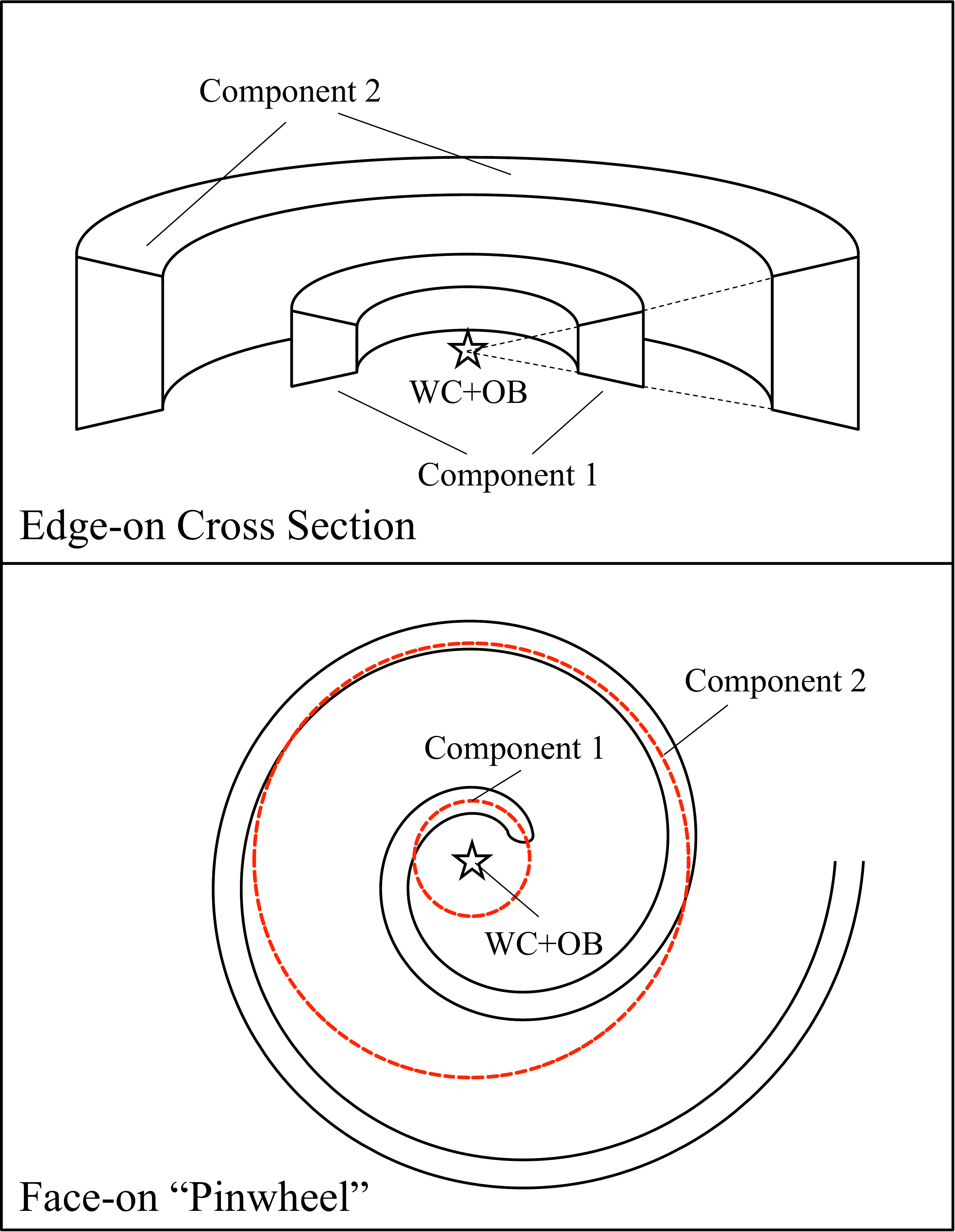}}
    \caption{(\textit{Top}) Two component dust model cross-section schematic (\textit{Bottom}) Face-on ``pinwheel" dust morphology schematic of continuous WC dustars overlaid with the approximated position of the two dust components in this model.}
    \label{fig:2D}
\end{figure}

In this section we describe the framework of the two component dust model. The ``pinwheel" dust morphology from these systems are
approximated as two concentric dust rings. Figure~\ref{fig:2D} presents a schematic illustration of the different components in this model and the comparison to the pinwheel dust, where the first dust component corresponds to the first dust spiral ``coil" and the second component corresponds to the second continuation of this coil. In these models, the inner dust ring attenuates the radiation impinging on the outer ring and is described as follows.  

The incident stellar radiative flux on dust components 1 and 2 at distances $r_1$ and $r_2$ (where $r_1<r_2$) from the central heating source, respectively, can be described as

\begin{equation}
    F_1 = \frac{L_*}{4\pi r^2_1}\,\,\mathrm{and}\,\,F_2 = \frac{L_*e^{-\tau}}{4\pi r^2_2},
    \label{eq:first}
\end{equation}
\noindent
where $\tau$ is the optical depth through component 1. Deriving the radiative heating of component 1 is therefore straightforward since it only requires information of the heating source radiation field and its distance to the heating source. However, the radiative heating of component 2 requires information on the attenuated stellar flux through component 1. This attenuation can be related to the total IR luminosity radiated by component 1, $L_\mathrm{IR,1}$:

\begin{equation}
    L_\mathrm{IR,1} = L_*(1-e^{-\tau})\,\Omega,
    \label{eq:LIR}
\end{equation}

\noindent
where $\Omega$ is the geometric coverage fraction of components 1 and 2 around the central heating source and it is assumed both components have the same coverage fraction (Fig.~\ref{fig:2D}). From Eq.~\ref{eq:LIR}, it follows that $e^{-\tau}=1-\frac{L_\mathrm{IR,1}}{L_*\,\Omega}$, and the incident flux on component 2 can then be expressed as 

\begin{equation}
    F_2 = \frac{1}{4\pi r^2_2} \left( L_*-\frac{L_\mathrm{IR,1}}{\Omega}\right).
    \label{eq:F2}
\end{equation}
\noindent

We assume that 100\% of the stellar flux impinging on components 1 and 2 is absorbed and re-radiated into the IR by the two components. This assumption is motivated by spatially resolved observations that reveal the IR emission from WC dustars are dominated by emitting regions within the first two pinwheel arcs (e.g. \citealt{Tuthill2008}). This implies that

\begin{equation}
    \Omega = \frac{L_\mathrm{IR,1} + L_\mathrm{IR,2}}{L_*},
\end{equation}

\noindent
from which it follows from Eq.~\ref{eq:F2} that $F_2$ can be re-expressed as 

\begin{equation}
    F_2 = \frac{L_*}{4\pi r^2_2}\left(1-\frac{1}{1+f^{-1}}\right),
    \label{eq:final}
\end{equation}
\noindent
where $f\equiv L_\mathrm{IR,1}/L_\mathrm{IR,2}$. With equations~\ref{eq:first} \&~\ref{eq:final}, we have therefore arrived at a simplified 2-component pseudo-radiative transfer model with $L_*$, $f$, $r_1$, and $r_2$ as the model parameters. 

In a test for limit cases, it is apparent that when component 1 absorbs all of the stellar flux ($f\mathrm{\rightarrow }\infty$), $F_2 = 0$. Conversely, when component 1 does not absorb any of the stellar flux ($f\mathrm{\rightarrow }0$), $F_2=\frac{L_*}{4\pi r^2_2}$.

This two component dust model is used for four dustars: WR48a, WR98a, WR104, and WR118.

\subsection{Adopted and Derived SED Model Parameters}

In our \dustem~SED modeling, we input the luminosity and radiation field of the heating source, the distance to the dust components, dust composition, and the dust grain size distribution as model inputs. For the two component models, we also input the IR luminosity ratio of the two dust components, $f$ (see Eq.~\ref{eq:final}). 

We adopt the appropriate radiation field that is representative of the spectral sub-type of each WC dustar from the Potsdam Wolf-Rayet Star (PoWR) model atmospheres (\citealt{Grafener2002,Sander2012,Sander2019}). The radiation field is then scaled by the luminosity of the heating source and the distance to the dust components assuming the incident flux decreases as $F\propto r^{-2}$ (Eq.~\ref{eq:first}). While the radiation field and luminosity are adopted from previous studies, we performed a logarithmic grid search for the dust component distances and (for the two component cases) the IR luminosity ratio of the two dust components. The grid search range and intervals for the different free parameters of each dustar model is provided in Table~\ref{tab:ModelDetails}. Table~\ref{tab:ModelDetails} also lists the observatories from which the IR archival data were obtained for each dustar in the sample.

The dust grain size distribution in WC dustars is still disputed and is one of the issues addressed by this work. Theoretical studies on WC dustars winds by \citet{Zubko1998} predict that dust grains only grow up to sizes of $a\sim0.01$ $\mu$m, which has been corroborated by dust SED analysis of observed IR emission from the dustar WR140 \citep{Williams2009a}. However, IR dust emission and extinction studies of WC dustars by different groups suggest the presence of large $a\gtrsim 0.5$ $\mu$m dust grains (\citealt{Chiar2001,Marchenko2002,Rajagopal2007}).

In order to address this grain size discrepancy, we test both small and large grain size distributions for the WC dustars in our sample that have been spatially resolved. Since heat capacity increases with increasing grain size, large grains need to be at a closer distance to the radiative heating source than small grains in order to exhibit a similar temperature and spectral shape. Therefore, we can compare which grain size distribution in our SED models provides dust shell distances consistent with the observed circumstellar dust morphology constraints. The small (large) grain size distributions includes dust grains ranging from $a = 0.01 - 0.1$ $\mu$m ($0.1-1.0$ $\mu$m) with a number density distribution proportional to $n(a)\propto a^{-3}$. We note that the grain size has a minimal effect on the derived dust mass since this quantity is fit from the longer wavelength IR emission ($\gtrsim10$ $\mu$m) in the SED where the dust opacity is not as sensitive to grain size as the shorter wavelength IR emission (e.g.~\citealt{Harries2004}).

We assume that the dust grains are composed purely of amorphous carbon, which is consistent with their featureless IR spectra and the C-rich environment in the vicinity of the WC star (\citealt{Williams1987,vdh1996}). We adopt the \dustem~ amorphous carbon (``amCBEx") grains with refractive indices derived by \citet{Zubko1996} and \citet{Compiegne2011} and assume a bulk density of $\rho_b=2.0$ g cm$^{-3}$.

By performing a least-squares fit of the \dustem~model to the WC dustar SEDs, we derive the dust mass ($M_d$), dust temperature ($T_d$), and distances between the dust components and central system ($r$). 
For the continuous dust-makers with a dust expansion velocity of $v_\mathrm{exp}$, the dust production rate can then be approximated by 

\begin{equation}
       \mathrm{(continuous)}\,\,\dot{M}_d \sim \frac{M_d \,v_\mathrm{exp}}{r},
       \label{eq:continuous}
\end{equation}
\noindent
where $v_\mathrm{exp}$ is assumed to be comparable to the stellar wind velocity $v_*$ or is derived directly from dust proper motion measurements in multi-epoch imaging.

For the periodic, non-continuous dust-makers with a recurring dust formation timescale $P$ associated with the orbital period of the central binary (e.g.~\citealt{Williams2009a,Monnier2011}), we approximate the dust production rate by

\begin{equation}
       \mathrm{(periodic)}\,\,\dot{M}_d \sim \frac{M_d}{P}.
       \label{eq:periodic}
\end{equation}

Eq.~\ref{eq:periodic} effectively calculates the dust production rate averaged over the orbital period, whereas Eq.~\ref{eq:continuous} is an ``instantaneous" dust production rate. For periodic, non-continuous dust makers, the instantaneous dust production rate calculation will vary depending on the phase the observations. The orbital period-averaged dust production rate calculation (Eq.~\ref{eq:periodic}) is therefore applied to the periodic, non-continuous dustars WR~140, WR~137, WR125, and WR~19, while Eq.~\ref{eq:continuous} is applied to the other 15 dustars. Note that for continuous dust makers, the dust production rate averaged over an orbital period will be the same as the instantaneous dust production rate. In the 2-component dust models, we take a conservative approach and estimate the dust production rates based only on the dust mass and the distance to the 1st dust component since cooler and more extended dust may contribute to the dust mass determined for the 2nd component.

Lastly, the fraction of carbon in the WC winds that condense into dust, $\chi_C$, is estimated based on the dust production rate, the adopted mass-loss rate for the WC wind, and an assumption that the WC wind is composed of $40\%$ carbon by mass \citep{Sander2019}. The adopted WC dustar properties are summarized in Table~\ref{tab:Adopted}.

\begin{figure*}[ht!]
    \includegraphics[width=.38\linewidth]{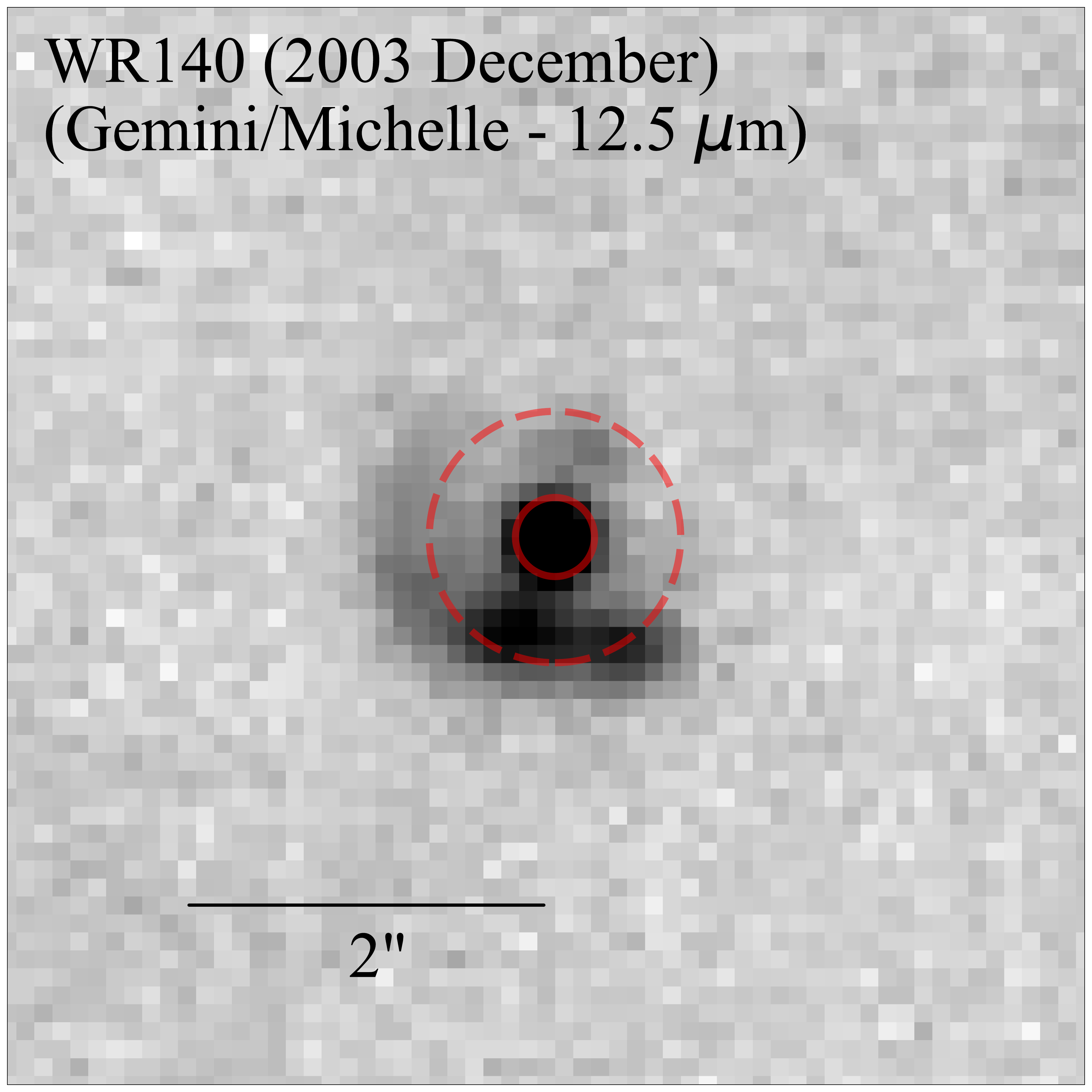}
    \hspace{25pt}
    \includegraphics[width=.53\linewidth]{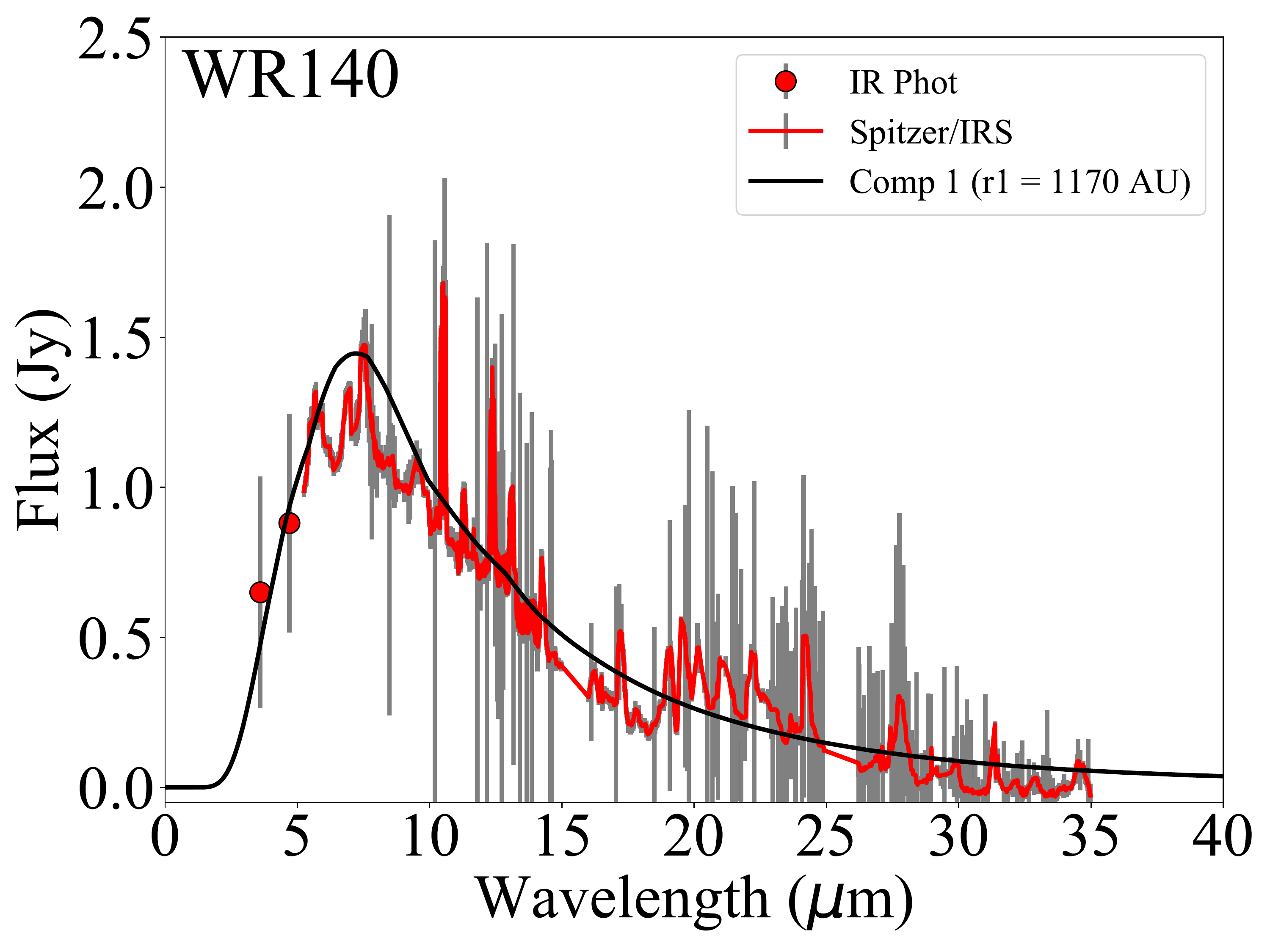}
    \caption{(Left) Mid-IR 12.5 $\mu$m image of WR140 taken by Gemini/Michelle \citep{Williams2009a} in 2003 Dec overlaid with circles centered on WR140 representing the fitted distances of emitting dust for the large (solid) and small (dashed) grain size distribution models. The small grain dust model best matches the observed location of the dust plume. (Right) Best-fit \dustem~SED model of WR140 using the small grain size distribution fit to UKIRT L- and M-band photometry taken in 2004 Jun \citep{Williams2009a} and \textit{Spitzer}/IRS spectroscopy taken in 2004 May. Gray bars correspond to the 1$\sigma$ flux uncertainty.}
    \label{fig:WR140}
\end{figure*}

\subsection{WC Dustar SED Modeling Results}
In this section we present the results of our \dustem~SED models for the \WRnum~WC dustars in our sample. First, as a verification of our modeling approach we highlight the results on the well-studied periodic dust-maker WR140, the continuous ``pinwheel" dust-maker WR104, and the 32.5-yr orbital period dustar WR48a. Then we present and describe the results on the remaining IR-luminous ($L_{IR}/L_*\gtrsim0.1$) dustars, a selection of the single component dustars, and additional systems with angular size constraints. Lastly, we discuss the overall results on dust formation and grain size properties, compare results from previous studies, and test our results against the theoretical model of dust formation in colliding winds by \citet{Usov1991}. The results from the \dustem~SED models are summarized in Table~\ref{tab:Results}.

\subsubsection{The Archetypal Periodic Dust-Maker WR140}

WR140 is a WC7+O5 system with a well-defined 7.94 yr orbital period,  a high orbital eccentricity of $e=0.90$, and a total luminosity of $L_*\sim1\times10^6$ L$_\odot$ \citep{Williams2009a, Monnier2011}. The distance to WR140 measured by Gaia parallax is $d=1640^{+110}_{-90}$ pc \citep{RC2020}, which is consistent with previous distance estimates \citep{Dougherty2005,Monnier2011}. 

Dust formation in WR140 only occurs when the binary system is at periapse. The proper motion of the resulting dust ``arc'' formed during periapse has been characterized by multi-epoch, spatially resolved mid-IR imaging as the arc propagates away from the central binary (Fig.~\ref{fig:WR140}, Left; \citealt{Williams2009a}). These images allows us to verify our SED model-derived separation distances between the dust component and central heating source. For the line-of-sight extinction we adopt $A_v=2.21$ \citep{RC2020}.

We perform single component \dustem~models to the stellar wind-subtracted 5 - 37 $\mu$m \textit{Spitzer}/IRS spectrum taken on 2004 May, UKIRT L- and M-band imaging taken on 2004 Jun, and JHK imaging taken on 1996 Aug where WR140 was at a nearly identical orbital phase ($\phi = 0.43$) as the UKIRT and \textit{Spitzer} observations (Fig.~\ref{fig:WR140}, Right). This orbital phase corresponds to 3.4 yr after IR maximum and periastron passage. The small (large) dust grain distribution SED fit provides a separation distance between the dust component and central heating source of $r = 1170^{+310}_{-340}$ AU ($360^{+160}_{-100}$ AU). Spatially resolved mid-IR imaging observations of WR140 at this orbital phase show that the extended dust arc is 700 - 900 mas from the central source \citep{Williams2009a}, which is closely consistent with the separation distance derived from our small grain dust distribution model, $1170$ AU $\approx 700$ mas (Fig.~\ref{fig:WR140}, Left). This small grain distribution is also consistent with the dust emission analysis by \citet{Williams2009a} who deduced a characteristic grain size of 0.01 $\mu$m. 

From the small grain dust SED, we derive a total dust mass of $M_d=(6.44^{+3.84}_{-3.30})\times 10^{-9}$ M$_\odot$, which implies a dust production rate of $\dot{M}_d=(8.1^{+4.8}_{-4.2})\times10^{-10}$ M$_\odot$ yr$^{-1}$ (Eq.~\ref{eq:periodic}) given the periodic 7.94 yr dust formation timescales in WR140. SED fits from ground-based observations by \citet{Williams2009a} indicate a total dust mass of less than $2\times10^{-8}$ M$_\odot$ at a phase of 0.56 around WR140, which consistently constrains our derived dust mass.  Adopting a total mass-loss rate of $\dot{M}\approx2\times10^{-5}$ M$_\odot$ yr$^{-1}$ derived by X-ray monitoring observations of shocked gas in the wind collision region between the WC and O stars \citep{Sugawara2015} and assuming a 40\% carbon composition by mass, the carbon dust condensation fraction is $\chi_C\approx 0.01\%$.

\begin{figure*}[t!]
    \includegraphics[width=.4\linewidth]{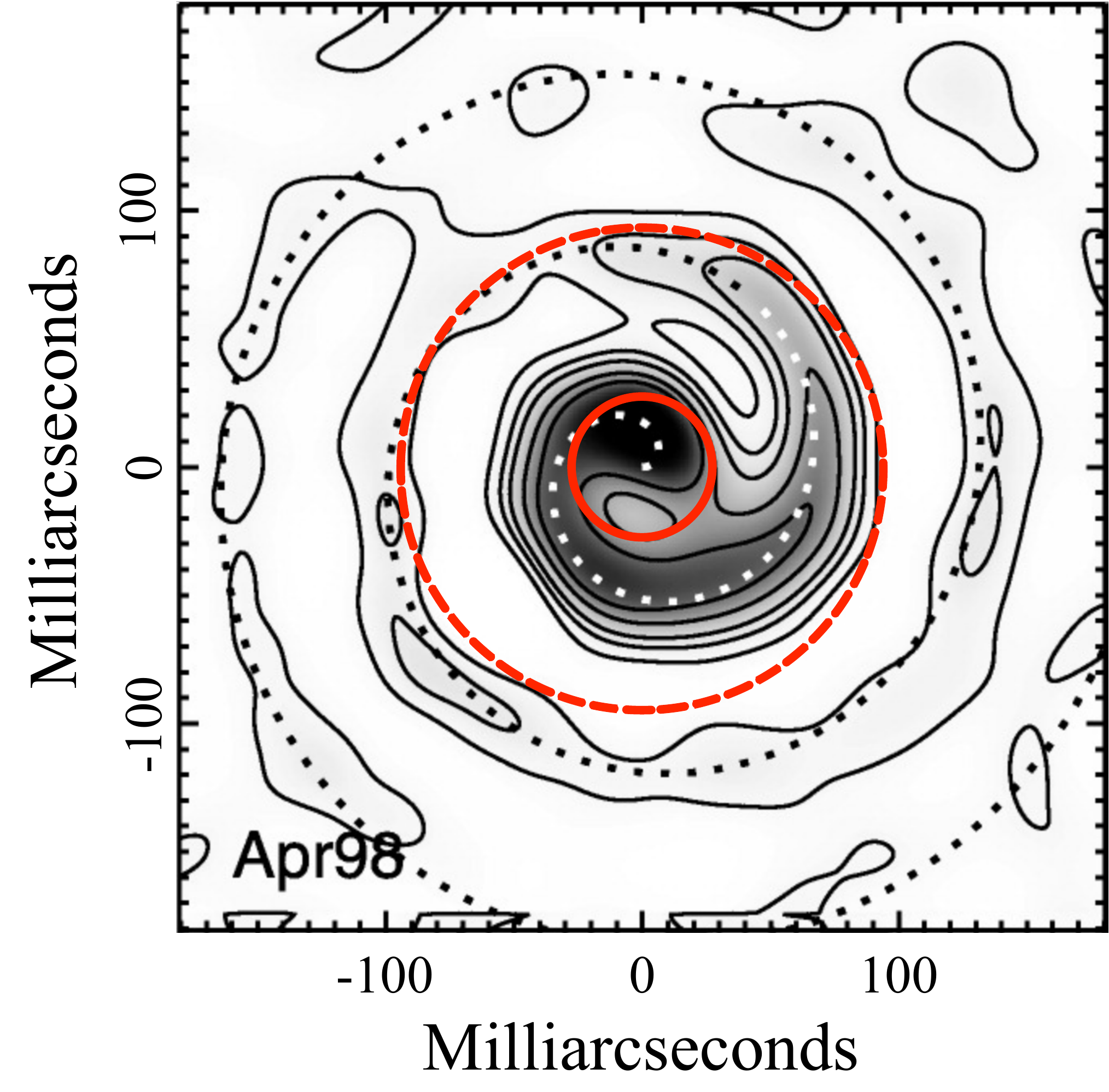}
    \hspace{20pt}
    \includegraphics[width=.54\linewidth]{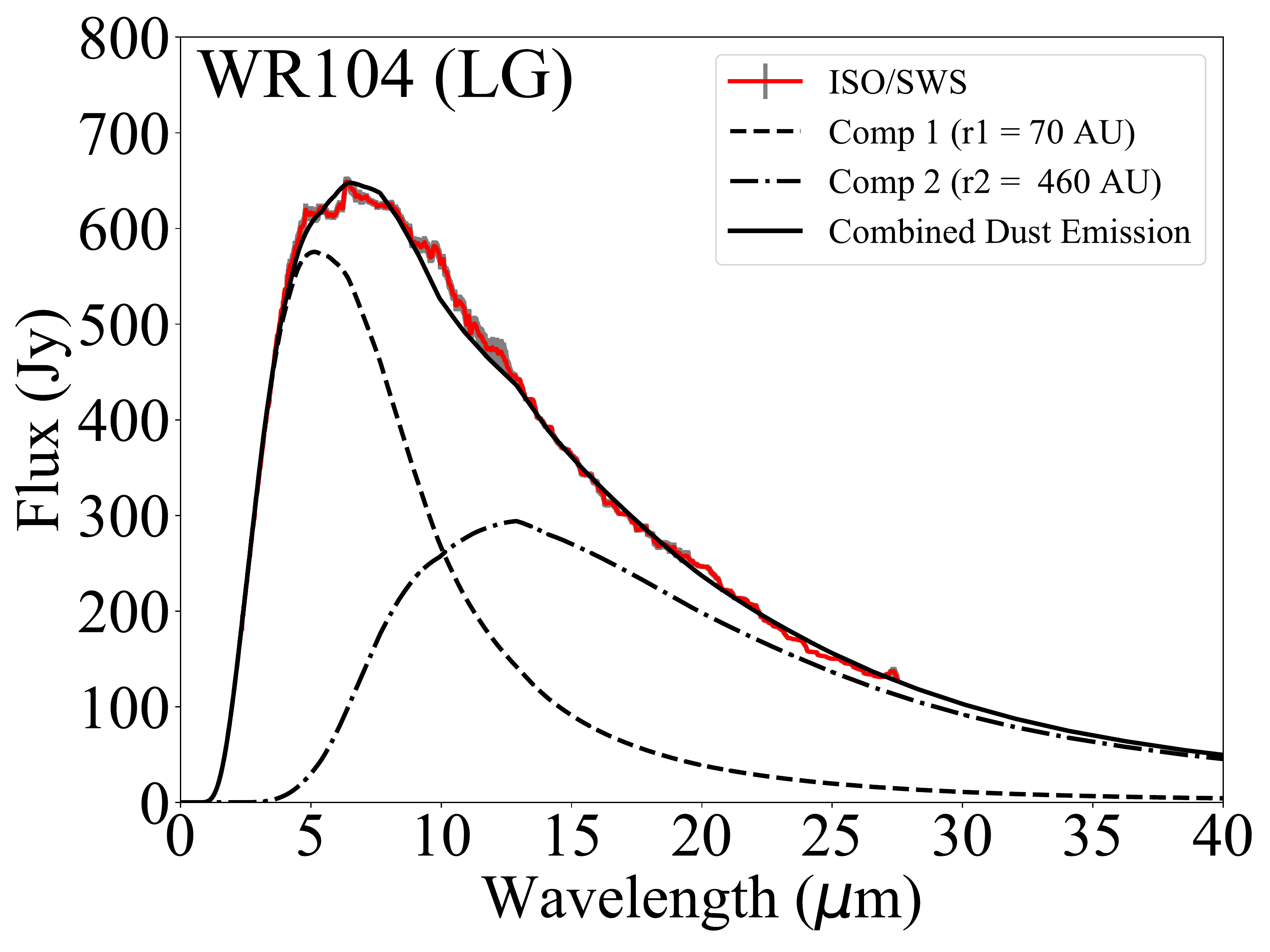}
    \caption{(Left) Near-IR reconstructed interferometric image of WR104 taken with the CH4 ($\lambda_c = 2.27$ $\mu$m) filter on the Keck I Telescope reproduced from Figure 1 in \citet{Tuthill2008} centered on the origin of their best-fit Archimedean spiral model (dotted line) and overlaid with red circles representing the location of the fitted dust component 1 for the large (solid) and small (dashed) grain models. Contour levels are 0.4$\%$, 1$\%$, 2$\%$, 5$\%$, 10$\%$, and 50$\%$ of the peak emission. The large grain dust model best matches the observed location of the centralized peak dust emission. (Right) Best-fit 2-component \dustem~SED model of WR104 using the large grain size distribution fit to ISO/SWS spectroscopy. rev{Gray bars correspond to the 1$\sigma$ flux uncertainty.}}
    \label{fig:WR104}
\end{figure*}

\subsubsection{The Continuous ``Pinwheel" Dust-Maker WR104}

 WR104 is the prototypical, continuous dust-maker that exhibits a nearly face-on pinwheel (Fig.~\ref{fig:WR104}, Left) with an orbital period of 242 days \cite{Tuthill1999,Tuthill2008}. The central binary is composed of a WC9+OB star and has a Gaia-derived distance of $d=2740^{+720}_{-550}$ pc. This is consistent with the well-defined distance determined from the location of its possible stellar association host, Sgr OB1, and observations of the proper motion of its circumstellar dust ($d = 2580 \pm120$ pc; \citealt{Soulain2018}). Given the tighter constraints from their high spatial resolution imaging, 
 we adopt the \citet{Soulain2018} distance for the dust SED model. The line-of-sight extinction towards WR104 is $A_v=6.67$ \citep{RC2020}. We adopt a total system luminosity of
 $L_*=2.5\times10^{5}$ L$_\odot$, consistent with \citet{Monnier2007} and the mean WC9 luminosity derived by \citet{Sander2019}.

Since single dust component models fail to fit the ISO/SWS spectrum of WR104 between 2 - 27.5 $\mu$m, we apply the double component SED models (Fig.~\ref{fig:WR104}, Right) and derive separation distances between the central binary and dust components of $r_1=240^{+60}_{-80}$ AU ($70^{+20}_{-10}$ AU) and $r_2 = 1300^{+1130}_{-740}$ AU ($460^{+110}_{-90}$ AU) for the small (large) dust distribution models. High spatial resolution IR imaging by \citet{Tuthill2008} show that the distance between the central binary and the onset of the spiral dust plume where the IR emission peaks is $\sim15$ mas (40 AU). Figure~\ref{fig:WR104} (Left) shows the approximate location of the first dust component for small and large grain models compared to the observed dust morphology. Given the closer agreement of the first dust component, we favor the large grain distribution for WR104. 

From our large grain dust SED model, we derive dust masses of $M_{d1}=(1.19^{+0.70}_{-0.44})\times 10^{-6}$ M$_\odot$ and $M_{d2} = (6.06^{+3.09}_{-2.06}) \times 10^{-5}$ M$_\odot$ for the two dust components. Adopting a dust expansion velocity of $v_\mathrm{exp}=1220$ km s$^{-1}$ \citep{Howarth1992}, the dust production rate based on component 1 is $\dot{M}_d= (4.39^{+1.27}_{-0.97})\times10^{-6}$ M$_\odot$ yr$^{-1}$. 

This dust production rate is a factor of $\sim5$ higher than the $\dot{M}_d = 8\pm1\times10^{-7}$ M$_\odot$ yr$^{-1}$ value derived by \citet{Harries2004} from 3D dust radiative transfer models. However, we can reconcile this discrepancy based on the following points:
\begin{itemize}

\item \citet{Harries2004} adopt a distance to WR104 of 1600 pc as opposed to our value of 2580 pc, which accounts for a factor of $(2580/1600)^{2}\approx 2.6$ higher for our derived dust mass. 

\item The large grain size distribution we adopt vs. the small grain distribution used by \citet{Harries2004} implies a closer/shorter distance of the dust component to the central binary and thus a higher dust production rate (see Eq.~\ref{eq:continuous}).

\end{itemize}

\noindent
When applying a small grain dust model and adopting a distance of 1600 pc we arrive at a consistent dust production rate of $\dot{M}_d\approx6\times10^{-7}$ M$_\odot$ yr$^{-1}$.

The dust production rate from our large grain SED model shows that WR104 is producing the most dust out of the all the dustars in our sample and is likely the highest dust maker amongst all known WC dustars (Tab.~\ref{tab:Results}). Assuming a WC mass-loss rate of $3\times10^{-5}$ M$_\odot$ yr$^{-1}$ \citep{Crowther1997,Harries2004} and a $40\%$ carbon mass fraction, we infer a carbon dust condensation fraction of $\chi_C\approx40\%$. This carbon dust condensation efficiency is notably higher than the $\sim$few percent derived by \citet{Harries2004} and implies that a substantial fraction of the available carbon in the WC wind forms dust.

\subsubsection{The Longest Period (32.5-yr) Dustar WR48a}
\label{sec:WR48a}

\begin{figure}[t!]
    \includegraphics[width=1\linewidth]{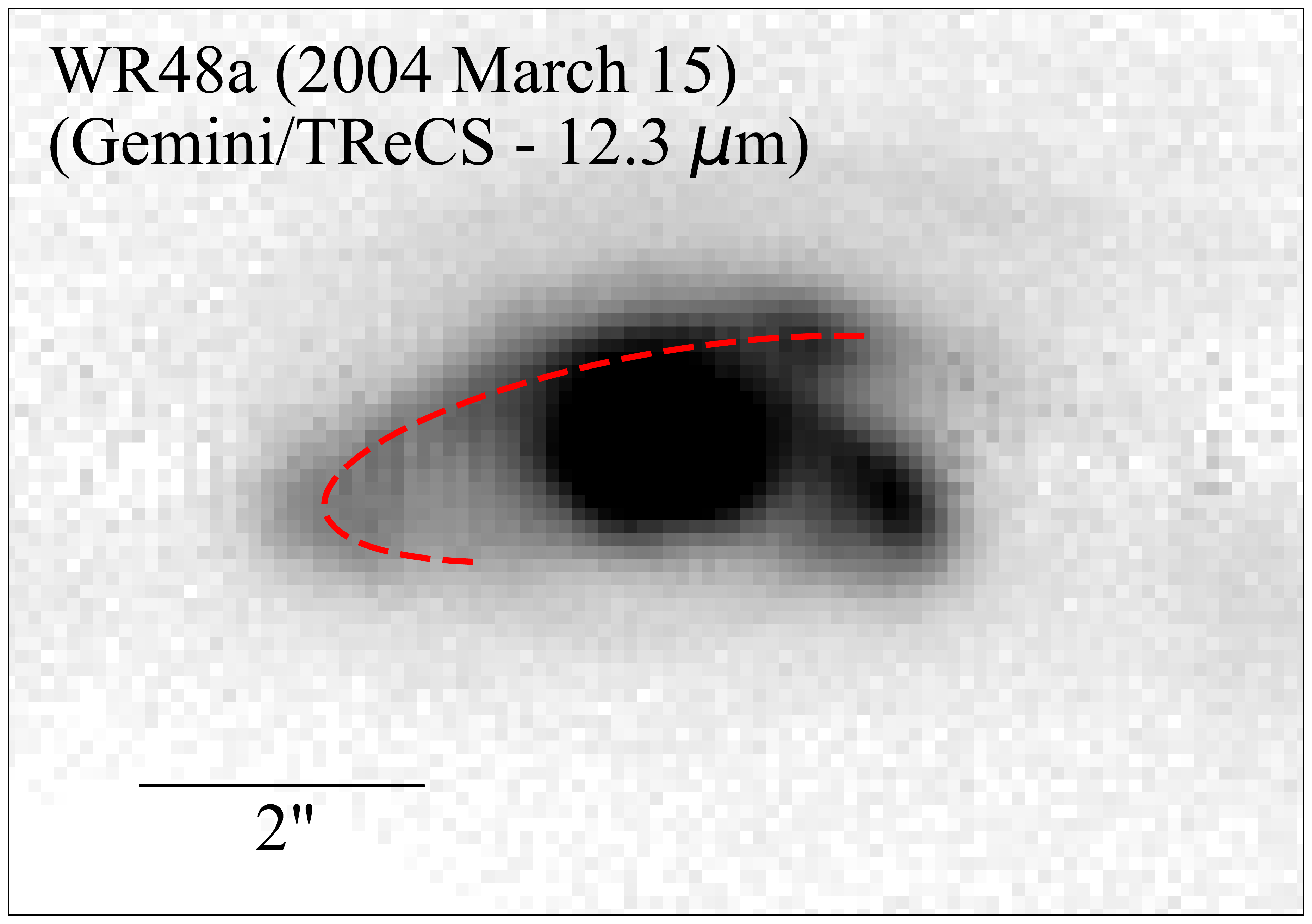} 
    \includegraphics[width=1\linewidth]{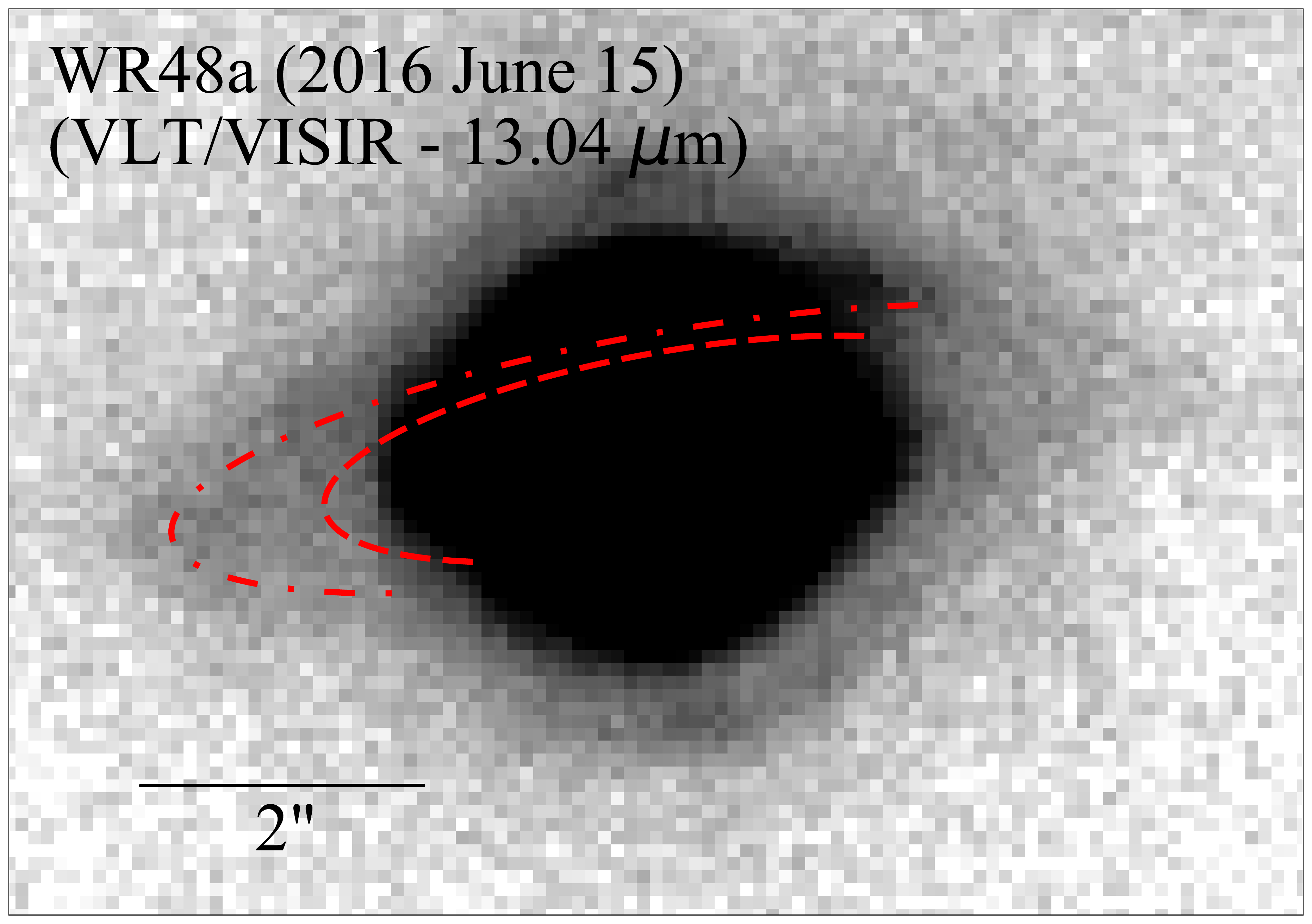}
    \caption{(Top) Mid-IR 12.3 $\mu$m image of WR48a taken by Gemini/TReCS on 2004 Mar 15 overlaid with a dashed arc showing the location of a prominent dust arc plume. (Bottom) 13.04 $\mu$m image of WR48a taken by VLT/VISIR on 2016 Jun 15 overlaid with the 2004 dust arc position (dashed) as above and its new position (dot-dashed) at the date of the VISIR observation. }
    \label{fig:WR48a}
\end{figure}
WR48a is a continuous but variable dust-maker hosting a WC8 star with a proposed O-star companion. WR48a is the longest period dustar known with an IR light curve-derived period of 32.5 yr and also exhibits episodes of brief dust-formation \citep{Williams2012}. The total luminosity of the system is $L_*\sim4\times10^5$ L$_\odot$, and its Gaia-derived distance of 
$d=2270^{+920}_{-570}$ pc is slightly less than previous distance estimates based on its suggested association with galactic clusters Danks 1 and 2 ($\sim3.3-4.6$ kpc, \citealt{Danks1984}). However, \citet{vdh2001} determine a distance of 1200 pc based on its WC8 spectral type. We therefore adopt the Gaia distance for our dust models since it falls between the two distance estimates. We also adopt a line-of-sight interstellar extinction towards WR48a of $A_v=8.28$ \citep{RC2020}.

In Figure~\ref{fig:WR48a}, we present mid-IR imaging observations taken by Gemini-South/TReCS (\citealt{Marchenko2007}; PID - GS-2004A-Q-63, PI - A.~Moffat) and VLT/VISIR at similar N-band wavelengths at 2004 March 15 and 2016 June 15, respectively. The overlaid arc segment in Fig.~\ref{fig:WR48a} shows the $\sim1.1''$ proper motion of eastern dust arc over the 12.3 years separating the two observations. Given the low signal-to-noise ratio of this dust arc in the VISIR images and its extended nature we assume position uncertainty of 0.23'', which is $0.5\times$ the PSF FWHM ($0.45''$). Adopting the Gaia distance of $2270$ pc, the dust expansion velocity of WR48a is
$v_\mathrm{exp}=1000^{+390}_{-240}\pm200$
km s$^{-1}$, where the first and second uncertainties are from the distance and proper motion, respectively. The derived expansion velocity is consistent with the $1200\pm170$ km s$^{-1}$ wind velocity measured for WR48a from its 10830-$\AA$ emission line width \citep{Williams2012}.

\begin{figure}[t!]
    \centerline{\includegraphics[width=1\linewidth]{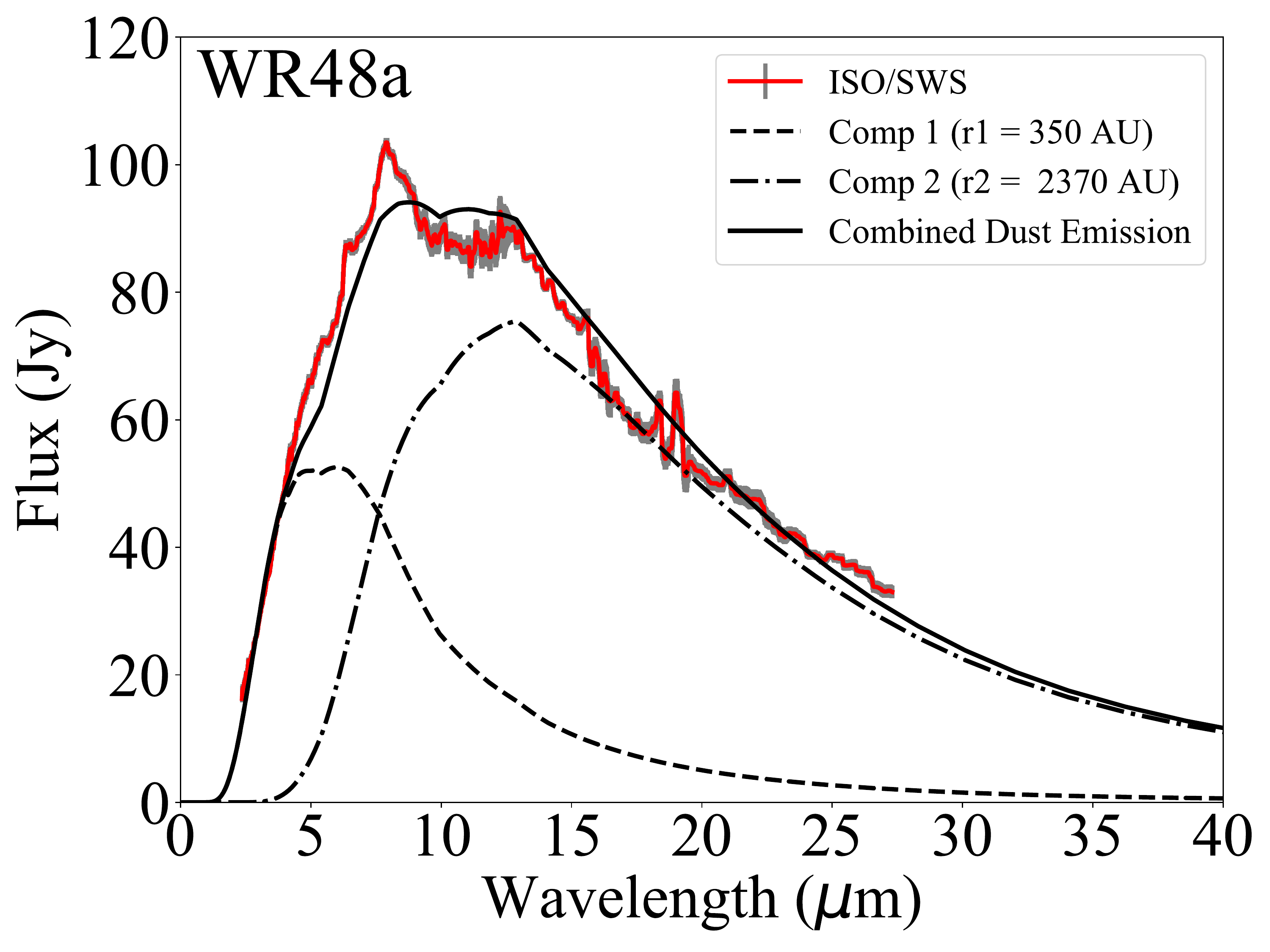}}
    \caption{Best-fit two-component \dustem~model of WR48a using the small grain size distribution fit to ISO/SWS spectroscopy taken in 1996 Feb.}
    \label{fig:48aSED}
\end{figure}

We apply the 2-component dust SED model to fit the ISO/SWS spectrum of WR48a between 2 - 27.5 $\mu$m (Fig.~\ref{fig:48aSED}). From the 2-component model, we derive separation distances between the central system and dust components of $r_1=350^{+100}_{-130}$ AU ($110^{+70}_{-40}$ AU) and $r_2 = 2370^{+640}_{-900}$ AU ($720^{+440}_{-150}$ AU) for the small (large) dust distribution models. Historic mid-IR light curves show an IR brightening consistent with enhanced dust formation around 1994.5 \citep{Williams2012}. Using the estimated dust expansion velocity of 1000 km s$^{-1}$, this dust formed in 1994.5 would be located at $\sim340$ AU from the central binary at time of the ISO/SWS observation (1996 Feb). This distance estimate is consistent with the small grain distribution, which we therefore infer for the dust composition.

Based on our small grain distribution SED model (Fig.~\ref{fig:48aSED}), we derive dust masses for components 1 and 2 of $M_{d1}=(1.42^{+1.12}_{-0.99})\times10^{-7}$ M$_\odot$ and $M_{d2} = (9.82^{+5.84}_{-5.93}) \times 10^{-6}$ M$_\odot$. Using the component 1 distance and mass and a dust expansion velocity of $v_\mathrm{exp}=1000$ km s$^{-1}$, we estimate a dust production rate of $\dot{M}_d=(8.46^{+3.48}_{-4.38})\times10^{-8}$ M$_\odot$ yr$^{-1}$. The carbon dust condensation fraction is therefore $\chi_C \approx0.1\%$ assuming the radio-derived mass-loss rate of $1.7\times10^{-4}$ M$_\odot$ yr$^{-1}$ \citep{Zhekov2014} and a $40\%$ carbon mass composition.

\subsubsection{IR Luminous Dustars}
\label{sec:OTD}

\textbf{\textit{WR98a.}} WR98a is a WC8 or WC9+OB dustar exhibiting continuous dust formation and appears like a ``rotating pinwheel" that ``revolves" with a 1.54 period \citep{Monnier1999}. The Gaia distance towards WR98a derived by \citet{RC2020} is $d=1260^{+870}_{-410}$ pc. Despite its close distance, WR98a is buried behind over 10 magnitudes of visual extinction from the ISM along the line-of-sight based on a deep silicate absorption feature \citep{Chiar2006}.
\citet{RC2020} derive a much lower visual extinction of $A_v = 1.25$, but assign it with a `b' flag that denotes it as highly implausible. Additionally, they note that the Gaia parallax measurement of WR98a exhibits large astrometric noise ($> 1$ mas).

Due to the uncertainties in the \citet{RC2020} interstellar extinction and distance measurements, we adopt $A_v = 13.79$ from \citet{vdh2001} and a distance of $d=1900$ pc derived by \citet{Monnier1999} that is consistent with the upper distance uncertainties from Gaia. \citet{Monnier1999} derive this distance to WR98a based on its dust proper motion and adopting an expansion velocity of 900 km s$^{-1}$ \citep{Williams1995}.
The stellar luminosity of WR98a is not well-characterized given the high extinction. We therefore adopt a system luminosity of $L_*=1.5 \times10^{5}$ L$_\odot$ that is consistent with the value adopted in the WR98a models by \citet{Hendrix2016}. Given its lower wind velocity, WC9 stellar properties are assumed for WR98a.

\begin{figure*}[t!]
    \includegraphics[width=.46\linewidth]{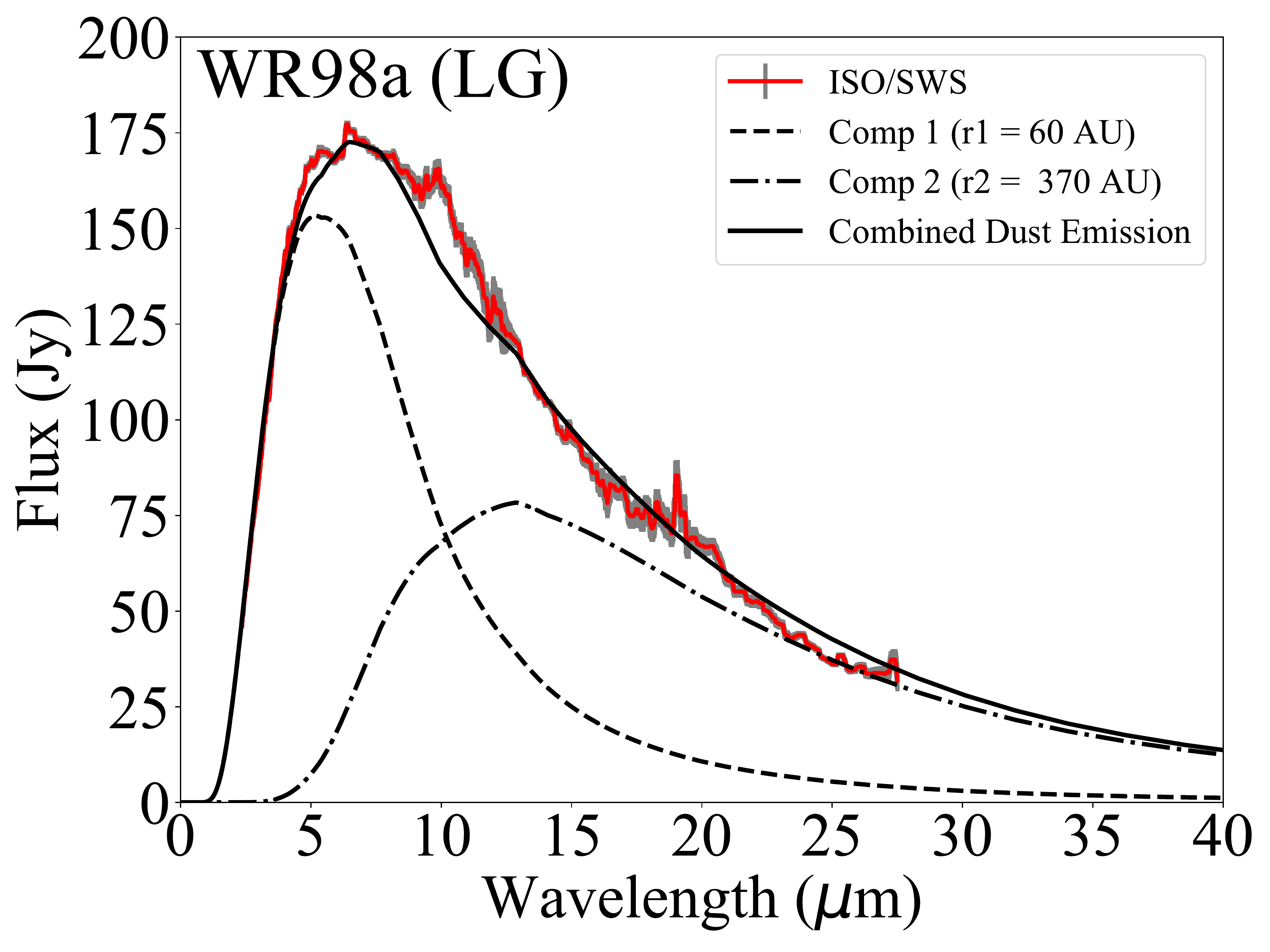} \hspace{10pt}
    \includegraphics[width=.46\linewidth]{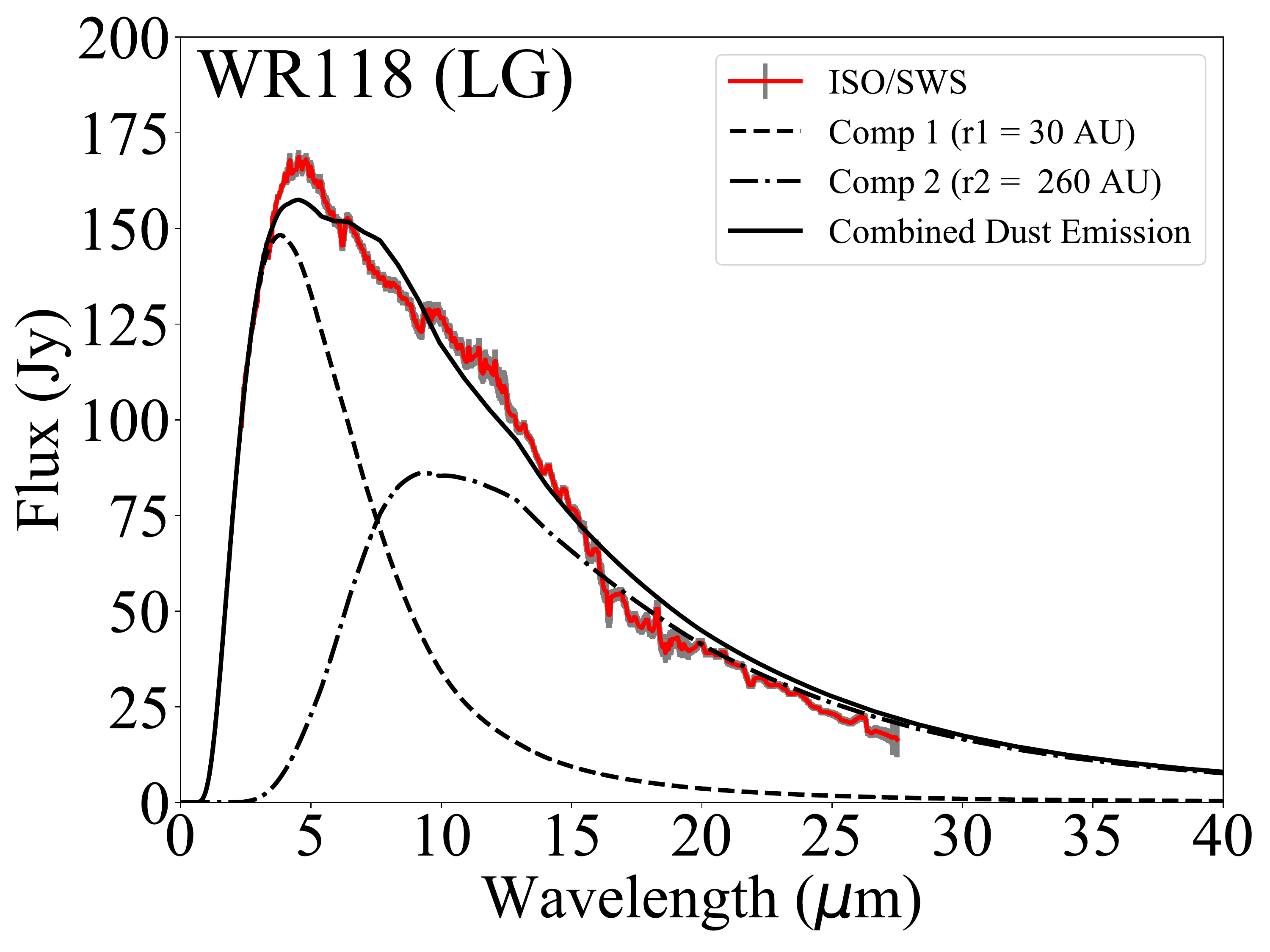}\\
    \includegraphics[width=.46\linewidth]{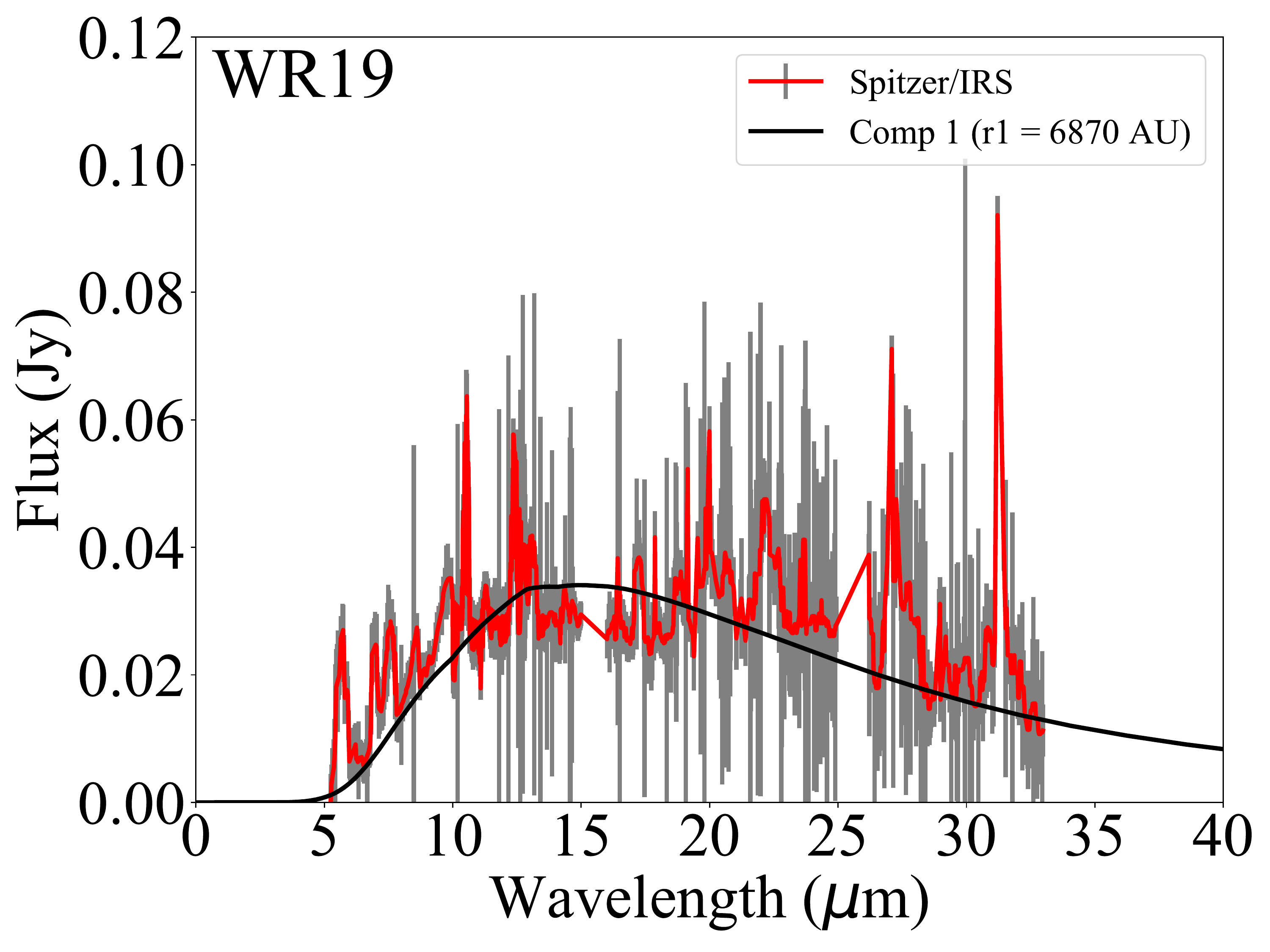} \hspace{10pt}
    \includegraphics[width=.46\linewidth]{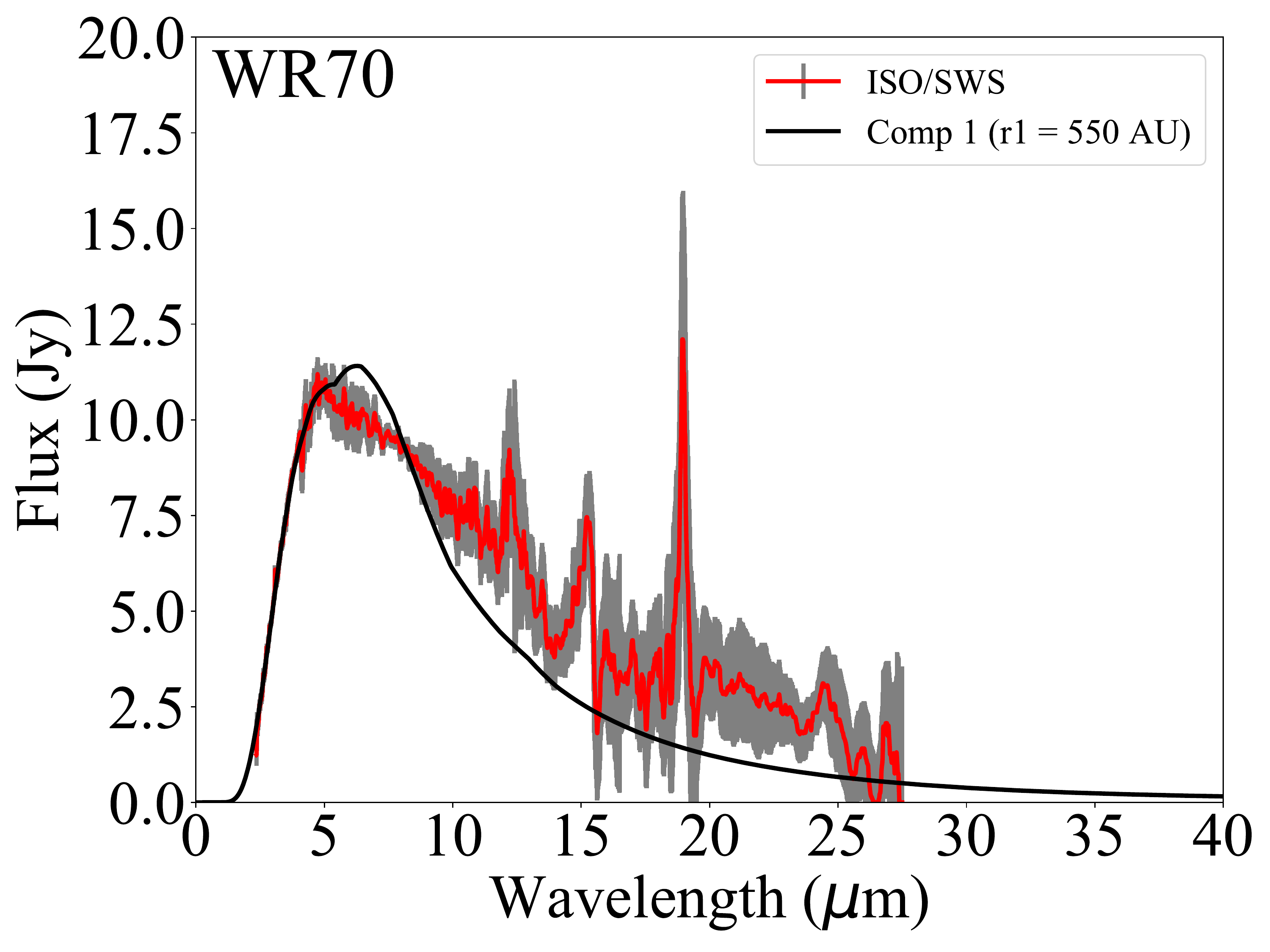}\\
    \includegraphics[width=.46\linewidth]{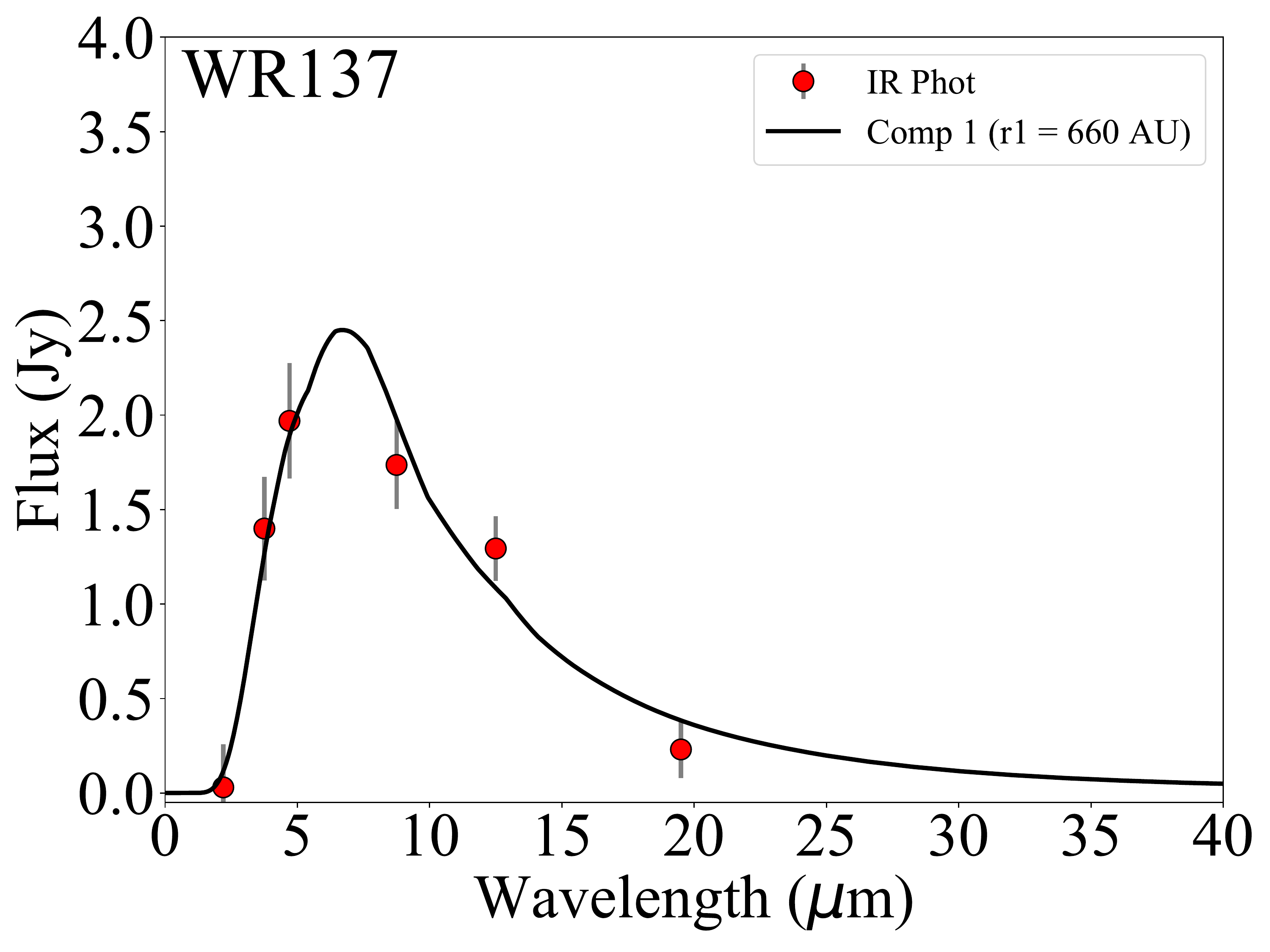} \hspace{10pt}
    \includegraphics[width=.46\linewidth]{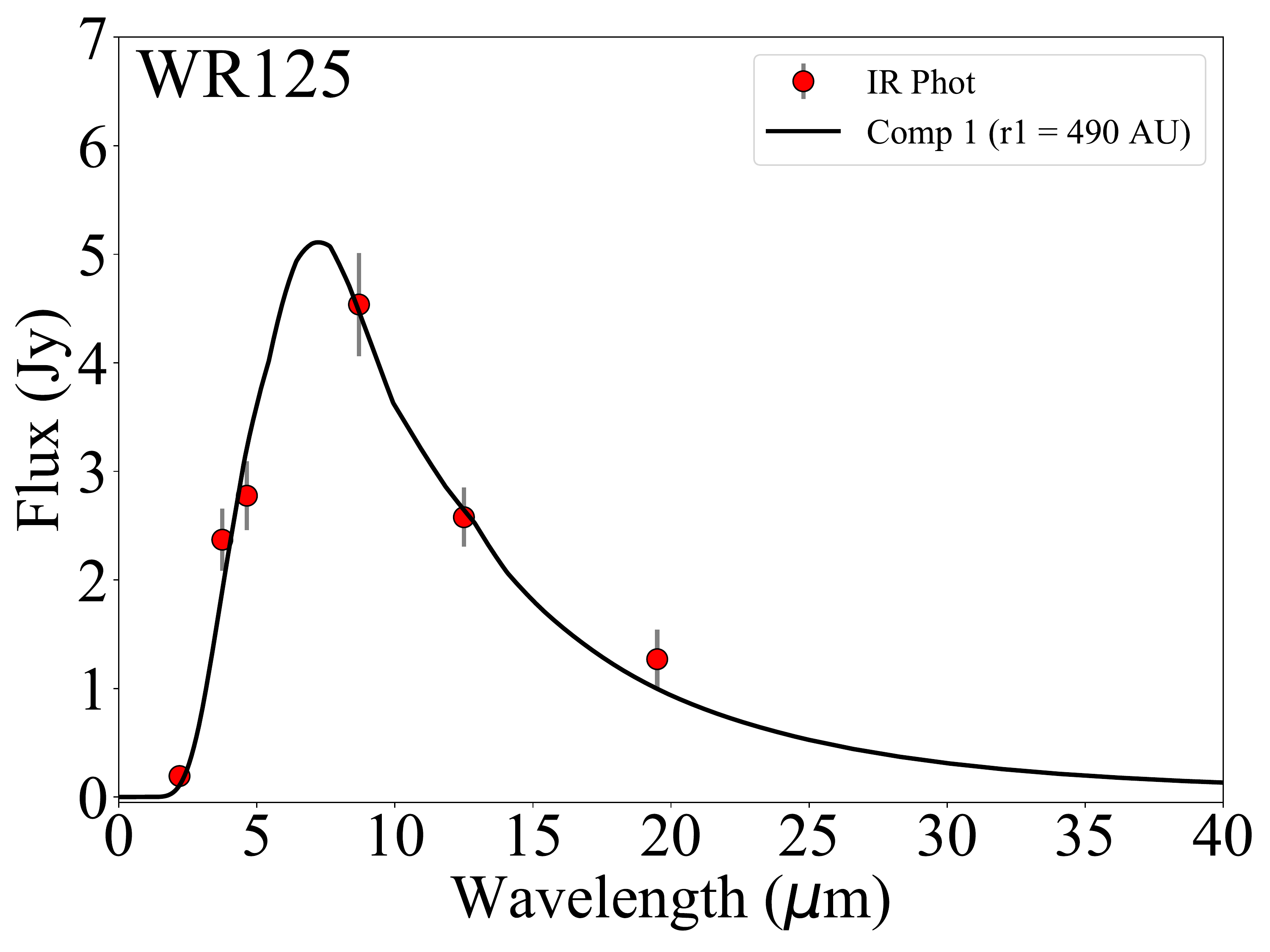}
    \caption{Best-fit \dustem~models of WR98a, WR118, WR19, WR70, WR137, and WR125. A large grain size distribution 2-component dust model was used for the WR98a and WR118, and a small grain size distribution 1-component dust model was used for the other four dustars.}
    \label{fig:TextSEDs}
\end{figure*}

We fit double component dust SED models to the ISO/SWS spectrum of WR98a between 2 - 27.5 $\mu$m (Fig.~\ref{fig:TextSEDs}, Top Left) and derive dust component distances of $r_1 = 200^{+50}_{-70}$ AU ($60^{+10}_{-10}$ AU) and $r_2 = 860^{+1120}_{-400}$ AU ($370^{+190}_{-130}$ AU)
for the small (large) grain size distribution models. Spatially resolved images of the circumstellar dust around WR98a demonstrate that the majority of the mid-IR dust emission is concentrated in the central $\sim70$ mas, or $\sim100$ AU, of the central binary \citep{Monnier1999, Monnier2007} with no prominent extended emission beyond $\sim150$ mas, or $\sim200$ AU. Similar to WR104, we prefer the large grain size distribution for WR98a. Notably, based on the dust component distances derived from the large grain SED model and adopting a dust expansion velocity of 900 km s$^{-1}$ \citep{Williams1995}, the expansion timescale between the two components is $\sim1.6$ yr, which is consistent with the $1.54$ yr orbital period. The separation between the two dust components is therefore consistent with the distance between the dust spirals in WR98a. 

From the large grain dust SED model, we derive dust masses for components 1 and 2 of $M_{d1}=(1.82^{+1.08}_{-0.68})\times10^{-7}$ M$_\odot$ and $M_{d2}=(9.25^{+11.73}_{-5.29})\times10^{-6}$ M$_\odot$. Using the distance and mass of component 1, the dust production rate is $\dot{M}_d=(6.10^{+1.77}_{-1.38})\times10^{-7}$ M$_\odot$ yr$^{-1}$. Adopting the mean WC9 star mass-loss rate of $\dot{M}=10^{-4.66}$ M$_\odot$ yr$^{-1}$ and a carbon mass fraction of $40\%$ \citep{Sander2019} for WR98a, we estimate a carbon dust condensation fraction of $\chi_C\approx7\%$.

We note that there is a discrepancy between our derived dust production rate and the value determined from the hydrodynamic simulations by \citet{Hendrix2016}, who obtain a dust production rate of $\sim10^{-8}$ M$_\odot$ yr$^{-1}$. This discrepancy is primarily due to their assumption of a fixed value of $0.2\%$ for the dust condensation fraction of the total WC mass loss rate.

\textbf{\textit{WR118.}} WR118 is a continuous WC9 dust-maker with a possible pinwheel morphology \citep{Millour2009}. The Gaia distance towards WR118 determined by \citet{RC2020} is $d=2490^{+780}_{-680}$ pc, which exhibited astrometric noise in excess $>1$ mas and large parallax uncertainty. However, this distance is consistent with previous estimates based on photometry and spectral type ($d = 3130$ pc, \citealt{vdh2001}). We therefore adopt the Gaia distance for the dust SED models. WR118 is also heavily reddened by interstellar extinction, where $A_v=13.68$ \citep{RC2020}. We adopt a combined stellar luminosity of $L_* = 2.5\times10^{5}$ L$_\odot$, consistent with the mean WC9 luminosity derived by \citet{Sander2019}.

We fit double component dust SED models to the ISO/SWS spectrum of WR118 between 2 - 27.5 $\mu$m (Fig.~\ref{fig:TextSEDs}, Top Right) and derive separation distances for components 1 and 2 of $r_1 = 120^{+30}_{-50}$ AU ($30^{+10}_{-10}$ AU) and $r_2=680^{+320}_{-300}$ AU ($260^{+120}_{-80}$ AU) for the small (large) grain size distributions. High spatial resolution interferometric IR observations of WR118 indicate that dust is located with $\sim15$ mas ($\sim40$ AU) of the central binary \citep{Yudin2001,Monnier2007,Millour2009}, which is comparable to size of component 1 in the large grain model. We therefore favor the large grain models for WR118, which is also consistent with the large grain composition trend for the other IR luminous WC9 dustars in our sample. 

From the large grain SED models of WR118, we determine dust masses of $M_{d1}=(5.58^{+6.24}_{-2.95})\times10^{-8}$ M$_\odot$ and $M_{d2}=(6.61^{+6.99}_{-3.42})\times10^{-6}$ M$_\odot$ for the components 1 and 2. Adopting an expansion velocity consistent with WC9 stars of $v_\mathrm{exp}=1390$ km s$^{-1}$, a WC9 mass-loss rate of $\dot{M}=10^{-4.66}$ M$_\odot$ yr$^{-1}$ \citep{Sander2019}, and a $40\%$ carbon fraction by mass, we find that the dust production rate and carbon dust condensation fraction is $\dot{M}_d=(6.27^{+2.78}_{-1.94})\times10^{-7}$ M$_\odot$ yr$^{-1}$ and $\chi_C\approx 7\%$, respectively.

\subsubsection{Selected Single Component Dustars}

\textbf{\textit{WR19.}} WR19 is a periodic WC5+O9 dustar with an orbital period of 10.1 yr and an eccentricity of $e=0.8$ \citep{Williams2009b}. This system hosts one of the earliest type WC stars amongst the known WC dustars. The Gaia distance to WR19 is $d=4330^{+780}_{-580}$ pc \citep{RC2020}, which is slightly larger than previous distance estimates based on optical photometry and spectral type ($d\sim3300$ pc, \citealt{vdh2001}). However, since Gaia observations of WR19 were taken before the onset of dust formation in 2017, it exhibited a mostly dust-free environment which likely contributed to the well-constrained distance. We therefore adopt the Gaia-derived distance for our dust SED modeling. WR19 is also affected by line-of-sight interstellar extinction, where $A_v = 5.19$ \citep{RC2020}. We adopt a combined stellar luminosity of $L_* = 4\times10^5$ L$_\odot$, the mean luminosity for WC5 stars as derived by \citet{Sander2019}.

We fit single component dust SED models to the stellar wind-subtracted 5-37 $\mu$m \textit{Spitzer}/IRS spectrum of WR19 (Fig.~\ref{fig:TextSEDs}, Center Left) and derive a dust separation distance from the central binary of $r = 6870^{+3130}_{-3630}$ AU ($2680^{+2030}_{-1530}$ AU) for the small (large) grain size models. Based on the orbital phase of WR19 when \textit{Spitzer} observed it (2005 Jun, $\phi\approx0.84$), we can estimate the extent of dust formed at periastron passage in early 1997 \citep{Williams2009b}. Adopting the mean WC5 wind velocity from \citet{Sander2019} of $v_\mathrm{exp}=2780$ km s$^{-1}$ for the dust expansion velocity, the dust formed in the 1997 periastron passage should be located $\sim5000$ AU away in 2005 Jun. We therefore favor the small grain composition given the consistency with the dust expansion distance estimate.

For the small grain SED models of WR19, we derive a dust mass of $M_d = (3.98^{+3.76}_{-2.84}) \times10^{-8}$ M$_\odot$, which implies a dust production rate of $\dot{M}_d = (3.94^{+3.72}_{-2.81}) \times10^{-9}$ M$_\odot$ yr$^{-1}$ (Eq.~\ref{eq:periodic}) given its orbital period of 10.1 yr. Adopting the mean WC5 mass-loss rate of $\dot{M}=10^{-4.39}$ M$_\odot$ yr$^{-1}$ \citep{Sander2019} and a $40\%$ carbon mass fraction, the carbon dust condensation fraction is $\chi_C\approx0.02\%$.

\textbf{\textit{WR70.}} WR70 is a continuous but variable WC9+B0I dustar with a proposed period of 2.8 yr derived from its IR light curve  \citep{Niemela1995,Williams2013a}. The Gaia-derived distance to WR70 is $d=3010^{+440}_{-340}$ pc, which is consistent with the recently revised extinction-based distance estimate of $d\approx3500$ pc \citep{Williams2013a}. We therefore adopt the Gaia distance for our dust SED models. The adopted line-of-sight extinction and combined stellar luminosity of the WR70 system is 
$A_v = 5.20$ and $7\times10^{5}$ L$_\odot$, respectively \citep{RC2020,Williams2013a}.

We fit single component dust SED models to the stellar wind-subtracted 2-27.5 $\mu$m ISO/SWS spectrum of WR70 (Fig.~\ref{fig:TextSEDs}, Center Right) and determine a separation distance between dust and the central binary of $r=550^{+80}_{-70}$ AU ($150^{+80}_{-50}$ AU) for the small (large) grain size models.
Extended emission around WR70 has not yet been resolved, so it is difficult to distinguish between a small or large grain size distribution. However, we choose to adopt the small grain size distribution to be consistent with the WC dust formation analysis by \citet{Zubko1998}. 

The small grain SED models of WR70 provide a dust mass of $M_d=(6.60^{+2.40}_{-1.78})\times10^{-8}$ M$_\odot$. Adopting the mean WC9 mass-loss rate and velocity of $\dot{M}=10^{-4.66}$ M$_\odot$ yr$^{-1}$ and $v_\mathrm{exp}=1390$ km s$^{-1}$ \citep{Sander2019}, we derive a dust production rate and carbon dust condensation fraction of $\dot{M}_d=(3.51^{+0.65}_{-0.56})\times10^{-8}$ M$_\odot$ yr$^{-1}$ and $\chi_C \approx 0.4\%$, where we have assumed the winds are composed of $40\%$ carbon by mass.

\begin{figure*}[t!]
    \includegraphics[width=.46\linewidth]{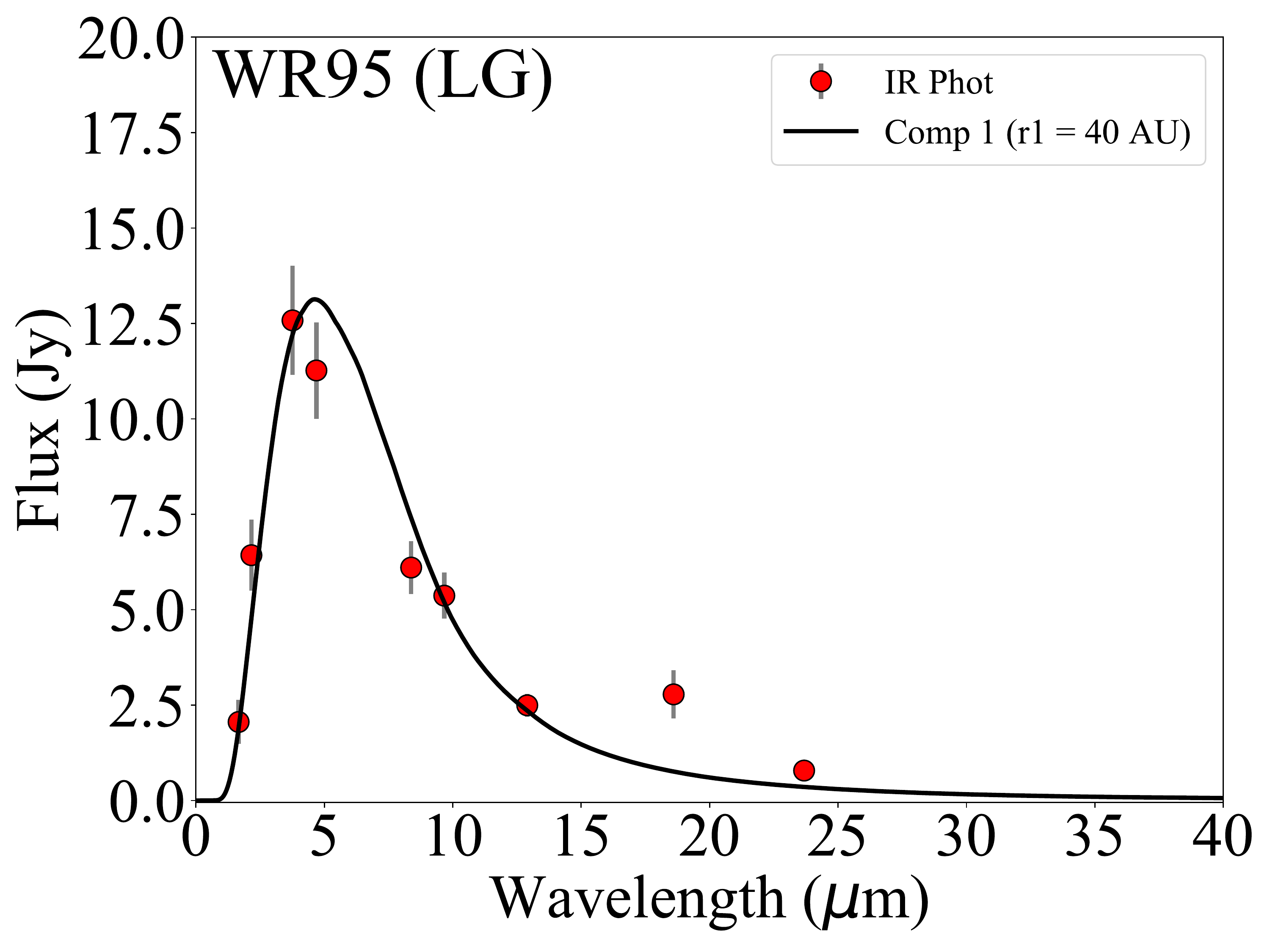} \hspace{10pt}
    \includegraphics[width=.46\linewidth]{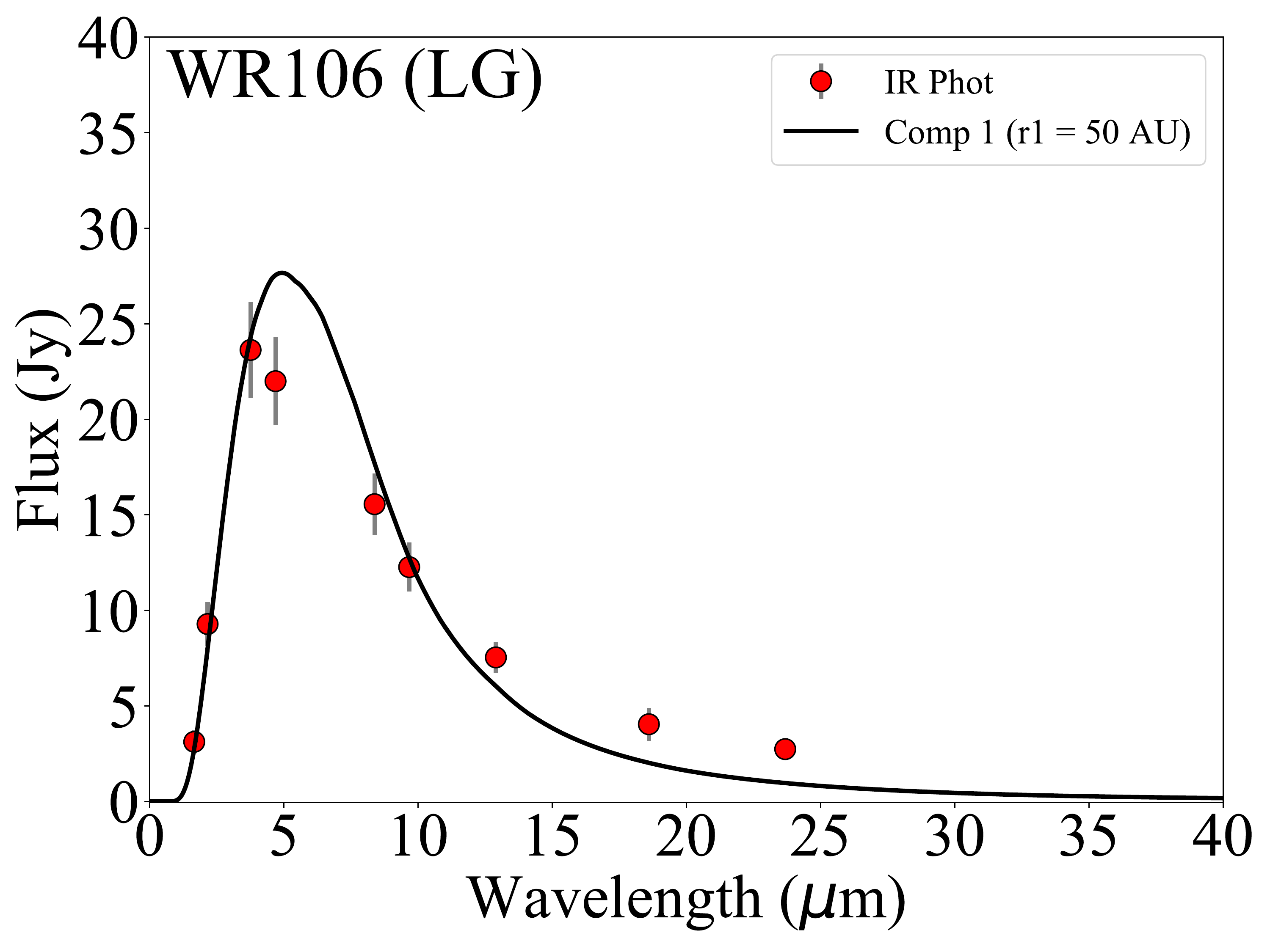}\\
    \caption{Best-fit \dustem~models of WR95 and WR106 with the large grain size distribution.}
    \label{fig:WR95106}
\end{figure*}

\textbf{\textit{WR137.}}
WR137 is a periodic WC7+O9 dustar with an orbital period of 13.05 yr derived from both radial velocity measurements and IR light curve variations \citep{Williams2001,Lefevre2005}. The Gaia-derived distance of $d=2100^{+180}_{-160}$ pc is consistent with previous distance estimates towards the system (e.g. $d\approx1820$ pc; \citealt{Nugis2000}). We therefore adopt the Gaia distances for our dust SED models of WR137. We also adopt an interstellar extinction and heating source luminosity of $A_v=1.70$ \citep{RC2020} and L$_*=5.4\times10^5$ L$_\odot$, respectively, where the stellar luminosity was derived by a spectral pseudo fit from the PoWR models by \citet{Sander2012}.

We fit single component dust SED models to the stellar wind-subtracted contemporaneous 1.2 - 20 $\mu$m photometry taken in 1985 May by \citet{Williams2001} (Fig.~\ref{fig:TextSEDs}, Bottom Left) and derive a separation distance between dust and the central binary of $r = 660^{+240}_{-220}$ AU ($210^{+110}_{-70}$  AU) for small (large) grain size models. The mid-IR emission from dust formation in WR137 peaked in mid-1984 \citep{Williams2001}; therefore, assuming the expansion velocity of 2000 km s$^{-1}$ \citep{Sander2012}, we expect the dust to be located $\sim400$ AU at the time of the WR137 observations. Since dust formation from WR137 started several years before the mid-1984 peak and may therefore have propagated further, we suggest that the dust is most likely composed of small grains. We therefore favor the small grain model. This is also supported by WR137's similarity in both orbital properties and spectral sub-type to WR140, which exhibits dust consistent with small grains.

Based on the small grain SED models and the 13.05 yr dust formation/orbital period of WR137, we derive a dust mass of M$_d=(1.20^{+0.64}_{-0.57})\times10^{-8}$ M$_\odot$ and a dust production rate of $\dot{M}_d=(9.21^{+4.90}_{-4.36})\times10^{-10}$ M$_\odot$ yr$^{-1}$ (Eq.~\ref{eq:periodic}). Adopting the mean WC7 mass-loss of $\dot{M}=2.7\times10^{-5}$ M$_\odot$ yr$^{-1}$ and a $40\%$ carbon composition by mass, the carbon dust condensation fraction is $\chi_C= 0.009\%$.

\textbf{WR125.} WR125 is a WC7+O9III system \citep{Williams1994,Midooka2019} that recently exhibited a second observed IR re-brightening implying a 28.3 yr period \citep{Williams2019}. The Gaia-derived distance towards WR125 is $3360^{+990}_{-650}$ pc \citep{RC2020}, which is closer than the previous distance estimate of 4700 pc \citep{Williams1992} based on a near-IR flux comparison to the similar system WR140 and its well-defined distance. Given the distance uncertainties arising from a relative comparison to WR140, we use the Gaia distance for our dust SED models of WR125. The adopted line-of sight extinction and combined stellar luminosity of WR125 is $A_v = 6.48$ \citep{RC2020} and L$_*=1.6\times10^5$ L$_\odot$, respectively, where the stellar luminosity was derived by a spectral pseudo fit from the PoWR models by \citet{Sander2012}.

We fit single component dust SED models to the stellar wind-subtracted 1.2 - 20 $\mu$m photometry of WR125 taken in 1993.47-1993.5 (Fig.~\ref{fig:TextSEDs}, Bottom Right), several years after the onset of observed dust formation between 1990 - 91 \citep{Williams1994}, and determine a dust separation distance between the central binary of $r = 490^{+170}_{-100}$ AU ($170^{+70}_{-40}$ AU) assuming small (large) grains. Assuming a dust expansion velocity of $2000$ km s$^{-1}$, consistent with the mean WC7 wind velocity derived by \citet{Sander2019}, the dynamical timescale of the modeled dust is 1.2 yr (0.4 yr) for the small (large) dust grains distributions. Since the small grain model timescale is consistent with formation during the observed peak IR emission in mid 1992 \citep{Williams1994}, we favor the small grain dust model.

For the small grain SED models, we derive a dust mass of M$_d=(1.00^{+0.61}_{-0.35})\times10^{-7}$ M$_\odot$ and a dust production rate of $\dot{M}_d=(3.54^{+2.15}_{-1.22})\times10^{-9}$ M$_\odot$ yr$^{-1}$ (Eq.~\ref{eq:periodic}) assuming an orbital period of 28.3 yr. Our derived dust mass is roughly consistent with the M$_d=1.64\times10^{-7}$ M$_\odot$ derived from the isothermal dust model by \citet{Williams1994} after correcting for different distance assumptions. We adopt a mass-loss rate of $1.6\times10^5$ M$_\odot$ yr$^{-1}$ from the spectral model fit by \citet{Sander2012} and a 40$\%$ carbon mass fraction and determine a carbon dust condensation fraction of $\chi_C\approx0.03\%$.

\begin{figure*}[t!]
    \includegraphics[width=.48\linewidth]{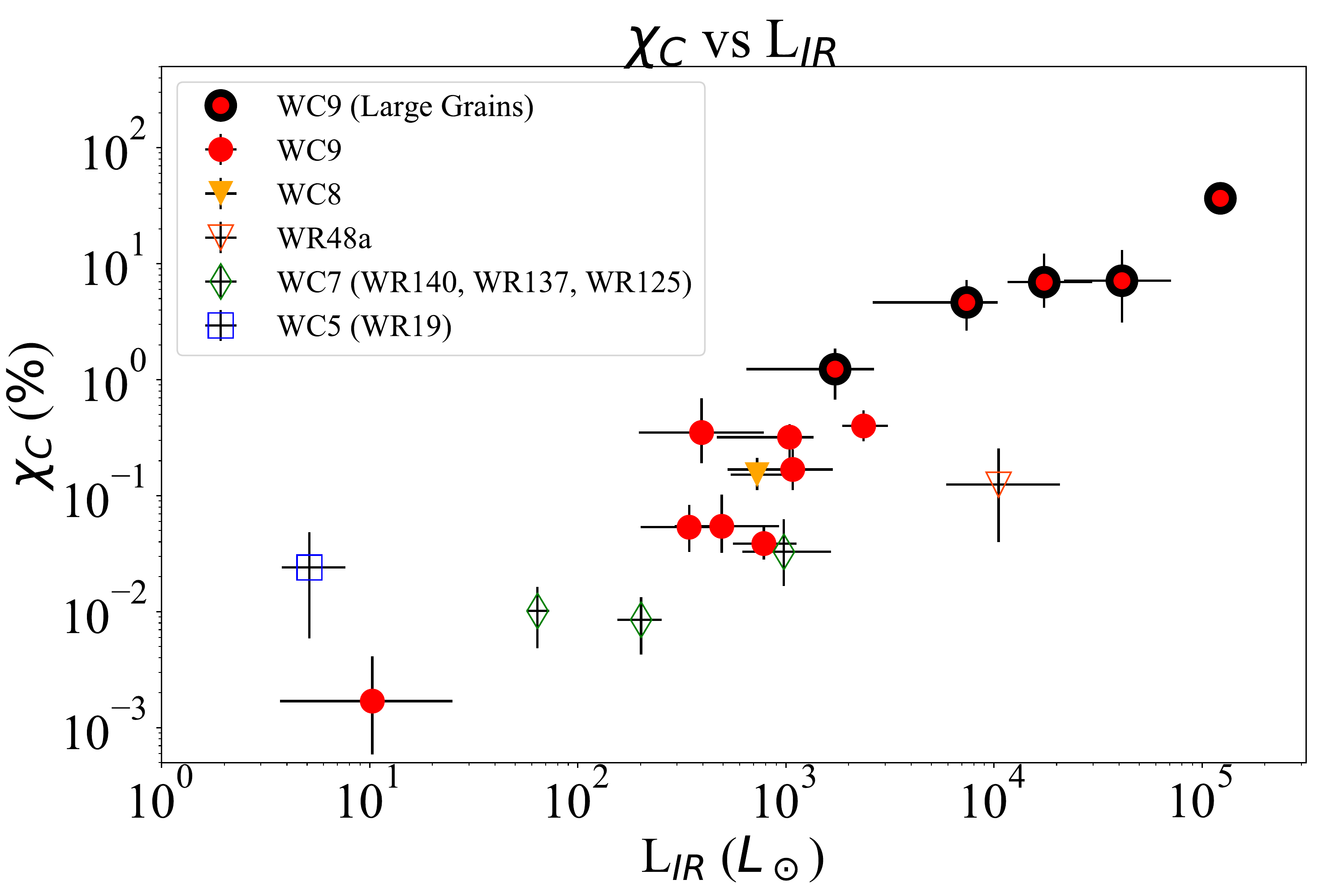} \hspace{10pt}
    \includegraphics[width=.48\linewidth]{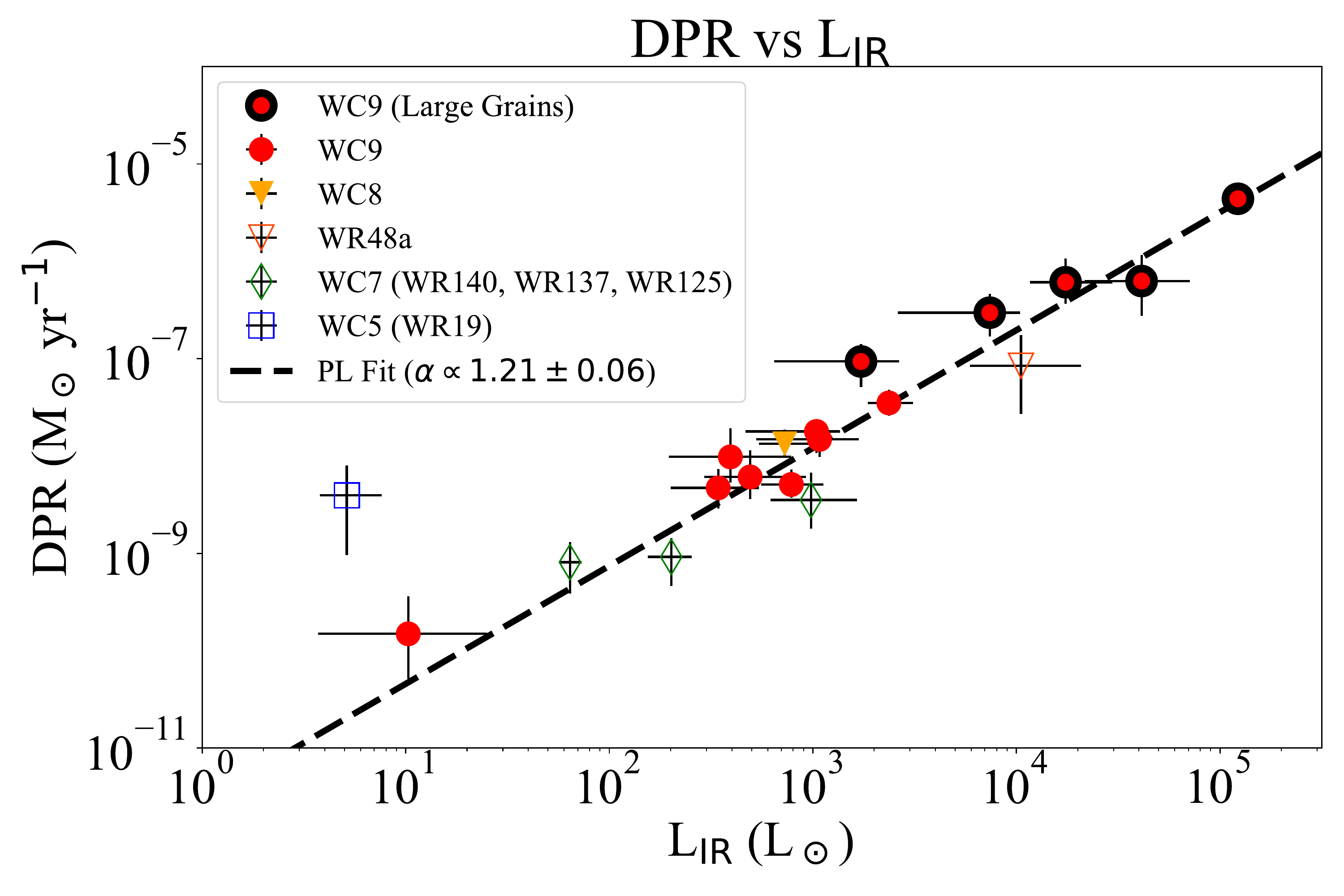}
    \caption{(Left) Distribution of derived dustar values for $\chi_C$ vs $L_{IR}$. The markers with thick border represent the dustars where the large grain model was favored, markers with thin borders represent dustars where the small grain model was favored, and markers without borders show the dustars where the small grain model was adopted by default. Open, unfilled markers show the highly variable or periodic dustars WR19, WR48a, WR125, WR137, and WR140. (Right) Distribution of DPR vs $L_{IR}$ overlaid with the best-fit power-law relation $DPR\propto L_\mathrm{IR}^{1.21\pm0.06}$. The highly variable or periodic dustars (WR19, WR48a, WR125, WR137, and WR140) were excluded from the fit since.}
    \label{fig:Results}
\end{figure*}

\subsubsection{Additional WC Dustars with Angular Size Constraints}
WR95 and WR106 are continuous WC9 dustars \citep{vdh2001} that have constraints on their extended dust emission from IR high spatial resolution observations \citep{Monnier2007,Rajagopal2007}. The Gaia-derived distances towards WR95 and WR106 are $2070^{+430}_{-310}$ pc and $3070^{+560}_{-430}$ pc, respectively \citep{RC2020}. The dust separation distances determined from our SED models are $r = 130^{+110}_{-40}$ AU ($40^{+70}_{-20}$ AU) for WR95 and $r = 170^{+220}_{-60}$ AU ($50^{+120}_{-20}$ AU) for WR106 assuming small (large) grains.

\citet{Rajagopal2007} measure a 28.4 and 45.0 mas Gaussian FWHM at 10.5 $\mu$m for WR95 and WR106, respectively, which is consistent with the FWHM measured by \citet{Monnier2007} for these systems at shorter IR wavelengths (2.2 and 3.08 $\mu$m) near the emission peaks in the SEDs (Fig.~\ref{fig:WR95106}). These angular sizes correspond to a linear size of 60 AU for WR 95 and 100 AU for WR106. Assuming a circular morphology, the extent of the dust shells should be $\sim$half of these linear sizes: $\sim30$ AU for WR 95 and $\sim50$ AU for WR106. This is consistent with our large grain size models, which agrees with the interpretation from \citet{Rajagopal2007} who also favor dust models with larger ($\sim1$ $\mu$m) grains for WR95 and WR106 and determine dust shell radii of a few tens of AU from the stars.

Assuming the large grain SED models, the dust production rates are $\dot{M}_d=(9.36^{+2.00}_{-3.36})\times10^{-8}$ M$_\odot$ yr$^{-1}$ and $\dot{M}_d=(2.97^{+1.14}_{-1.01})\times 10^{-7}$ M$_\odot$ yr$^{-1}$ for WR95 and WR106, respectively. Given the adopted mass-loss rates from PoWR models by \citet{Sander2012} (Tab.~\ref{tab:Adopted}) and assuming a $40\%$ carbon composition in the winds by mass, the carbon dust condensation fraction is $\sim1.2\%$ for WR95 and $\sim4.6\%$ for WR106.

\begin{figure*}[t!]
    \includegraphics[width=.46\linewidth]{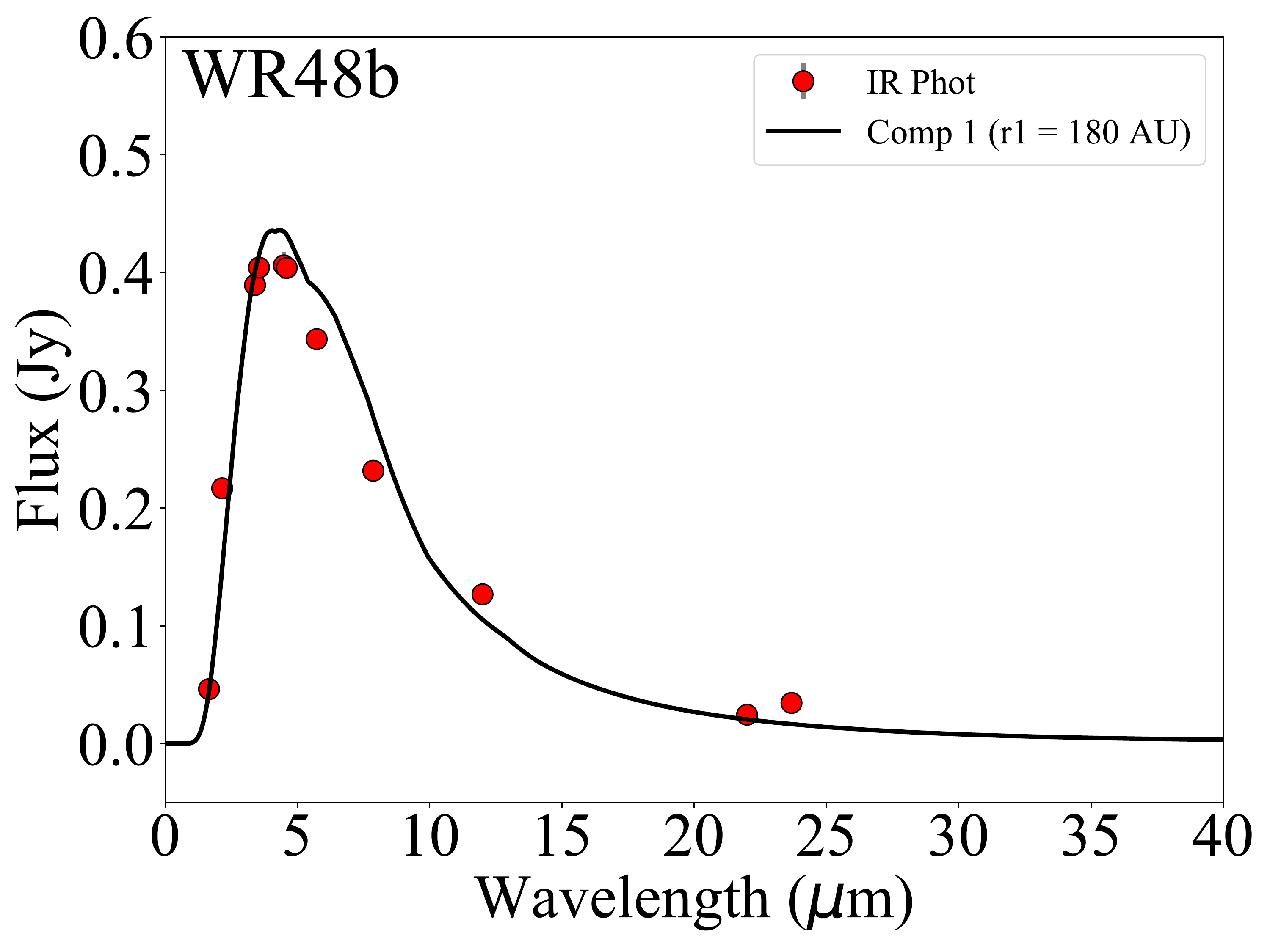} \hspace{10pt}
    \includegraphics[width=.46\linewidth]{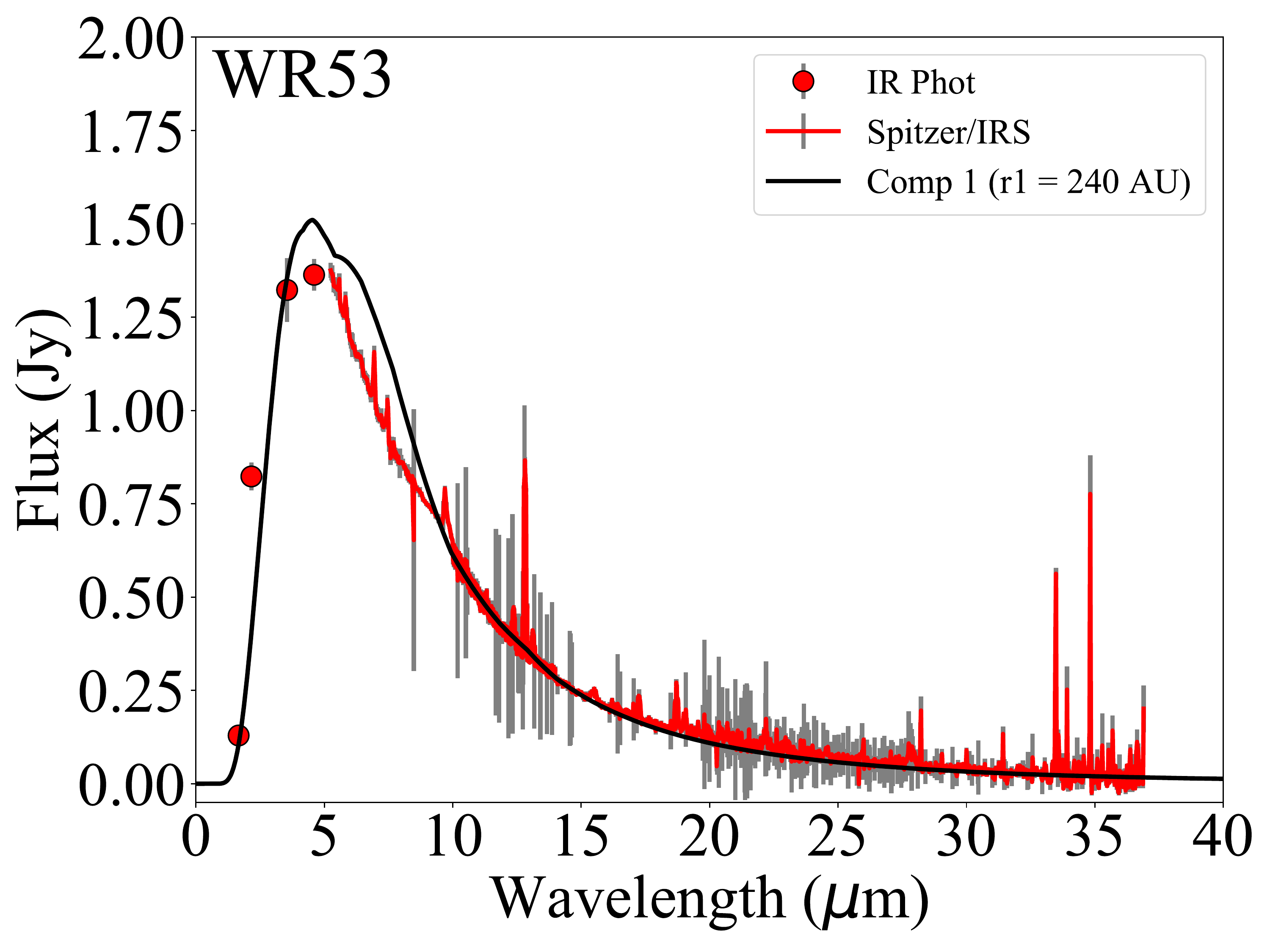}\\
    \includegraphics[width=.46\linewidth]{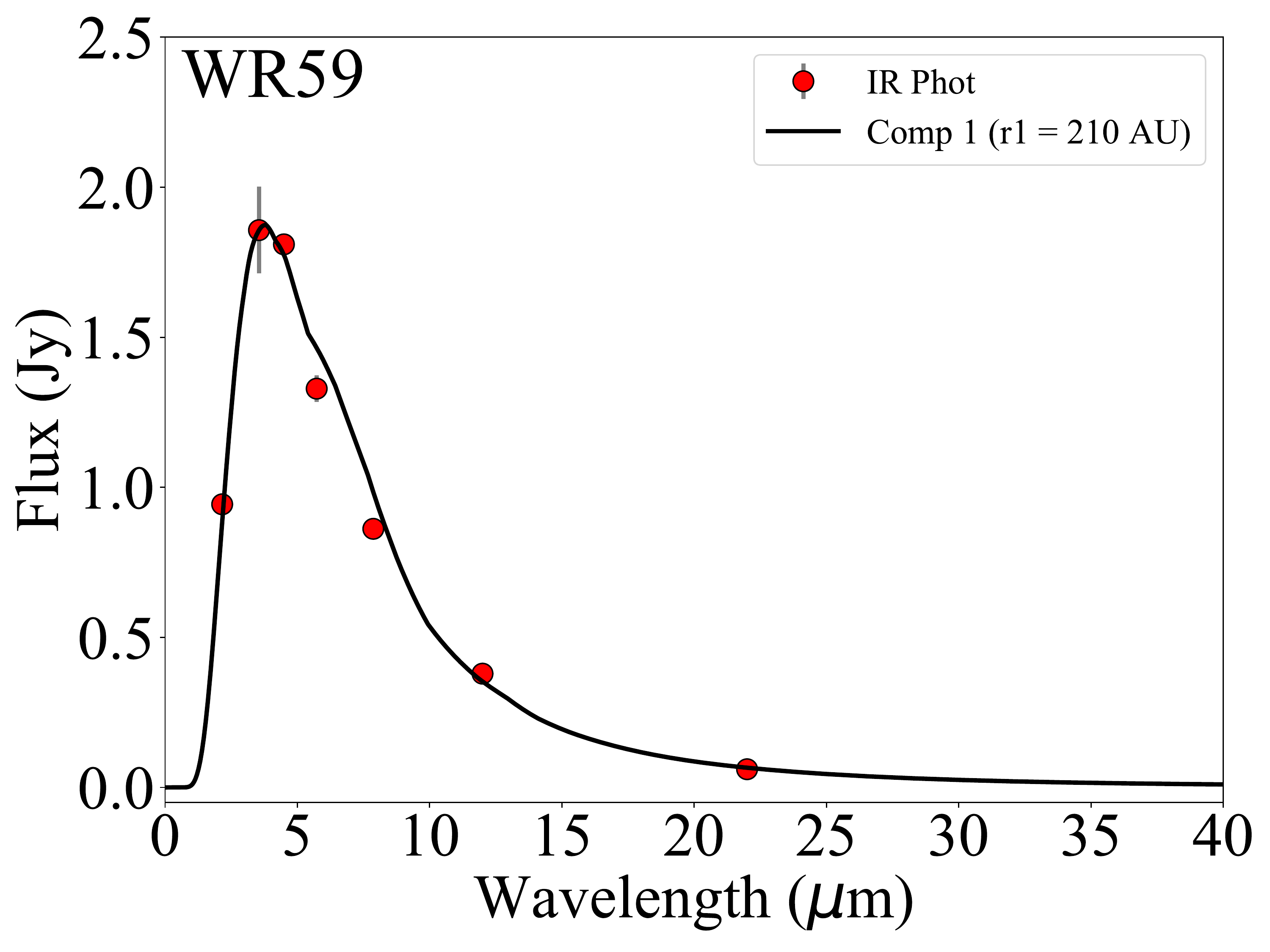} \hspace{10pt}
    \includegraphics[width=.46\linewidth]{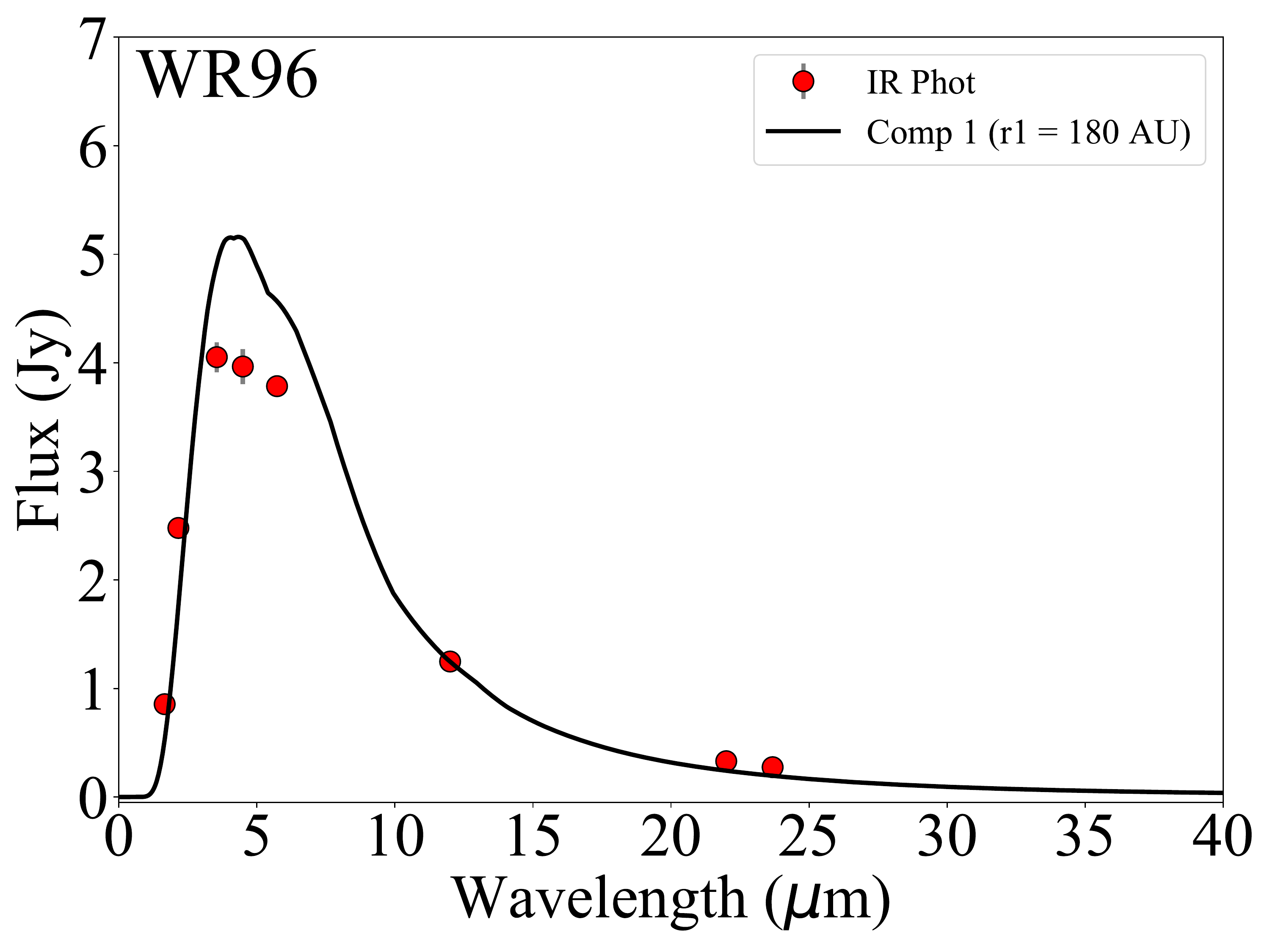} \hspace{10pt} \\
    \includegraphics[width=.46\linewidth]{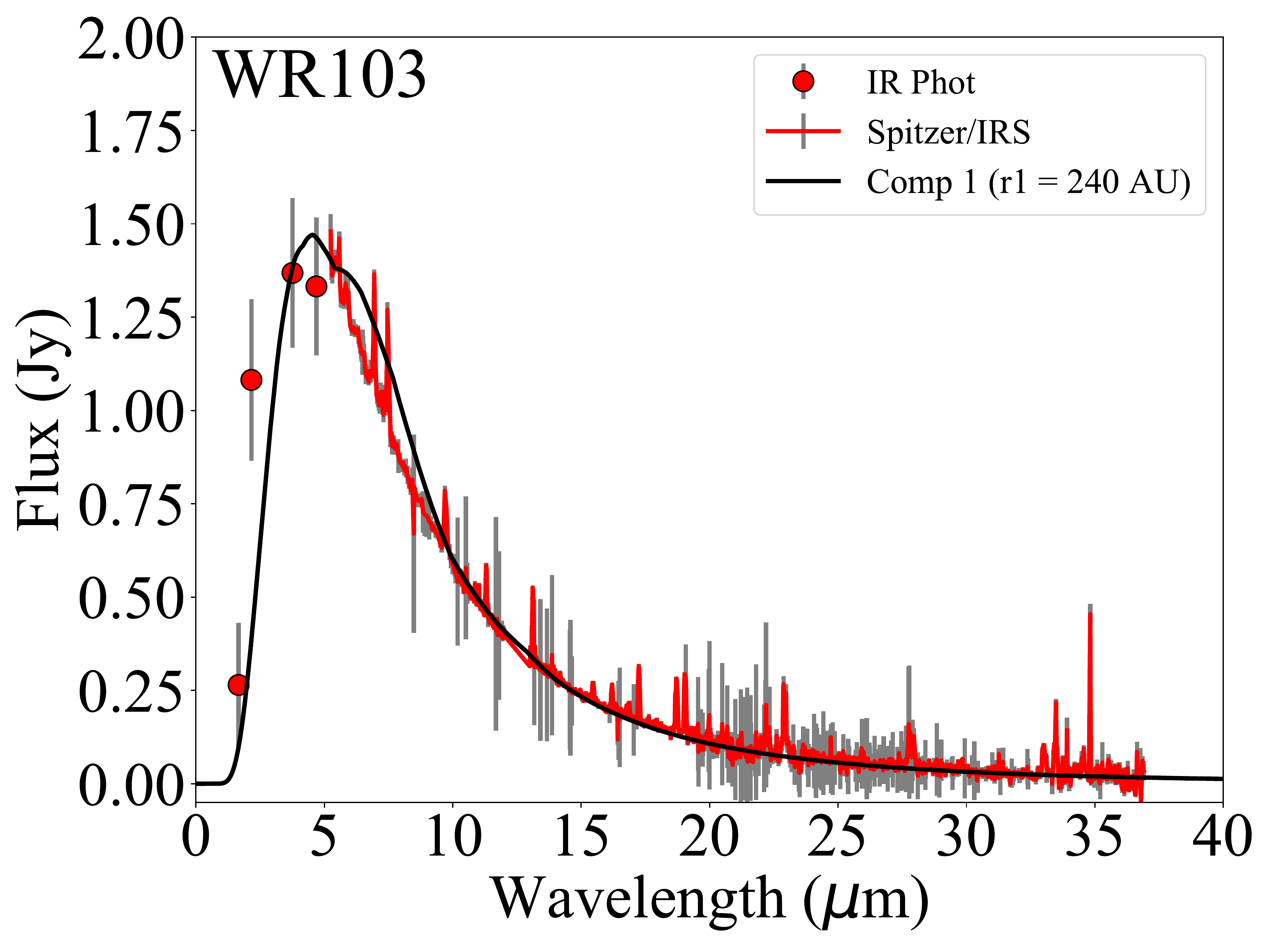}
    \includegraphics[width=.46\linewidth]{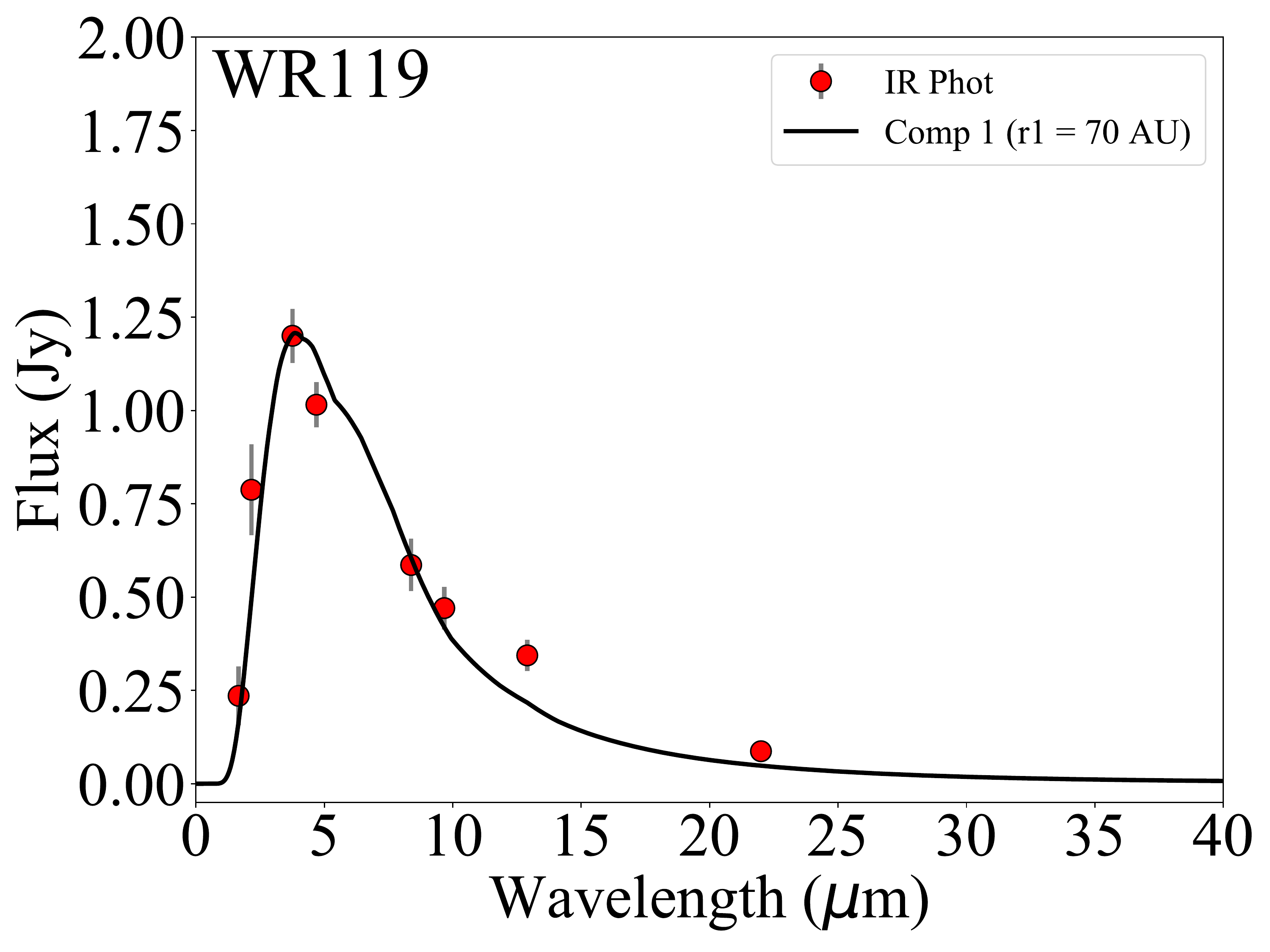}
    \caption{Best-fit \dustem~models of WR48b, WR53, WR59, WR96, WR103, and WR119.}
    \label{fig:SED1}
\end{figure*}

\begin{figure*}[t!]
    \includegraphics[width=.46\linewidth]{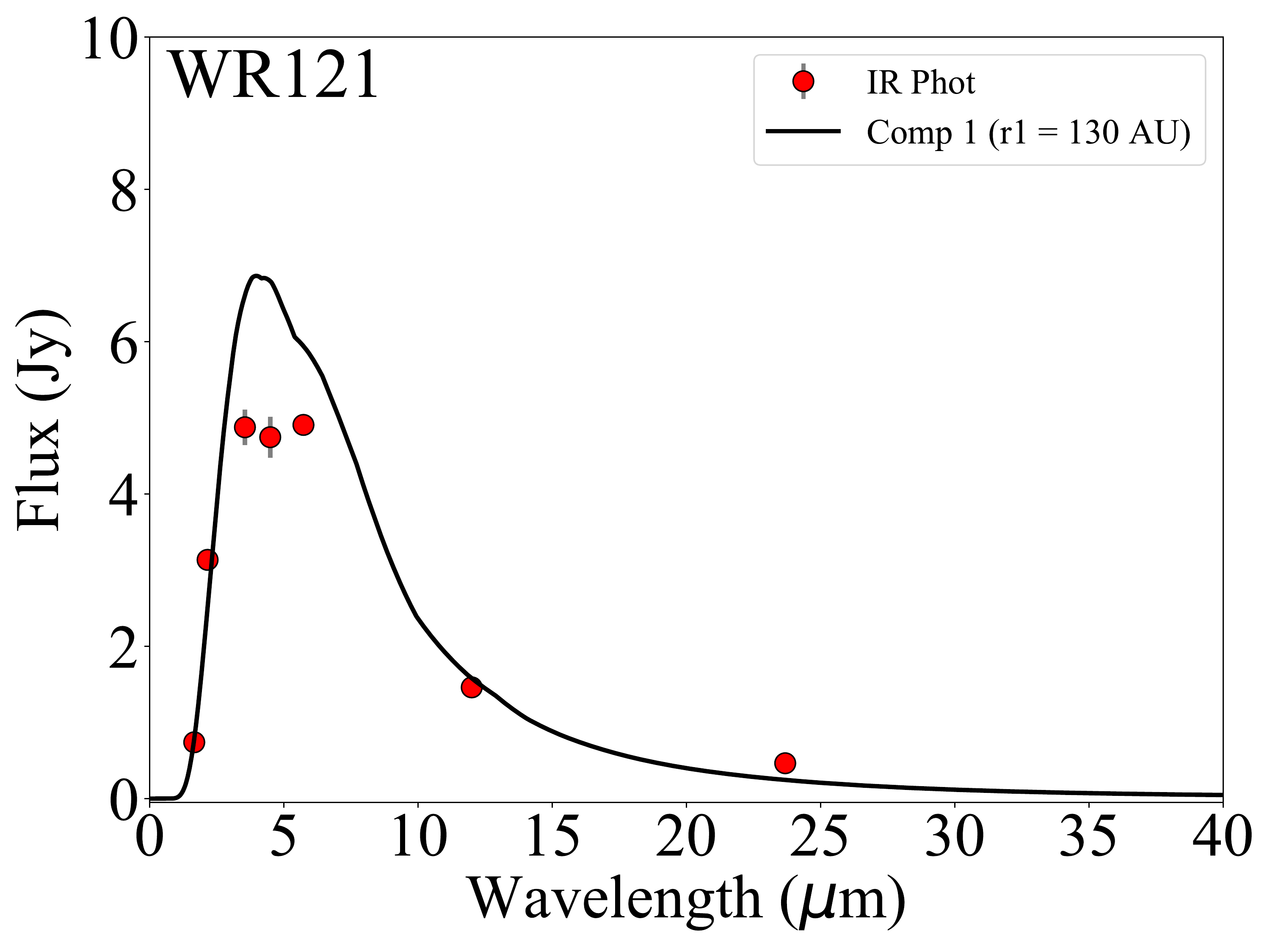} \hspace{10pt}
    \includegraphics[width=.46\linewidth]{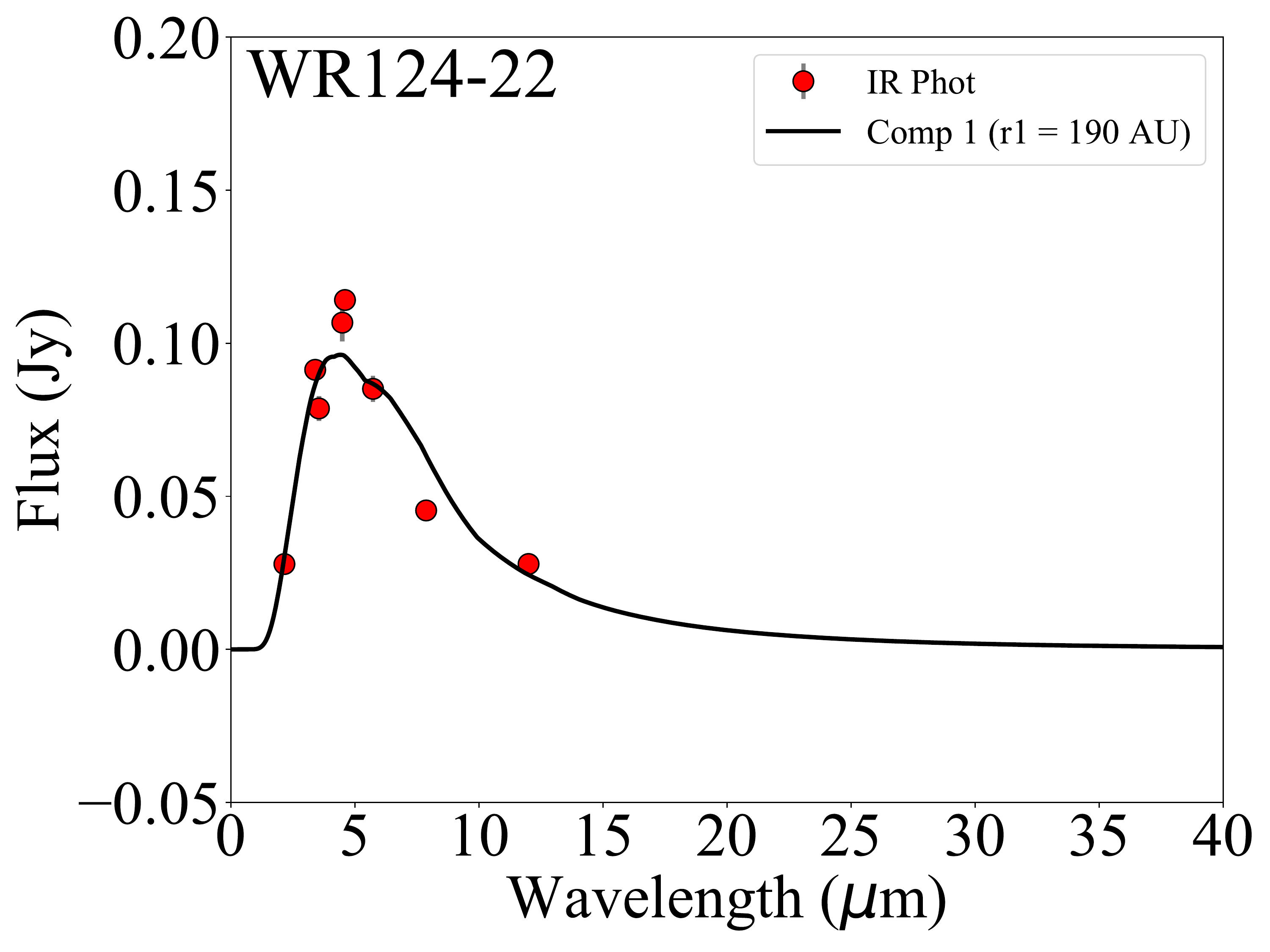}
    \caption{Best-fit \dustem~models of WR121 and WR124-22.}
    \label{fig:SED2}
\end{figure*}

For the dustars without dust radius constraints (e.g.~WR70), we assume a small grain size distribution. In addition to WR70, the 8 other dustars that fall into this category are WR48b, WR53, WR59, WR96, WR103, WR119, WR121, and WR124-22. \dustem~SED models of these 8 dustars are shown in Figures~\ref{fig:SED1} and~\ref{fig:SED2}.

\subsubsection{Dust Properties and Comparison to Previous Studies}

The results of the \dustem~SED models to the \WRnum~WC dustars are presented in Table~\ref{tab:Results}. In Figure~\ref{fig:Results}, the dust condensation fraction $\chi_C$ and the DPR is plotted against the total IR luminosity of each dustar. Both plots show that $\chi_C$ and the DPR increase with increasing with IR luminosity. The periodic or highly variable dust producers (WR19, WR48a, WR125, WR137, and WR 140) show slight discrepancies from this trend since their IR luminosity varies depending on the orbital phase. These results demonstrate that WC dustars exhibit a range of DPRs and $\chi_C$ values ranging from $1\times10^{-10}$ - $4\times10^{-6}$ M$_\odot$ yr$^{-1}$ and $\sim0.002 - 40\%$, respectively.

The most recent and relevant study to compare our results to is that of \citet{Zubko1998}, who conducted a homogeneous modelling analysis of the SEDs and dust shells of 17 Galactic WC dustars incorporating grain physics and dynamics. A majority of their sample overlap with ours. \citet{Zubko1998}, however, modelled the dust formation physics and dynamics under the assumption of spherically symmetric winds since the complex, colliding-wind morphology of WC dustars was not known at that time. 
Despite the difference in analytical techniques, \citet{Zubko1998} derive a range mass loss rates and carbon dust fractions consistent with our results. They determine condensed carbon fractions of 0.002 - 20$\%$ and dust production rates of $5\times10^{-10}$ - $6\times10^{-6}$ M$_\odot$ yr$^{-1}$.

\subsubsection{IR Luminosity, Dust Production Rate, and Grain Size}
\label{sec:WCDPP}
The distribution of L$_\mathrm{IR}$ and the DPR of WC dustars provides a valuable empirical relation for estimating DPRs from IR luminosities that can be directly measured from IR observations. Excluding the periodic and highly variable dustars, the DPR vs L$_\mathrm{IR}$ (Fig.~\ref{fig:Results}, Right) relation can be characterized by the following power-law fit:

\begin{equation}
    \mathrm{Log}\left(\frac{\dot{M_d}}{\mathrm{M}_\odot \, \mathrm{yr}^{-1}}\right)=1.21^{+0.06}_{-0.06}\,\mathrm{Log}\left(\frac{L_{IR}}{L_\odot}\right)-11.55^{+0.32}_{-0.18}.
    \label{eq:DPR}
\end{equation}

In Fig.~\ref{fig:Results} (Left) it is apparent that the efficient dust makers, where $\chi_C\gtrsim 1\%$, are consistent with large grain dust models. This suggests that there is no homogeneous grain size distribution exhibited by all WC dustars, but rather that larger grain sizes ($\gtrsim 0.1$ $\mu$m) are produced by dustars with higher dust condensation fractions. However, if the WC subtype influences the grain size, it is possible that WC9 dustars preferentially form large grains. Regardless, the results suggest that efficient WC9 dustars form larger dust grain that will be more robust against destruction and sputtering via SN shocks and survive longer in the ISM.

\subsection{Investigating the Colliding Wind Dust Formation Model}
\label{sec:DustFormation}

\citet{Usov1991} investigated a theoretical model of dust formation in the wind collision between WR and OB binaries. They derived an analytical relation for the fraction of the WR wind which is strongly cooled and compressed in the shock layer between the two stars:

\begin{equation}
\alpha\propto \eta^2(1 + \eta^{1/2})^2 \dot{M}_\mathrm{WR}^2\,v_\mathrm{WR}^{-6}\,D^{-2},
\label{eq:Usov}
\end{equation}

\noindent
where $\eta$ is the wind momentum ratio between the OB and WR stars $\left(\eta = \frac{\dot{M}_\mathrm{OB}v_\mathrm{OB}}{\dot{M}_\mathrm{WR}v_\mathrm{WR}}\right)$, $\dot{M}_\mathrm{WR/OB}$ is the mass-loss rate of the WR/OB star, $v_\mathrm{WR/OB}$ is the terminal wind velocity of the WR/OB star, and $D$ is the separation distance between the WR and OB star. \citet{Usov1991} assume that most of the carbon from the WC wind in this cold, compressed shock layer condenses into dust, which implies that
\begin{equation}
\alpha\propto\chi_C.
\label{eq:alpha}
\end{equation}

\noindent
We can therefore test the theoretical model from \citet{Usov1991} by investigating the relation between $\chi_C$ and the WC binary mass-loss rates, wind velocities, and orbital separation. 

We focus on the 8 WC dustars in our sample with known orbital periods ($P_\mathrm{orb}$): WR19, WR48a, WR70, WR98a, WR104, WR125, WR137, and WR140. Using Kepler's Third Law and assuming a similar total mass for each dustar system, it follows that the orbital separation $D$ is proportional to $P_\mathrm{orb}^{2/3}$. Eccentricity is also an important parameter that regulates the dust formation and orbital separation for the periodic, non-continuous dustars (e.g.~WR19, WR125, WR137, and WR140); however, their dust production rates and thus their carbon condensation efficiencies were calculated as an average over the orbital period (Eq.~\ref{eq:periodic}). Since the time-averaged separation distance is $a\,(1+e^2/2)$ (See \citealt{Williams2003}), where $a$ is the semi-major axis and $e$ is the eccentricity of the system, the eccentricity should only contribute variations from $a$ by a factor of $1.5$ at most. Given the minor effect of eccentricity in this relation and the eccentricity uncertainty in most dustars, we assume that $D\propto P_\mathrm{orb}^{2/3}$ for all 8 systems.

Due to uncertainties in the mass-loss rate and wind velocity of the binary OB companions of the WC stars, we make the assumption that $\eta$ is similar for the dustars. If the observed dustar properties are consistent with the \citet{Usov1991} colliding-wind model (Eq.~\ref{eq:Usov}), they should exhibit the following relation with our assumptions:
\begin{equation}
\chi_C\,\dot{M}_\mathrm{WR}^{-2}\,v_\mathrm{WR}^{6}\propto P_\mathrm{orb}^{-4/3}.
\label{eq:Usov2}
\end{equation}

\begin{figure}[t!]
    \centerline{\includegraphics[width=1\linewidth]{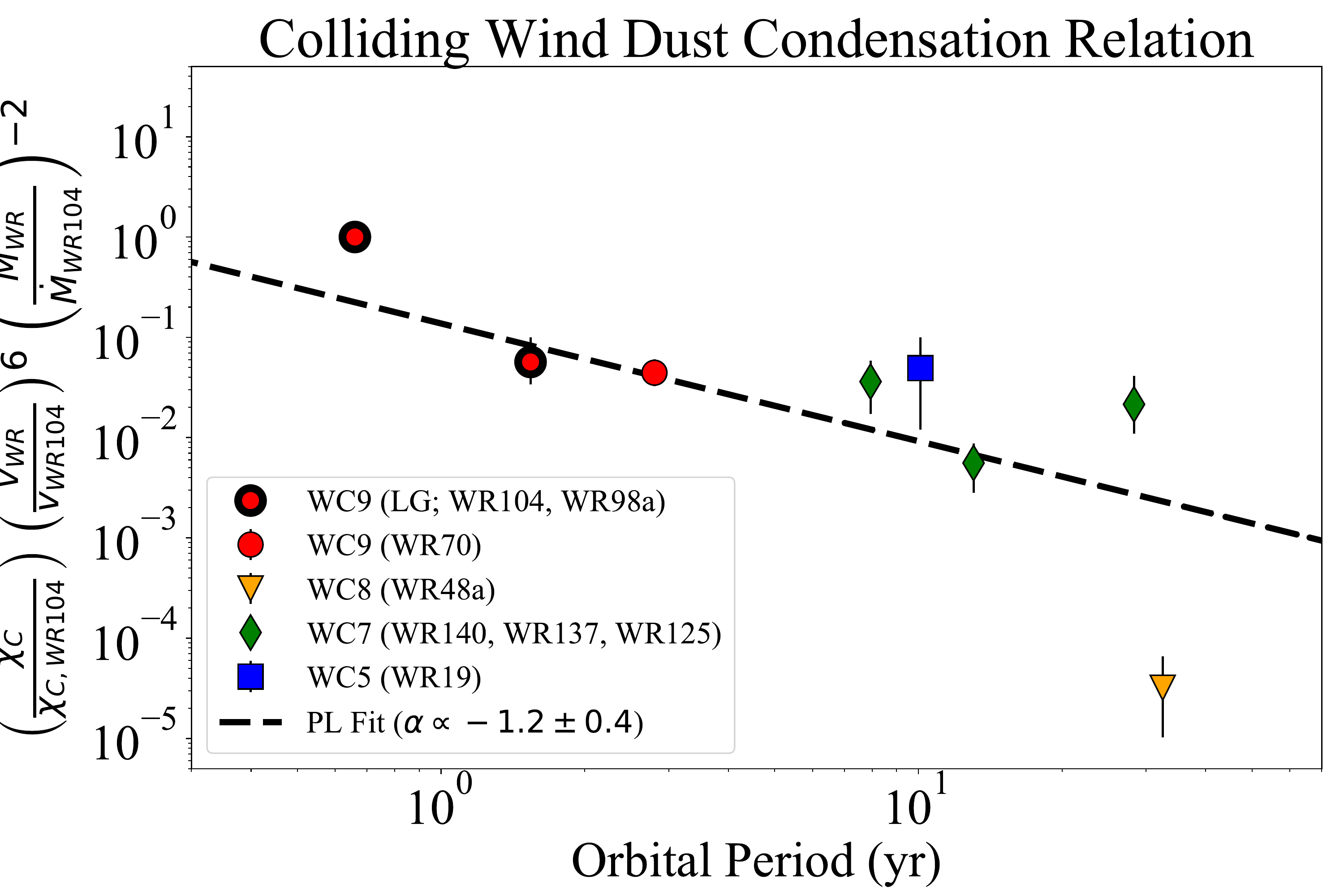}}
    \caption{$\chi_C\,\dot{M}_\mathrm{WR}^{-2}\,v_\mathrm{WR}^{6}$ normalized to the $\chi_C$, velocity, mass-loss rate of WR104 plotted against $P_\mathrm{orb}$ for the 8 dustars in our sample with orbital periods. The dashed line shows the best-fit power-law relation $\chi_C\,\dot{M}_\mathrm{WR}^{-2}\,v_\mathrm{WR}^{6}\propto P_\mathrm{orb}^{-1.2\pm0.4}$, where the 32.5 yr orbital period dustar WR48a is the outlier.}
    \label{fig:PerChiC}
\end{figure}

In Fig.~\ref{fig:PerChiC} we plot $\chi_C\,\dot{M}_\mathrm{WR}^{-2}\,v_\mathrm{WR}^{6}$ vs $P_\mathrm{orb}$ for the 8 dustars normalized to the carbon dust condensation efficiency, wind velocity, and mass-loss rate of WR104. $\dot{M}_\mathrm{WR}$, $v_\mathrm{WR}$, and $P_\mathrm{orb}$ are provided from Tab.~\ref{tab:Adopted} and $\chi_C$ is provided from Tab.~\ref{tab:Results}. We note that the power-law relation is insensitive to the dustar system used to provide the normalization factors. We determine a $P_\mathrm{orb}^{-1.2\pm0.4}$ power-law fit, where WR48a is excluded as an outlier since it exhibits a higher mass-loss rate than the other systems (See Tab.~\ref{tab:Adopted}). This derived power-law fit is consistent with the theoretical power-law relation from the \citet{Usov1991} model with our assumption of constant $\eta$ across the dustars (Eq.~\ref{eq:Usov2}). The consistency of the observational results and theoretical model supports the colliding-wind interpretation of WC dustars and demonstrates the effect of the binary orbital parameters on dust formation efficiency.

\section{BPASS Dust Production Model Results}
\label{sec:4}
\subsection{Binary Population and Spectra Synthesis Models}
\label{sec:BPASSdesc}
The Binary Population and Spectral Synthesis (BPASS; \citealt{BPASS}) suite of binary stellar evolution and stellar population models provides a valuable tool for investigating dust production from stellar populations. The presence of a binary companion is a key factor that influences both the formation of WC stars and their dust production via colliding winds, which underscores the utility of the binary evolution tracks utilized in BPASS for modeling the dust contribution from WC binaries in stellar populations. Notably, BPASS evolutionary models may also be used to study the formation history of individual WC binary system (e.g.~Thomas et al., in preparation). This work presents the first use of BPASS to model dust production from stellar populations. 

We use the BPASS v2.2.1 binary models as described in \citet{S2018} with the default ``$135\_300$" BPASS initial mass function (IMF). This IMF is based on \citet{Kroupa1993} with a power-law slope from 0.1 to 1.0 M$_\odot$ of -1.30 that steepens to -2.35 above this to an upper stellar mass of 300 M$_\odot$.
Stellar mass-loss is determined by the difference between the stellar mass in the current and previous timesteps.

In addition to the WC binaries, three more sources of dust are considered in our BPASS models based on the current understanding of leading dust producers: dust condensation in type II supernova (SN) ejecta and in stellar winds of AGB and RSG stars \citep{Matsuura2009,Dwek2011,Boyer2012,Temim2015,Srinivasan2016}. We do not include the dust input from H-poor type Ia or Ib/c SNe since type II SNe are considered to be the most efficient and prolific source of SN dust (e.g.~\citealt{Sarangi2018} and ref.~therein).

Luminous Blue Variables (LBVs), massive evolved stars that exhibit giant outbursts of extreme mass-loss \citep{Humphreys1994,Smith2014}, may also be important sources of dust. Based on observations of their circumstellar dusty nebulae, LBVs are capable of forming between $\sim0.001 - 0.1$ M$_\odot$ of dust (\citealt{Kochanek2011} and ref.~therein). This corresponds to a DPR of $\sim10^{-7} - 10^{-5}$ M$_\odot$ yr$^{-1}$ for an LBV assuming a lifetime of $\sim10^4$ yr \citep{Smith2014}, which is comparable to the upper DPR range of WC dustars. However, given their rarity, the evolutionary formation pathways for LBVs and the physics of their eruptive mass-loss is not well understood. Due to these uncertainties it is difficult to identify LBVs and quantify their dust contribution via eruptive mass-loss in current BPASS models \citep{BPASS}. We therefore we omit them from the BPASS models in this study but acknowledge their potential as important dust producers that warrant further investigation.

The dust formation prescriptions and the assumed parameters for identifying the four dust inputs sources (type II SNe, AGBs, RSGs, and WC binaries) in the BPASS models are described as follows:

\begin{itemize}

\item \textit{Type II SNe.}~For stars with a final mass greater than 2 M$_\odot$, CO core mass greater than 1.38 M$_\odot$, and H surface mass fraction greater than $1\%$, the terminal explosion is considered a Type II SN. A $10\%$ dust condensation efficiency by mass is adopted for the dust-forming elements in the SN ejecta \citep{Dwek2014,Sarangi2018}. Dust forming elements include carbon, oxygen, magnesium, silicon, and iron. The hydrogen, helium, nitrogen, and neon synthesized in the SN ejecta are not considered as dust-forming elements.

\item \textit{Type II SNe Dust Destruction.}~When a SN explodes its shock interacts with the surrounding ISM and destroys the ISM dust grains. Given a total ISM mass that the SN clears of dust, $m_g$, the destroyed quantity of dust per SN explosion can be estimated as

\begin{equation}
    m_d = m_g \, \chi_{d/g},
\end{equation}

\noindent
where $\chi_{d/g}$ is the gas-to-dust mass ratio of the ISM \citep{Dwek2014,Temim2015}. Note that this does not include destruction of SN ejecta dust by the reverse shock, which is not considered in our work. 
The following fitting formula derived by \citet{Yamasawa2011} is adopted for determining $m_g$:

\begin{equation}
    m_g = 1535\, n_\mathrm{ISM}^{-0.202} ((Z/Z_\odot)+0.039)^{-0.298}\,\mathrm{M}_\odot,
    \label{eq:mg}
\end{equation}

where $n_\mathrm{ISM}$ is the density of the surrounding ISM and $Z_\odot$ is the solar metallciity ($Z = 0.02$). Given the weak dependence on $n_\mathrm{ISM}$, an ISM density of $n_\mathrm{ISM} = 1$ g cm$^{-3}$ is assumed for all BPASS models. Total SN dust input rates are determined with and without incorporating this dust destruction prescription.

\item \textit{AGBs.}~An AGB star is identified as a star with an effective temperature less than $10^{3.66}$ K, a luminosity less than $10^{4.4}$ L$_\odot$, a CO core mass greater than 0.5 M$_\odot$, and a difference between the He and CO core masses less than 0.1 M$_\odot$. This luminosity cut off is consistent with the most luminous AGB in the Large Magellanic Cloud \citep{Riebel2012}. The AGB dust production is calculated assuming a $0.5\%$ condensation fraction of the total mass-loss and is scaled by the metallicity ratio of the model and the solar metallicity, Z/Z$_\odot$ \citep{vanLoon2000}. 

\item \textit{RSGs.}~A star is identified as an RSG if its effective temperature is less than $10^{3.66}$ K and its luminosity is greater than $10^{4.4}$ L$_\odot$. The empirically derived M$_\mathrm{bol}$-DPR relation from \citet{Massey2005} is adopted and scaled by the metallicity ratio of the model and the solar metallicity, Z/Z$_\odot$, to quantify the dust input from the RSGs. An upper limit was set on the RSG dust production requiring the dust input of each time step to be less than the total metals in the mass-loss. 

\item \textit{WC Binaries.}~WC+OB binaries are identified as systems where the WC star has no hydrogen, a combined C and O mass fraction higher than N, a He mass fraction higher than $40\%$, a C mass fraction higher than $25\%$, and a companion with an effective temperature greater than $10^{4.38}$ K. The metallicity dependence for the total mass-loss rates from WR stars in BPASS are described as a power-law relation $\dot{M}\propto Z^\alpha$, where $\alpha=0.5$ \citep{BPASS}. 
The dust production rate from WC binaries is $\dot{M}_d = \dot{M} \,X_C\, \chi_C$, where $X_C$ is the surface mass fraction of carbon that is output from BPASS. The $\chi_C$ relation (Eq.~\ref{eq:Usov2}) that we verified in Sec.~\ref{sec:DustFormation} is adopted and scaled to the properties of WR104:

\begin{equation}
    \chi_C \approx 40\%  \left(\frac{\dot{M}}{\dot{M}_\mathrm{WR104}}\right)^2 \left(\frac{v_\infty}{v_\mathrm{WR104}}\right)^{-6}
    \left(\frac{P_\mathrm{orb}}{P_\mathrm{WR104}}\right)^{-4/3}.
\end{equation}

The dust condensation fraction is therefore treated as a function of the mass-loss rate and terminal wind velocity of the primary star (i.e.~the WC star) and the orbital period of the system. The terminal wind velocity ($v_\infty$) is defined by the stellar mass-loss rate ($\dot{M}$), luminosity ($L_*$), and the momentum transfer efficiency ($\eta_\mathrm{mom}$) from the relation $\eta_\mathrm{mom}=\dot{M}v_\infty /(L_*/c)$, where $c$ is the speed of light. $\dot{M}$ and $L_*$ are output parameters from BPASS, and we adopt a fixed value of $\eta_\mathrm{mom} = 5.3$, the median value of the 11 Galactic dusty WC stars from the optical spectroscopic analysis by \citet{Sander2019}. In the BPASS models, we only consider the dust input from WC binaries with orbital periods consistent with the range of orbital periods in our sample: $P_\mathrm{orb} = 0.66 - 32.5$ yr. Lastly, since not all WC binaries are dust producers, we conservatively assume that $28\%$ of the BPASS WC systems that meet these criteria are dustars, which is based on the observed WC-dustar/WC ratio in the Galaxy \citep{RC2015}. This is percentage is likely an underestimate given the recent mid-IR variability analysis by \citet{Williams2019}.

\end{itemize}

Since the shocks from end-of-life SN explosions may destroy the newly formed circumstellar dust from RSGs and WC binaries, their dust input is not included if they end their lives as SNe. It is assumed that RSGs and WC stars end their lives as explosive SNe if their remnant core mass is less than 3 M$_\odot$, whereas a core with a mass greater than $3$ M$_\odot$ leads to a direct collapse into a black hole with minimal ejecta energy and circumstellar dust survival.

\subsection{BPASS Dust Models at Z = 0.001, 0.008, and 0.020}

\begin{figure*}[t!]
    \includegraphics[width=.45\linewidth]{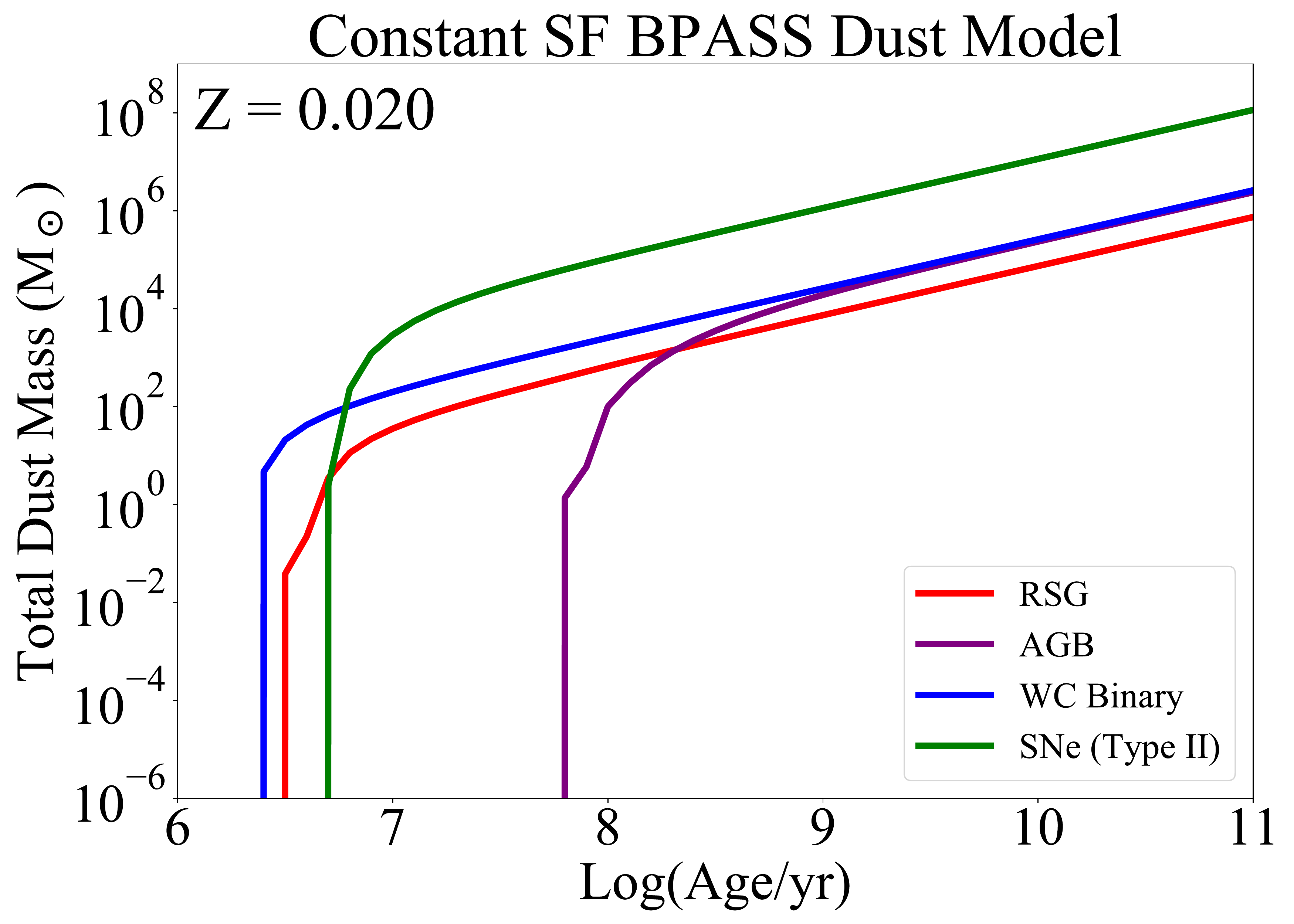} \hspace{10pt}
    \includegraphics[width=.45\linewidth]{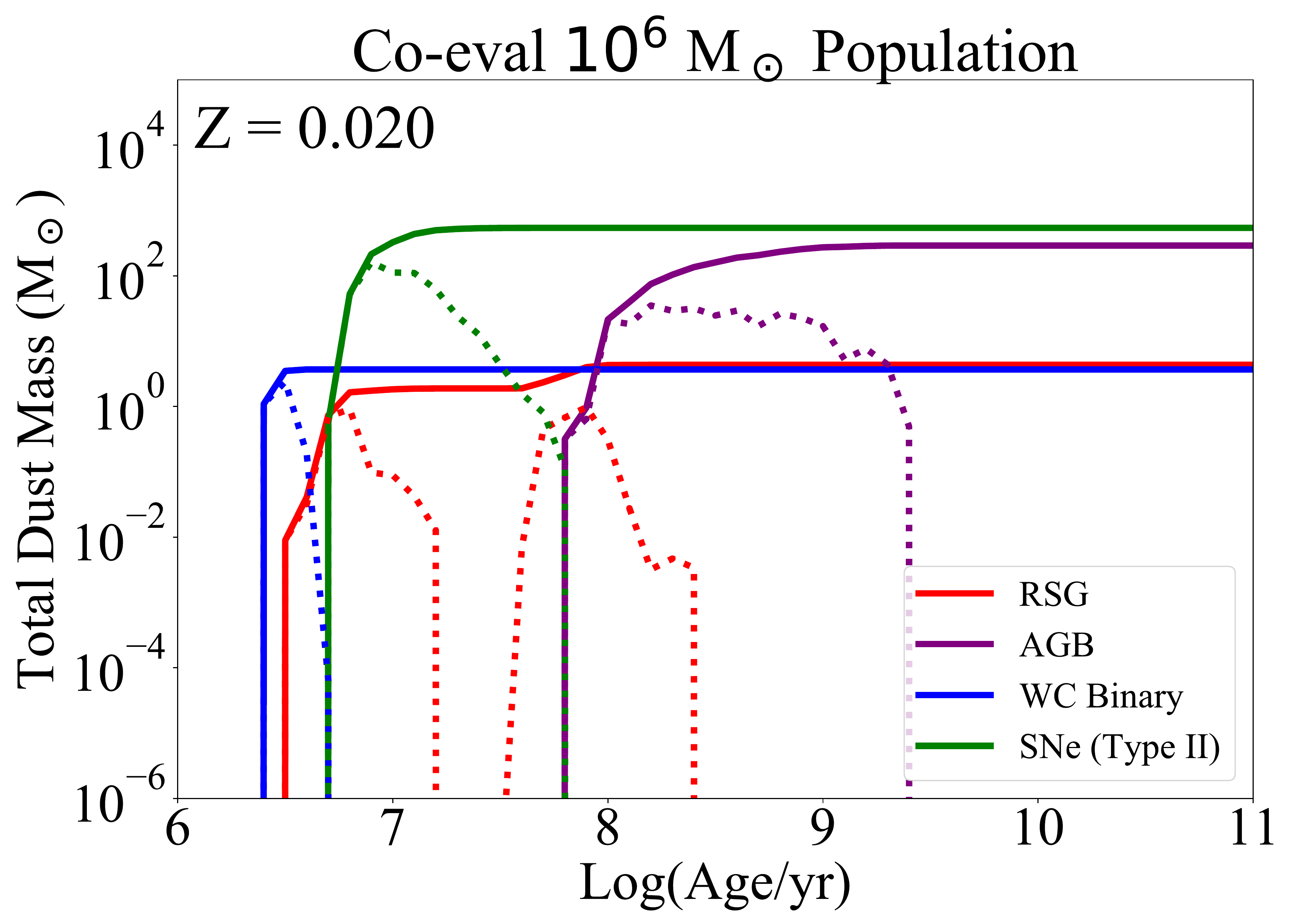}\\
    \includegraphics[width=.45\linewidth]{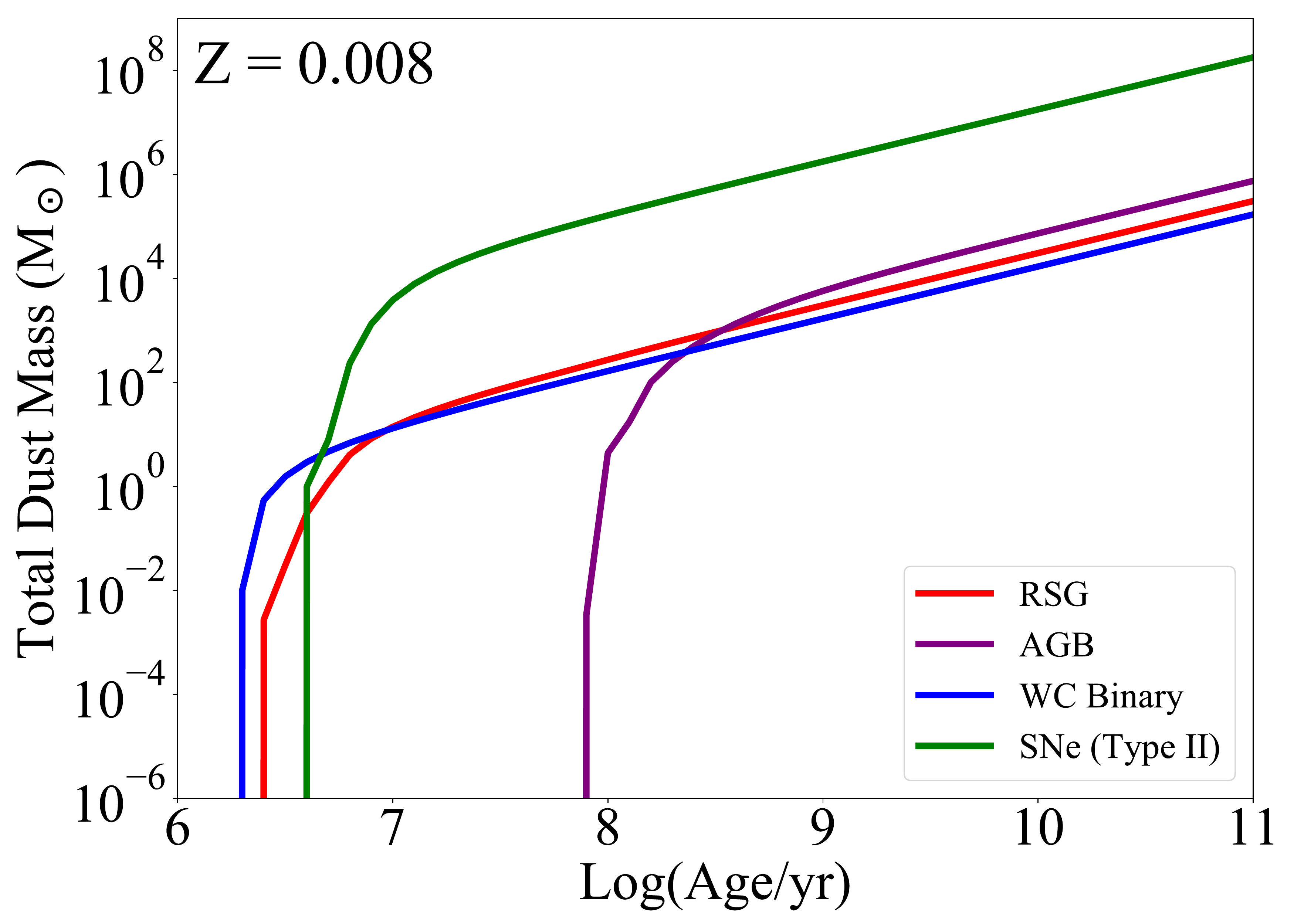} \hspace{10pt}
    \includegraphics[width=.45\linewidth]{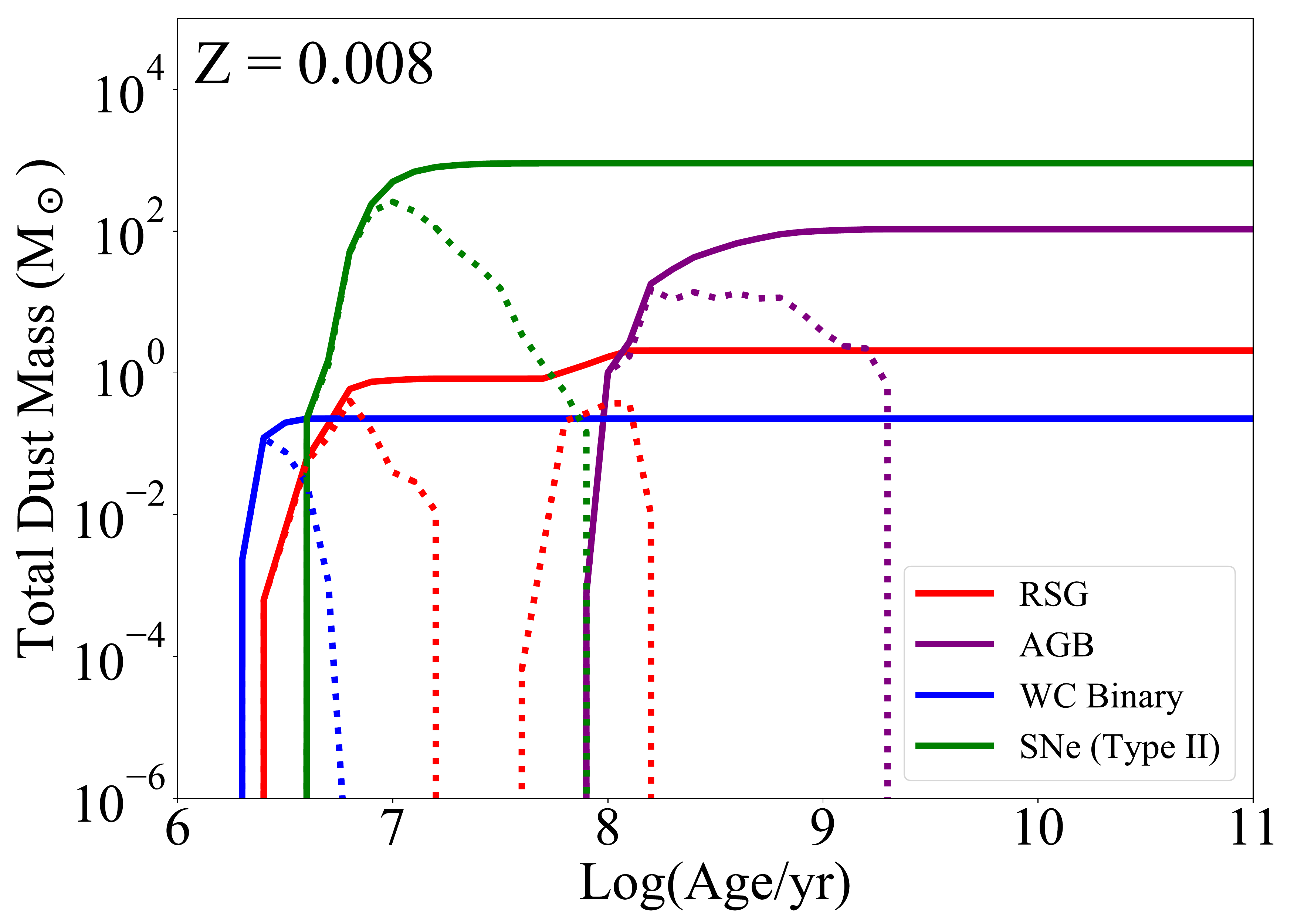}\\
    \includegraphics[width=.45\linewidth]{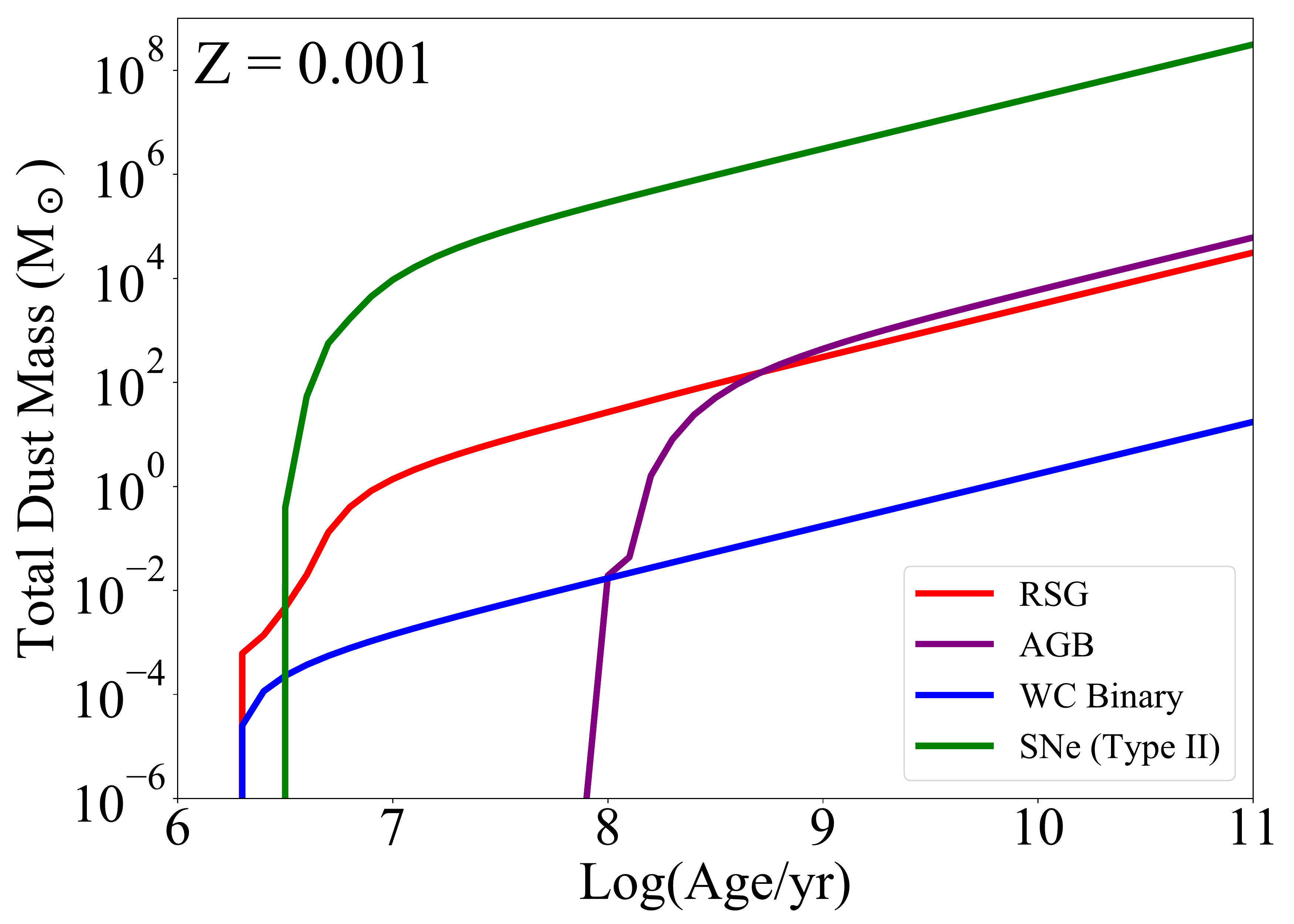} \hspace{10pt}
    \includegraphics[width=.45\linewidth]{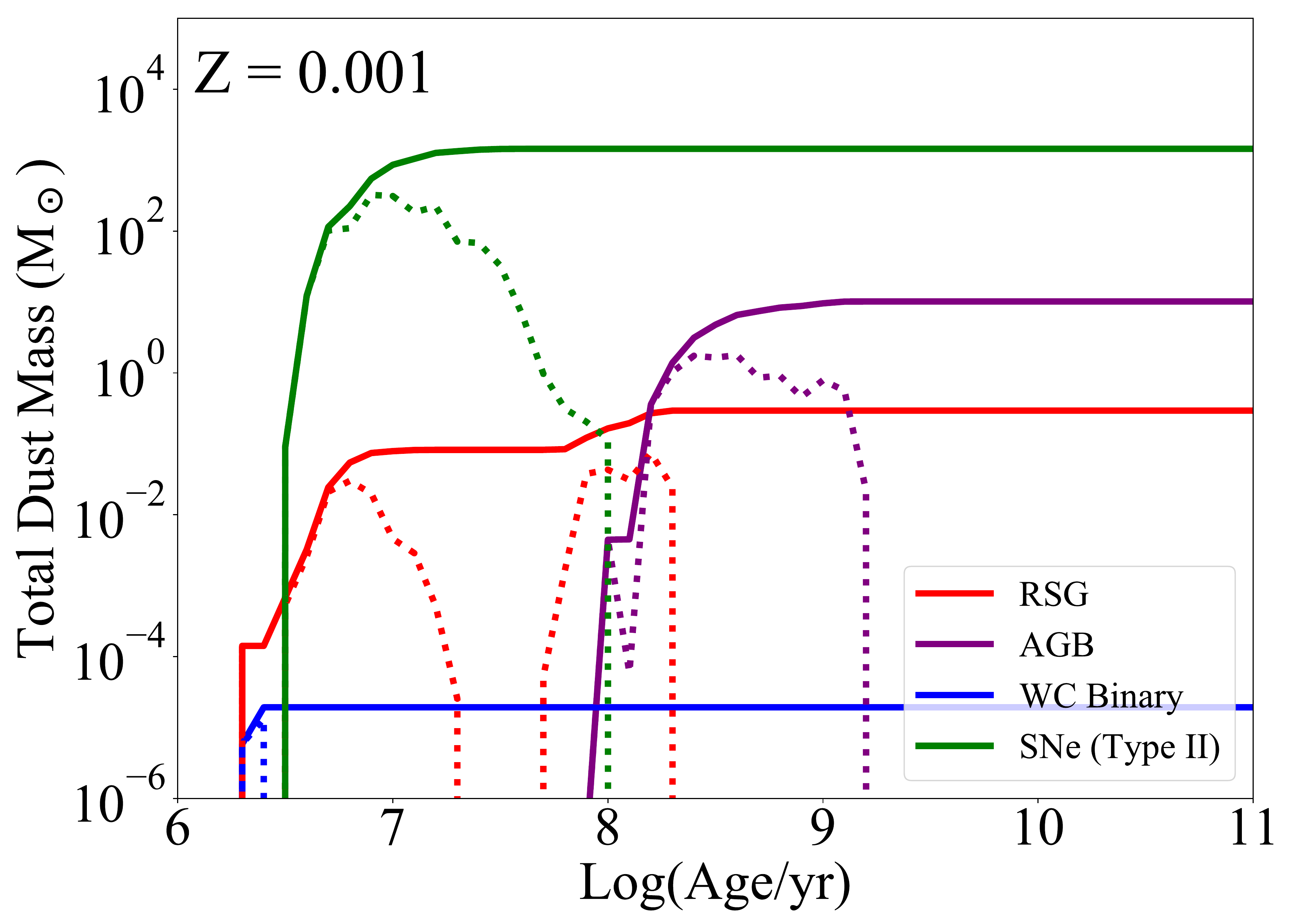}\\
    \caption{(Left Column) Constant SF BPASS models at low (Z = 0.001), LMC-like (Z = 0.008), and solar (Z = 0.020) metallicities showing the cumulative dust mass input as a function of time. The SNe models with dust destruction do not appear in any of the plots since SNe are net destroyers of dust at all timesteps. (Right Column) Co-eval $10^6$ M$_\odot$ stellar population BPASS models at low, LMC-like, and solar metallicities showing the total cumulative dust mass input (solid lines) as a function of time. The dotted lines show the dust input in a single time bin for each dust source. The SNe + dust destruction model does not appear for the same reason as described above.}
    \label{fig:BPASS}
\end{figure*}

BPASS dust models were run at low ($Z = 0.001$), LMC-like ($Z = 0.008$), and solar ($Z = 0.020$) metallicities in order to explore the relative dust contribution of WC binaries, AGBs, RSGs, and SNe at different epochs in cosmic time as characterized by different metallicities. The star formation history will impact the stellar populations and their dust input. For example, in a scenario with constant star formation (SF), the dust sources associated with massive stars have short lifetimes but will be continuously replenished by on-going star formation. Two star formation scenarios were therefore considered for the BPASS models at each metallicity: a co-eval $10^6$ M$_\odot$ stellar population and constant SF. These two different star formation scenarios represent contrasting star forming histories, i.e.~a single starburst event vs. continuous star formation. 

The dust-to-gas mass ratios, which are relevant for the SN destruction rates, are assumed to be $\chi_{d/g} = $ 0.0008, 0.0018, and 0.007 \citep{Zubko2004}
for the Z = 0.001, 0.008, and 0.020 metallicity models, respectively. The low-Z and LMC-like dust-to-gas mass ratios are adopted from the values derived for the SMC and LMC, respectively, by \citet{Temim2015}. Based on Eq.~\ref{eq:mg}, the total ISM mass swept up by SNe in the low-Z, LMC-like, and solar metallicity models is $m_g =$ 3088, 1947, and 1518 M$_\odot$, respectively.

The cumulative dust production rates in for the constant SF and co-eval $10^6$ M$_\odot$ stellar population models are shown in Fig.~\ref{fig:BPASS}. 
Results from the constant SF and $10^6$ M$_\odot$ stellar population models at a time $t = 1.0$ Gyr are provided in Tab.~\ref{tab:BPASSTab} and~\ref{tab:BPASSTab106}, respectively.

\subsubsection{Constant SF Models}
\label{sec:CSF}

The results from the constant SF models are shown in Fig.~\ref{fig:BPASS} (Left Column) and Tab.~\ref{tab:BPASSTab}. In order to check the DPR prescriptions of the four dust sources, we compare the model results against observationally measured values and previous studies. The average DPRs for AGB and RSG stars in the LMC-like metallicity model are $\left<\dot{M}_{d,AGB}\right>=1.1\times10^{-9}$ M$_\odot$ yr$^{-1}$ and $\left<\dot{M}_{d,RSG}\right>=1.7\times10^{-9}$ M$_\odot$ yr$^{-1}$, which are consistent within factors of a few to the average DPRs measured from mid-IR observations of the LMC \citep{Riebel2012,Srinivasan2016}, where $\left<\dot{M}_{d,AGB}\right>=6.9\times10^{-10}$ M$_\odot$ yr$^{-1}$ and $\left<\dot{M}_{d,RSG}\right>=4.0\times10^{-10}$ M$_\odot$ yr$^{-1}$. \citet{Temim2015} calculate a theoretical IMF-averaged SN dust yield of 0.65 M$_\odot$ for a $100\%$ dust condensation efficiency in SNe. A 10$\%$ condensation efficiency, which we adopt for the BPASS models, applied to the \citet{Temim2015} SN dust yields is therefore consistent within a factor of two to the BPASS SN dust yields ($\sim0.1$ M$_\odot$). The average dust mass destroyed per SN in the LMC measured by \citet{Temim2015} is $6.1\pm2.6$ M$_\odot$ for silicates and $1.6\pm0.7$ M$_\odot$ for carbon grains, which is consistent within factors of a few to the average dust mass destroyed per SN in the LMC-like BPASS models ($3.5$ M$_\odot$). We discuss the comparisons to the LMC in further detail in Sec.~\ref{sec:LMC}.

The average dust production rate from WC binaries in the solar metallicity models is $\left<\dot{M}_{d,WC}\right>=9.9\times10^{-7}$ M$_\odot$ yr$^{-1}$, which is slightly greater than the average DPR from our sample of \WRnum~Galactic dustars $\left<\dot{M}_{d,WC}\right>=3.3\times10^{-7}$ M$_\odot$ yr$^{-1}$. However, we find that the WC binary dust production from the BPASS models is largely dominated by systems with the shortest orbital period: $\sim50\%$ of the total WC binary dust input in the solar metallicity model originates from 3 systems with P$_\mathrm{orb}$ between 0.66 and 1.0 yr. These 3 systems have an average DPR of $3.7\times10^{-6}$ M$_\odot$ yr$^{-1}$, which is consistent with our measured DPR of the P$_{orb}=0.66$ yr system WR104, $\dot{M}_d=4.4\times10^{-6}$ M$_\odot$ yr$^{-1}$. We therefore conclude that our BPASS dust prescriptions provide reasonable estimates for the dust input from the four sources.

The constant SF BPASS models show that dust production from SNe without dust destruction dominates AGBs, RSGs, and WC binaries by over an order of magnitude for all 3 metallicities. However, SNe are consistently net dust destroyers at all time steps when incorporating SN dust destruction, and the SN dust destruction rate exceeds the production rate by over an order of magnitude. The average dust produced per SN is $\sim0.1$ M$_\odot$ for all 3 metallicities. We note that dust formation in SN ejecta is still uncertain and measurements range from $\sim10^{-6} - 1$ M$_\odot$ of dust per SN \citep{Temim2015,Gall2018, Sarangi2018}, but SNe would have to form dust beyond the highest mass range of the observed values in order to compensate for the dust destruction rate.

In the solar metallicity models, WC binaries are the dominant source of dust for $\sim60$ Myr until the onset of the AGB star population that forms dust at a similar rate to the WC binaries. We note that the WC binaries are significant dust producers in the solar metallicity model despite almost half of the WC stars ending their lives as SNe with their dust input removed (Tab.~\ref{tab:BPASSTab}).

The LMC-like metallicity model shows slightly reduced dust input from AGB and RSG stars compared to the solar metallicity models. The WC binaries show a much greater reduction in dust production due to a smaller population by a factor two and a lower average dust production rate by over an order of magnitude from the solar metallicity model. AGB stars lead the dust production in the LMC-like metallicity models but only by factors of $2-3$ over RSGs and WC binaries. WC binaries notably form within the first 2 Myr after the onset of star formation and are the first sources of dust in the LMC-like (and solar) models.

In the low Z models, RSGs, AGBs, and WC binaries show smaller populations and form dust at a much lower rate than the LMC-like and solar models. WC binaries exhibit the most substantial decrease in dust production and population. This is because only the most massive and luminous stars evolve to the WC phase at lower metallicities. Such luminous stars drive faster winds and therefore form dust much less efficiently given the sensitivity of $\chi_C$ to wind velocity (Eq.~\ref{eq:Usov}). At low Z, if SNe are indeed net dust destroyers, AGBs and RSGs are likely the dominant dust sources with dust input rates exceeding WC binaries by over 3 orders of magnitude.

\subsubsection{Co-Eval $10^6$ M$_\odot$ Stellar Population Models}

Figure~\ref{fig:BPASS} (Right Column) shows the dust mass input in each time bin and the cumulative dust mass input as a function of time for WC binaries, AGBs, RSGs, and SNe. The most notable difference in the dust production between the co-eval population and the constant SF models is the relatively higher dust contribution by AGBs. For example, the total dust mass from AGB stars at a time $t = 1.0$ Gyr is over an order of magnitude greater than the total dust mass input from RSGs and WC binaries in all three metallicity models (Tab.~\ref{tab:BPASSTab106}). Similar to the constant SF models, SNe are consistently net dust destroyers at all 3 metallicites.
 The RSG dust input shows two peaks, where the secondary peak around $\sim100$ Myr is consistent with the onset of AGBs. This secondary, late-time peak arises due to the difficulties in distinguishing the population of luminous AGB stars from faint RSGs \citep{BPASS}.
SN models with dust destruction show that the AGBs dominate the dust input at all three metallicities.

The dust production timescales, which we define as the length of time the sources are forming dust, are provided in Tab.~\ref{tab:BPASSTab106} and shows that WC binaries only produce dust for $\Delta t_{WC}=0.5 - 4.3$ Myr. Note that the RSG and WC binary timescales in Tab.~\ref{tab:BPASSTab106} only include sources that do not end their lives as SNe. When including sources that end their lives as SNe, the first RSG peak is broader and bridges the second peak in all 3 metallicity models, and the WC binary dust formation timescale in the solar metallicity model is broader by factor of $\sim2$ ($\Delta t_{WC}=5.4$ Myr) than the $2.5$ Myr timescale reported in Tab.~\ref{tab:BPASSTab106}. 

As expected, the WC dust production timescale is substantially shorter than the dust production timescales of AGB stars, where $\Delta t_{AGB}=1.5 - 2.4$ Gyr.  The short dust production timescale of WC binaries explains the difference in their cumulative dust input rate relative to the longer-lived AGB stars between the constant SF and co-eval population models. Given their short timescales, the presence of WC binaries should also reflect the intensity of recent star formation.

The results from the co-eval $10^6$ M$_\odot$ stellar population BPASS models indicates that for a ``bursty" SFH, AGB stars will dominate the dust budget at late times ($t\gtrsim70$ Myr) over WC binaries and RSGs with SNe as net dust destroyers. However, the LMC-like and solar models show that WC binaries are the first and dominant source of dust for $\sim1$ Myr before the RSG dust production starts to increase. Notably, in the solar metallicity mode, the WC binaries are the leading dust source until the onset of AGB stars.

\subsection{LMC BPASS Dust Model and Massive Star Population Discrepancy}
\label{sec:LMC}

The well-studied dust and supernova remnant properties of the nearby LMC ($d\sim50$ kpc) highlights its utility as a laboratory to test dust input rates in its ISM (e.g. \citealt{Matsuura2009,Temim2015}). Importantly, the stellar populations and dust input from AGBs, RSGs, and SNe have been previously measured or estimated observationally and thus provide a comparison for the BPASS model output (Sec.~\ref{sec:CSF}). 
In this section, we apply the $Z = 0.008$ constant star formation BPASS models to the LMC and demonstrate the discrepancy in the massive star populations due to its complex and variable star formation history (SFH; \citealt{HZ2009}).
The LMC dust-to-gas mass ratio, which is relevant for the SN dust destruction rates, is assumed to be $\chi_{d/g} = $ 0.0018 \citep{Temim2015}. The total ISM mass swept up per SN adopted in the BPASS models based on Eq.~\ref{eq:mg} is $m_g=1947$ M$_\odot$ and is consistent with the LMC value measured by \citet{Temim2015} of $m_g\sim2000$ M$_\odot$.

In order to model the dust input the LMC, the dust production and stellar population output from the constant SF BPASS models at a metallicity of $Z = 0.008$ were scaled by a constant factor such that the total number of AGB stars at a time $t=12.6$ Gyr matches the current observed number of AGBs in the LMC, $N_{AGB}\approx20000$ \citep{Riebel2012,Zhukovska2013, Srinivasan2016}. The current age of the LMC is assumed to be $t=12.6$ Gyr.

In Table~\ref{tab:BPASSLMCTab}, we compare the dust input, populations, and SN rates from the LMC BPASS models with values measured from mid-IR observations of the LMC by \citet{Riebel2012,Srinivasan2016} and the Magellanic Cloud SN remnant analysis by \citet{Temim2015}. The total dust input rate from the AGB stars at time $t=12.6$ Gyr is closely consistent to the observed rate. However, the SN rate and the total WC binary and RSG dust input rates from the models are larger than the observed values.

Both the BPASS models and results from previous studies show an agreement that the dust production from SNe without dust destruction dominates the other three sources by over an order of magnitude. However, when incorporating the effects of ISM dust destruction from SN shocks, SNe are net destroyers of dust (\citealt{Temim2015}). 

It is apparent that the population of sources associated with massive stars is over-predicted by the constant SF BPASS models: the total number of RSGs at $t=12.6$ Gyr ($N_{RSG}=11887$), which includes RSGs that eventually will explode as SNe and not input dust, is a factor of $\sim3$ greater than the observed number of RSGs, and both the SN rate and the number of dust-producing WC binaries exceed the observed value by over an order of magnitude. 

For the WC binaries, there are only 2 currently known dust producers in the LMC as opposed to the 52 predicted by the models. These two systems, HD 36402 and HD 38030, both host WC4 stars and exhibit dust formation periods of 5.1 and $>20$ yr, respectively \citep{Williams2019}. The estimated lower limit on the current observed WC binary dust input rate in the LMC, $3\times10^{-8}$ M$_\odot$ yr$^{-1}$ (Tab.~\ref{tab:BPASSLMCTab}) is derived by the measured amorphous carbon dust mass of $M_d\sim1.5\times10^{-7}$ M$_\odot$ around HD 36402 by \citet{Williams2011} divided by its 5.1 yr orbital period. Interestingly, the DPR of HD 36402 is consistent within a factor of two to the mean dust input rate for the WC binaries in the BPASS model where $\left<\dot{M}_{d,WC}\right>=6.2\times10^{-8}$ M$_\odot$ yr$^{-1}$. The dust input rate from the longer period HD 38030 WC system is not yet known since it was only recently identified as a periodic/episodic dust former. However, given their longer orbital period ($>5$ yr) and lower IR luminosities \citep{Williams2013b}, these two LMC WC systems are unlikely to be significant dust producers like WR104 where $\dot{M}_d\sim10^{-6}$ M$_\odot$ yr$^{-1}$ (Tab.~\ref{tab:Results}).

We attribute the number discrepancy of the massive star population to the assumption of a constant SF in the BPASS models since the SFH of the LMC is complex and variable \citep{HZ2009}. Analysis of the SFH in the LMC by \citet{HZ2009} indicates that the current total stellar mass is significantly composed of stars formed in an initial epoch of star formation that ceased 10 Gyr ago; this was followed by a quiescent period until 5 Gyr ago with recent star formation peaks occurring 12 Myr, 100 Myr, 500 Myr, and 2 Gyr ago. Normalizing the number of sources from the constant SF BPASS models to the observed AGB stars in the LMC would therefore over-predict the number of RSGs and WC binaries and SN rates, all of which are products of shorter-lived massive stars. Notably, the lifetime of the WC binary dust producers in the BPASS models is $<5\times10^6$ Myr and is shorter than one of the most recent star formation peaks in the LMC that occurred 12 Myr ago \citep{HZ2009}.

The results from this LMC analysis highlight the importance of considering the SFH in modeling the dust input rates, especially from massive stars. Although this is beyond the scope of the work presented here, future work with BPASS dust models incorporating variable SFH is crucial for obtaining a more complete understanding of galactic dust production rates.

\section{Discussion and Conclusions}
\label{sec:5}
\subsection{Astrophysical Implications}
The WC dustar SED analysis and the predicted dust input from BPASS models indicate that WC binaries
are leading sources of dust, comparable to AGB stars, in environments with constant star formation at solar metallicities if SNe are indeed net dust destroyers. For a stellar population formed in an instantaneous starburst at solar metallicities, WC binaries will dominate the dust contribution until the onset of AGBs. WC binaries are also among the earliest sources of dust in LMC and solar metallicities, which implies they are an early reservoir of carbon-rich dust.

Due to the net dust destruction rates from SNe, additional dust input is still necessary to account for the observed dust in galactic ISM \citep{Dwek2014}. For example, in the LMC an additional dust source with an input rate more than an order of magnitude greater than the combined input from AGB stars and SNe is needed to balance out the SN dust destruction \citep{Temim2015}. The net SN dust destruction rate for all three metallicity models also shows that the dust destruction exceeds the input from AGBs, RSGs, and WC binaries by several orders of magnitude. 

Although WC dustars may not provide enough dust to account for this deficit, the large 0.1 - 1.0 $\mu$m-sized grains produced by the heavy WR104-like dust makers would be more robust against destruction from SN shocks. Such large grains would exhibit grain lifetimes a factor of $\sim3$ longer relative to small $\sim100$ $\AA$-sized grains \citep{Jones1996}. Therefore, the destruction timescales of the ISM grains would be lengthened if a significant fraction of the ISM was composed of dust from WC binaries. 

In the context of cosmochemistry, the formation of large dust grains supports the hypothesis of WR stars as source of short-lived radionuclides (SLRs) in the Solar System. The longer destruction timescales of larger grains suggest WR winds were able to transport and seed SLRs like Al$^{26}$ in the pre-solar nebula \citep{Dwarkadas2017}.

Given the impact of WC dustars on the ISM, it is important to understand their dust chemistry and the grain formation physics in their hostile wind collision environment. The C-rich dust that forms in the H-deficient and fast winds ($\gtrsim1000$ km s$^{-1}$) of WC binaries likely follows a different chemical formation pathway than that of C- and H-rich AGBs \citep{LeTeuff1998,Cherchneff2000}. Distinguishing emission features of the molecular precursors of C-rich dust condensed in WC binaries from other sources could provide a means of observationally tracing their contribution to galactic ISM. Interestingly, the ISO/SWS spectroscopy of WC dustars re-analyzed in this work also show evidence of hydrocarbon features \citep{Marchenko2017}.

Our analysis of the carbon dust condensation fraction (Sec.~\ref{sec:DustFormation}) strengthens our understanding of dust formation in colliding-wind binaries; however, it is important to consider that not all WC+OB binaries form dust. For example, the well-studied colliding-wind system $\gamma^2$ Velorum (a.k.a WR11) hosts a WC8+O7.5III binary with a 78.5-day orbital period and does not exhibit any dust formation \citep{Schmutz1997,Lamberts2017}. Systems with shorter orbital periods and thus shorter orbital separations may be affected by ``sudden radiative braking" of the WR winds due to deceleration as winds approach the luminous OB companion \citep{Gayley1997}. In their analysis, \citet{Gayley1997} indeed suggest that the wind-wind collision $\gamma^2$ Velorum is significantly affected by radiative braking. \citet{Tuthill2008} also explore the impact of radiative braking in WR104 and find that it may be play an important role in the wind-wind interaction and shock cone geometry, which are both related to the dust formation. Continued high spatial-resolution observations combined with hydrodynamic simulations of the wind-wind collisions are therefore crucial for understanding the conditions leading to dust formation in colliding-wind systems.

From a dust chemistry perspective, it is interesting to consider why dust formation via colliding winds has only been observed for \textit{carbon}-rich WRs but not for nitrogen-rich WR (WN) stars. As pointed out by \citet{Mauerhan2015}, a C-rich chemistry is not a requirement for dust formation in a colliding-wind binary. For example, the N-rich LBV system $\eta$ Car exhibits episodic dust formation via colliding winds similar to WR140, but dust around $\eta$ Car is likely composed of metal-rich silicates, corundum (Al$_2$O$_3$), and/or iron as opposed to amorphous carbon \citep{Smith2010,Morris2017}. However, the high mass-loss rate of $\sim10^{-3}$ M$_\odot$ yr$^{-1}$ from the primary star in $\eta$ Car \citep{Hillier2001} exceeds the mass-loss rates of WN and WC stars by over an order of magnitude. We therefore speculate that the density of refractory elements such as Al and O may be too low in colliding-wind WN binaries to facilitate dust formation, whereas C in WC winds is readily abundant ($40\%$ by mass; \citealt{Sander2019}).

\subsection{Conclusions}
We conducted an SED analysis of a sample of \WRnum~Galactic WC dustars with archival ground- and space-based IR photometry and spectroscopy obtained over the past several decades and revised Gaia distance estimates by \citet{RC2020}. The SEDs were modeled using \dustem~with single or double dust component models.

Results from the SED modeling show that WC dustars can produce small ($0.01 - 0.1$ $\mu$m) or large ($0.1 - 1.0$ $\mu$m) dust grains, where large grains are favored for the efficient ($\chi_C\gtrsim 1\%$) dust makers WR 95, WR98a, WR104, WR106, and WR118. We find that WC dustars exhibit a range of dust production rates and carbon dust condensation fractions, which increases with increasing IR luminosity (Sec.~\ref{sec:WCDPP}). We also fit the empirical relation between the DPR and IR luminosity that applies to the non-periodic dustars (Eq.~\ref{eq:DPR}). For the WC dustars with known periods and under the assumption of consistent wind momentum ratios ($\eta$), we successfully demonstrate the predicted power-law relation between $\chi_C$, $\dot{M}_\mathrm{WR}$, $v_\mathrm{WR}$, and P$_\mathrm{orb}$ (Fig.~\ref{fig:PerChiC}) for theoretical models of dust-forming colliding wind WR systems by \citet{Usov1991}.

In order to study the dust contribution of WC dustars in comparison to other leading stellar dust producers (AGBs, RSGs, SNe), we performed binary population synthesis models using BPASS incorporating dust production perscriptions for the four sources in addition to dust destruction from SNe. For the WC binaries, the dust production rate in the BPASS models was defined from the verified dust condensation relation in Eq.~\ref{eq:Usov2} from the SED analysis results and scaled to the properties of WR104.

We performed two sets of BPASS dust models at three different metallicities Z = 0.001, 0.008, and 0.020 representing different phases of cosmic time assuming constant star formation and a co-eval $10^6$ M$_\odot$ stellar population. Both the constant SF and co-eval models show that SNe produce the most dust out of all the sources, but are net dust destroyers at all metallicities when considering the ISM dust destruction from SN shocks. At low (Z = 0.001) metallicities, AGB stars input a comparable amount of dust to RSGs and outproduce WC binaries by over 3 orders of magnitude. At LMC-like (Z = 0.008) metallicities, the AGB stars slightly outproduce WC binaries and RSGs by factors of 2-3, whereas at solar metallicities WC binaries are the dominant source of dust until the onset of AGBs, which form dust at a comparable rate (Fig.~\ref{fig:BPASS}, Tab.~\ref{tab:BPASSTab}).

Results from the co-eval population models indicate that for a ``bursty" star formation history, AGB stars become significant sources of dust at late times after their onset ($t>100$ Myr) but still do not produce as much dust as SNe without considering dust destruction. However, at LMC-like and solar metallicities, the AGBs dominate the dust production over WC binaries and RSGs as well as SNe if they are indeed net dust destroyers at these metallicities (Fig.~\ref{fig:BPASS}, Tab.~\ref{tab:BPASSTab106}). 

Constant star formation BPASS models of the LMC (Z = 0.008) and comparisons to the observationally derived populations of AGBs, RSGs, WC binaries, and SNe demonstrated the inconsistencies between the observed and model massive star populations. The SN rates and the population of WC binaries and RSGs from the constant SF BPASS model over-predicted the observed values, which we attribute to the complex and variable star forming history of the LMC \citep{HZ2009}.

Lastly, our conclusions on WC binaries as important but overlooked dust sources strongly motivates further pursuits of BPASS dust models with variable SFH and future studies addressing the open questions on WC dust formation physics and dust chemistry. With the combination of high sensitivity and spatial resolution in the mid-IR, observations with the upcoming \textit{James Webb Space Telescope} will transform our understanding of WC dustars and dust formation in hostile environments.

\acknowledgments

\textit{Acknowledgements.} We thank T.~Onaka, I.~Endo, T.~Bakx, M.~Buragohain, E.~Lagadec, T.~Nozawa, A.~Soulain, T.~Takeuchi, and M.~Corcoran for the enlightening discussions on dust, SNe, AGBs, RSGs, and WR binaries. 
We also thank Greg Sloan for the guidance on assessing the quality of ISO/SWS data.
We would also like to thank the anonymous referee for their careful review and insightful comments that have improved the focus and clarity of our work. 
PMW thanks the Institute for Astronomy for continued hospitality and 
access to the facilities of the Royal Observatory.
RML acknowledges the Japan Aerospace Exploration Agency's International Top Young Fellowship.
This work is based in part on observations made with the Spitzer Space Telescope, obtained from the NASA/IPAC Infrared Science Archive, both of which are operated by the Jet Propulsion Laboratory, California Institute of Technology under a contract with the National Aeronautics and Space Administration.
Based in part on observations collected at the European Organisation for Astronomical Research in the Southern Hemisphere under ESO programme 097.D-0707(A).
Based in part on observations with ISO, an ESA project with instruments funded by ESA Member States (especially the PI countries: France, Germany, the Netherlands and the United Kingdom) and with the participation of ISAS and NASA.
This publication makes use of data products from the Wide-field Infrared Survey Explorer, which is a joint project of the University of California, Los Angeles, and the Jet Propulsion Laboratory/California Institute of Technology, funded by the National Aeronautics and Space Administration

\vspace{5mm}
\facilities{Spitzer(IRS, IRAC, and MIPS), WISE, ISO(SWS), VLT(VISIR)}


\software{ 
          BPASS, \dustem, MIRA
          }

\begin{deluxetable}{lllllll}
\tablecaption{WC Dustar Table}
\tablewidth{0pt}
\tablehead{WR Num &Ref & Alt Name & RA &Dec &Spec. Type& Spec. Ref  }
\startdata
	19 & VII & LS 3 & 10 18 05.02 & -58 16 25.90 & WC5d & CDB98, W09b\\ 
	48b & VII & SMSNPL 8 & 13 11 27.45 & -63 46 00.80 & WC9d & SMS99, VII \\ 
	48a & VII & D83 1 & 13 12 39.65 & -62 42 55.80 & WC8vd+WN8 & Z14 \\ 
	53 & VII & HD 117297 & 13 30 53.26 & -62 04 51.80 & WC8d & WH00 \\ 
	59 & VII & GSC 9004-3553 & 13 49 32.66 & -61 31 42.20 & WC9d & HH88, VII \\ 
	70 & VII & HIP 75863 & 15 29 44.70 & -58 34 51.20 & WC9vd+B0I & N95, W13a \\ 
	95 & VII & He3-1434 & 17 36 19.76 & -33 26 10.90 & WC9d & VII \\ 
	96 & VII & LSS 4265 & 17 36 24.20 & -32 54 29.00 & WC9d & HH88, VII \\ 
	98a & VII & IRAS 17380-3031 & 17 41 12.90 & -30 32 29.00 & WC8-9vd & W95 \\ 
	103 & VII & HIP 88287 & 18 01 43.14 & -32 42 55.20 & WC9d & VII \\ 
	104 & VII & Ve2-45 & 18 02 04.07 & -23 37 41.20 & WC9d+B0.5V & WH00 \\ 
	106 & VII & IC14-8 & 18 04 43.66 & -21 09 30.70 & WC9d & VII \\ 
	118 & VII & GL 2179 & 18 31 42.30 & -09 59 15.00 & WC9d & CV90, VII \\ 
	119 & VII & The 2 & 18 39 17.91 & -10 05 31.10 & WC9d & WH00 \\ 
	121 & VII & AS 320 & 18 44 13.15 & -03 47 57.80 & WC9d & VII \\ 
	125 & VII & V378 Vul & 19 28 15.61 & +19 33 21.4 & WC7ed+O9III & W94, W19 \\
	137 & VII & HIP 99769 & 20 14 31.77 & +36 39 39.60 & WC7pd+O9 & SS90, W01 \\ 
	140 & VII & HIP 100287 & 20 20 27.98 & +43 51 16.30 & WC7pd+O5 & W90, W09a \\ 
	124-22 & KSF15 & 1695-2B7 & 19 27 17.98 & 16 05 24.6 & WC9d & KSF15, W19 \\ 
\enddata
\tablecomments{Name, coordinates, and spectral sub-type of the \WRnum~WC dustars in our sample obtained from V 1.23 of the Galactic Wolf-Rayet Catalogue. The references correspond to the following: VII - \citet{vdh2001}, KSF15 - \citet{Kanarek2015}, RC18 - \citet{Rosslowe2018}, CDB98 - \citet{Crowther1998}, W09b - \citet{Williams2009b}, SMS99 - \citet{Shara1999}, Z14 - \citet{Zhekov2014}, WH00 - \citet{Williams2000}, HH88 - \citet{vdh1988}, N95 - \citet{Niemela1995}, W13a - \citet{Williams2013a}, W95 - \citet{Williams1995}, CV90 - \citet{Conti1990}, SS90 - \citet{Smith1990}, W01 - \citet{Williams2001}, W90 - \citet{Williams1990}, W94 - \citet{Williams1994}, W09a - \citet{Williams2009a}, W19 - \citet{Williams2019}  }
\label{tab:DustarTab}
\end{deluxetable}
\clearpage


\begin{deluxetable}{lllll}
\tablecaption{WC Dustar SED Model Details}
\tablewidth{0pt}
\tablehead{WR Num & $r_1$ range (AU) & $r_1$ intervals & IR Archival Data & Ref.}
\startdata
	19 & (100, 10000) & 50 & \textit{Spitzer}/IRS & MM17 \\ 
	48b & (50, 3000) & 30 & 2MASS, WISE, \textit{Spitzer}/IRAC+MIPS & Cu03, Cu13,Ch09, Ca09 \\ 
	48a & (10, 3000) & 25 & ISO/SWS & vdH96 \\ 
	53 & (50, 3000) & 30 & 2MASS, \textit{Spitzer}/IRAC+IRS & Cu03, Ch09, Ar10 \\ 
	59 & (50, 3000) & 30 & 2MASS, WISE,\textit{Spitzer}/IRAC & Cu03,Cu13, Ca09 \\ 
	70 & (50, 3000) & 30 & ISO/SWS & S03  \\ 
	95 & (10, 1000) & 30 & 2MASS, ESO-3.8m, \textit{Spitzer}/MIPS & Cu03, W87, Ca09 \\ 
	96 & (50, 3000) & 30 & 2MASS, WISE, \textit{Spitzer}/IRAC+MIPS & Cu03, Cu13,Ch09, Ca09 \\
	98a & (20, 3000) & 25 & ISO/SWS & vdH96 \\ 
	103 & (10, 1000) & 30 & 2MASS, ESO-3.8m, \textit{Spitzer}/IRS & Cu03, W87, Ar10 \\
	104 & (20, 3000) & 25 & ISO/SWS & vdH96 \\ 
	106 & (10, 1000) & 30 & 2MASS, ESO-3.8m, \textit{Spitzer}/MIPS & Cu03, W87, Ca09 \\ 
	118 & (10, 1000) & 25 & ISO/SWS & vdH96 \\ 
	119 & (50, 1000) & 30 & 2MASS, ESO-3.8m, WISE & Cu03, W87, Cu13 \\ 
	121 & (10, 1000) & 30 & 2MASS, WISE, \textit{Spitzer}/IRAC+MIPS & Cu03, Cu13, Ch09, Ca09 \\ 
	125 & (50, 1000) & 30 & UKIRT & W94 \\ 
	124-22 & (20, 3000) & 30 & 2MASS, WISE, \textit{Spitzer}/IRAC+MIPS & Cu03, Cu13, Ch09, Ca09 \\ 
	137 & (50, 1000) & 30 & UKIRT & W01 \\ 
	140 & (100, 3000) & 30 & TCS, UKIRT, \textit{Spitzer}/IRS* & W09a \\ 
\enddata
\tablecomments{The range of $r_1$ values (in AU) and number of logarithmic intervals is shown for each dustar model. For the two-component dust models, the $r_2$ fitting parameters is the same as that of $r_1$, and $f$ ( = $L_\mathrm{IR,1}/L_\mathrm{IR,2}$) is searched in 20 logarithmic intervals between $(0.1,100)$. Archival IR photometry taken by 2MASS, WISE, and \textit{Spitzer} were adopted from the MIPSGAL 24 $\mu$m point source catalog \citep{Gutermuth2015}. The following references are associated with the archival data of each dustar and is indicated in the rightmost column: MM17 - \citet{Marchenko2017}, Cu03 - \citet{Cutri2003}, Cu13 - \citet{Cutri2013}, Ch09 - \citet{Churchwell2009}, Ca09 - \citet{Carey2009}, vdH96 - \citet{vdh1996}, Ar10 - \citet{Ardila2010}, S03 - \citet{Sloan2003}, W87 - \citet{Williams1987}, W94 - \citet{Williams1994}, W01 - \citet{Williams2001}, W09a - \citet{Williams2009a}. *\textit{Spitzer/IRS} spectroscopy of WR140 is unpublished.}
\label{tab:ModelDetails}
\end{deluxetable}
\clearpage

\begin{deluxetable}{llllllllllllll}
\tablecaption{Adopted Dustar Parameters}
\tablewidth{0pt}
\tablehead{WR & L$_*$  & Ref & T$_*$ & Log(R$_t$) & v$_\mathrm{exp}$& Ref & $A_v$ &  $\dot{M}$ & Ref & d & Ref & P$_\mathrm{orb}$ & Ref \\
& (L$_\odot$) & & (K) & (R$_\odot$) & (km s$^{-1}$) & & & (M$_\odot$ yr$^{-1}$) & & (kpc) & & (yr) &}
\startdata 
	19 & $4\times10^5$ & S19 & 79400 & 0.3 & 2780 & S19 & 5.19 &  $4.1\times10^{-5}$ & S19 & $4.33^{+0.78}_{-0.58}$ & RC20 & 10.1 & W09b \\   
	48a & $4\times10^5$ & W12 & 50000 & 0.9 & 1000 & this work & 8.28 &  $1.7\times10^{-4}$ & Z14$^\mathrm{a}$ & $2.27^{+0.92}_{-0.57}$ & RC20 & 32.5 & W12 \\   
	48b & $2.5\times10^5$ & S19 & 40000 & 1 & 1390 & S19 & 5.73 &  $2.2\times10^{-5}$ & S19 & $5.12^{+1.25}_{-0.92}$ & RC20 & \  & \  \\   
	53* & $3.3\times10^5$ & S19 & 50000 & 0.9 & 1800 & S12 & 3.25 &    $2.2\times10^{-5}$ & S19 & $4.14^{+0.74}_{-0.56}$ & RC20 & \  & \  \\   
	59* & $5.8\times10^5$ & S19 & 40000 & 1 & 1300 & S12 & 6.43 &    $3.3\times10^{-5}$ & S19 & $3.57^{+0.69}_{-0.51}$ & RC20 & \  & \  \\   
	70 & $7\times10^5$ & W13a & 40000 & 1 & 1390 & S19 & 5.20 &  $2.2\times10^{-5}$ & S19 & $3.01^{+0.44}_{-0.34}$ & RC20 & $\sim2.8$ & W13a \\   
	95* & $1.7\times10^5$ & S19 & 45000 & 0.9 & 1900 & S12 & 6.63 &    $1.9\times10^{-5}$ & S19 & $2.07^{+0.43}_{-0.31}$ & RC20 & \  & \  \\   
	96* & $2.5\times10^5$ & S19 & 40000 & 1 & 1390 & S19 & 5.48 &    $2.2\times10^{-5}$ & S19 & $2.64^{+0.58}_{-0.43}$ & RC20 & \  & \  \\   
	98a & $1.5\times10^5$ & H16 & 45000 & 0.9 & 900 & W95 & 13.79& $2.2\times10^{-5}$ & S19 & $1.9^{+0.58}_{-0.35}$ & M99 & 1.54 & M99 \\   
	103* & $3.2\times10^5$ & S19 & 45000 & 0.8 & 1190 & S12 & 1.40 &  $2.8\times10^{-5}$ & S19 & $3.46^{+1.28}_{-0.77}$ & RC20 & \  & \  \\   
	104 & $2.5\times10^5$ & S19,M07 & 40000 & 1 & 1220 & HS92 & 6.67 &  $3\times10^{-5}$ & C97 & $2.58^{+0.12}_{-0.12}$ & So18 & 0.66 & T08 \\   
	106* & $1.7\times10^5$ & S19 & 45000 & 0.8 & 1100 & S12 & 4.61 &   $1.6\times10^{-5}$ & S19 & $3.07^{+0.56}_{-0.43}$ & RC20 & \  & \  \\   
	118 & $2.5\times10^5$ & S19 & 40000 & 1 & 1390 & S19 & 13.68 &   $2.2\times10^{-5}$ & S19 & $2.49^{+0.78}_{-0.68}\dagger$ & RC20 & \  & \  \\   
	119* & $0.5\times10^{5}$ & S19 & 45000 & 0.8 & 1300 & S12 & 3.91 &   $0.7\times10^{-5}$ & S19 & $3.22^{+1.24}_{-0.73}$ & RC20 & \  & \  \\   
	121* & $1.4\times10^5$ & S19 & 45000 & 0.8 & 1100 & S12 & 5.31 &    $1.4\times10^{-5}$ & S19 & $2.23^{+0.30}_{-0.24}$ & RC20 & \  & \  \\
	125* & $1.6\times10^5$ & S12 & 55000 & 1.1 & 2000 & S12 & 6.48 &   $2.7\times10^{-5}$ & S19 & $3.36^{+0.99}_{-0.65}$ & RC20 & 28.3  & W19  \\
	137* & $5.4\times10^5$ & S12 & 56000 & 1 & 2000 & S12 & 1.70 &  $2.7\times10^{-5}$ & S19 & $2.10^{+0.18}_{-0.16}$ & RC20 & 13.05 & L05 \\   
	140 & $1\times10^6$ & W09a & 56000 & 1 & $\sim2400$ & W09a & 2.21 &   $2\times10^{-5}$ & Su15 &$1.64^{+0.11}_{-0.09}$ & RC20 & 7.94 & M11 \\   
	124-22 & $2.5\times10^5$ & S19 & 40000 & 1 & 1390 & S19 & 14.77$^\mathrm{b}$ &  $2.2\times10^{-5}$ & S19 & $1.91^{+1.07}_{-0.72}\dagger$ & RC20 & \  & \  \\   
\enddata
\tablecomments{PoWR models are used for the radiation field of the dust heating source with the following parameters: the stellar heating source luminosity, L$_*$, the effecting temperature of the heating source, T$_*$, the ``transformed radius" of the WC star \citep{Sander2012,Sander2019}, R$_t$ in Log units of R$_\odot$. Additional model parameters are the wind velocity/dust expansion velocity, v$_\mathrm{exp}$, the visual extinction towards the dustar, $A_v$, the mass-loss rate of the WC star, $\dot{M}$, the distance towards the dustar, $d$, and the dustar orbital period, P$_\mathrm{orb}$. All of the visual extinctions were adopted from \citet{RC2020} except for WR98a whose extinction value is from \citet{vdh2001}. WR names with *'s indicate systems whose stellar spectra were explicitly modeled with the PoWR grid by \citet{Sander2012}, which provides L$_*$, T$_*$, R$_t$, v$_\mathrm{exp}$, and $\dot{M}$. The $\dagger$ symbol indicates distance values that were flagged for astrometric excess noise $> 1$ mas and large parallax uncertainty by RC20. The provided references correspond to the following: S19 - \citet{Sander2019}, W12 - \citet{Williams2012}, Z14 - \citet{Zhekov2014},
W13a - \citet{Williams2013a}, H16 - \citet{Hendrix2016}, M07 - \citet{Monnier2007}, So18 - \citet{Soulain2018}, S12 - \citet{Sander2012}, W95 - \citet{Williams1995}, HS92 - \citet{Howarth1992}, W09b - \citet{Williams2009b}, W19 - \citet{Williams2019}, C97 - \citet{Crowther1997}, Su15 - \citet{Sugawara2015},
RC20 - \citet{RC2020}, M99 - \citet{Monnier1999}, T08 - \citet{Tuthill2008}, L05 - \citet{Lefevre2005}, M11 - \citet{Monnier2011}. \\ (a) WR48a mass-loss rate from Z14 was adjusted for the revised RC20 distance.\\ (b) Derived from $A_{Ks}=1.58$ (RC20) and assuming $A_v = (0.107)^{-1}A_{Ks}$.}
\label{tab:Adopted}
\end{deluxetable}
\clearpage

\begin{turnpage}

\begin{deluxetable}{lllllllllllll}
\tablecaption{WC Dustar SED Model Results}
\tablewidth{0pt}
\tablehead{WR & r$_1$ (AU) & r$_2$ (AU) & T$_{d1}$ (K) & T$_{d2}$ (K) & $d_\mathrm{unc}$ Factor & L$_{IR}$ (L$_\odot$) & L$_{IR}$/L$_{*}$ ($\%$)& L$_{IR,1}$/L$_{IR,2}$ & M$_{d1}$ (M$_\odot$) & M$_{d2}$ (M$_\odot$) & $\dot{M}_d$ (M$_\odot$ yr$^{-1}$) & $\chi_C$ ($\%$)}
\startdata
19 (SG) & $6870^{+3130}_{-3630}$ &  & $250^{+80}_{-30}$ &  & $(0.75, 1.39)$ & $(5.13^{+1.48}_{-0.43})\mathrm{e+}0$ & $0.00128^{+0.00037}_{-0.00011}$ &  & $(3.98^{+3.76}_{-2.84})\mathrm{e-}8$ &  & $(3.94^{+3.72}_{-2.81})\mathrm{e-}9$ & $0.024^{+0.023}_{-0.017}$\\
48a (SG) & $350^{+100}_{-130}$ & $2370^{+640}_{-900}$ & $770^{+150}_{-70}$ & $310^{+90}_{-40}$ & $(0.56, 1.97)$ & $(1.05^{+0.07}_{-0.01})\mathrm{e+}4$ & $2.63^{+0.18}_{-0.03}$ & $1.83^{+0.80}_{-0.95}$ & $(1.42^{+1.12}_{-0.99})\mathrm{e-}7$ & $(9.82^{+5.84}_{-5.93})\mathrm{e-}6$ & $(8.46^{+3.48}_{-4.38})\mathrm{e-}8$ & $0.12^{+0.05}_{-0.06}$\\
48b & $180^{+90}_{-60}$ &  & $910^{+150}_{-130}$ &  & $(0.67, 1.55)$ & $(3.43^{+0.66}_{-0.87})\mathrm{e+}2$ & $0.14^{+0.03}_{-0.03}$ &  & $(2.89^{+1.96}_{-1.36})\mathrm{e-}9$ &  & $(4.71^{+0.56}_{-0.96})\mathrm{e-}9$ & $0.053^{+0.006}_{-0.011}$\\
53 & $240^{+30}_{-30}$ &  & $860^{+40}_{-40}$ &  & $(0.75, 1.39)$ & $(7.27^{+0.34}_{-0.29})\mathrm{e+}2$ & $0.22^{+0.01}_{-0.01}$ &  & $(8.43^{+1.81}_{-1.67})\mathrm{e-}9$ &  & $(1.33^{+0.11}_{-0.11})\mathrm{e-}8$ & $0.152^{+0.012}_{-0.013}$\\
59 & $210^{+30}_{-30}$ &  & $1010^{+60}_{-50}$ &  & $(0.73, 1.42)$ & $(7.84^{+0.96}_{-0.79})\mathrm{e+}2$ & $0.14^{+0.02}_{-0.01}$ &  & $(3.91^{+0.68}_{-0.69})\mathrm{e-}9$ &  & $(5.10^{+0.14}_{-0.19})\mathrm{e-}9$ & $0.039^{+0.001}_{-0.001}$\\
70 & $550^{+80}_{-70}$ &  & $730^{+40}_{-40}$ &  & $(0.79, 1.31)$ & $(2.36^{+0.07}_{-0.07})\mathrm{e+}3$ & $0.34^{+0.01}_{-0.01}$ &  & $(6.60^{+2.40}_{-1.78})\mathrm{e-}8$ &  & $(3.51^{+0.65}_{-0.56})\mathrm{e-}8$ & $0.399^{+0.074}_{-0.063}$\\
95 (LG)& $40^{+70}_{-20}$ &  & $1110^{+300}_{-340}$ &  & $(0.72, 1.46)$ & $(1.72^{+0.48}_{-0.96})\mathrm{e+}3$ & $1.01^{+0.28}_{-0.57}$ &  & $(0.93^{+2.18}_{-0.63})\mathrm{e-}8$ &  & $(9.36^{+2.00}_{-3.36})\mathrm{e-}8$ & $1.232^{+0.263}_{-0.443}$\\
96 & $180^{+130}_{-60}$ &  & $910^{+150}_{-170}$ &  & $(0.70, 1.49)$ & $(1.08^{+0.29}_{-0.45})\mathrm{e+}3$ & $0.43^{+0.12}_{-0.18}$ &  & $(9.09^{+6.67}_{-3.97})\mathrm{e-}9$ &  & $(1.48^{+0.03}_{-0.23})\mathrm{e-}8$ & $0.168^{+0.003}_{-0.026}$\\
98a (LG) & $60^{+10}_{-10}$ & $370^{+190}_{-130}$ & $960^{+70}_{-70}$ & $340^{+80}_{-80}$ & $(0.67, 1.70)$ & $(1.74^{+0.05}_{-0.07})\mathrm{e+}4$ & $11.63^{+0.32}_{-0.46}$ & $5.46^{+5.83}_{-1.66}$ & $(1.82^{+1.08}_{-0.68})\mathrm{e-}7$ & $(9.25^{+11.73}_{-5.29})\mathrm{e-}6$ & $(6.10^{+1.77}_{-1.38})\mathrm{e-}7$ & $6.93^{+2.02}_{-1.57}$\\
103 & $240^{+40}_{-40}$ &  & $860^{+50}_{-50}$ &  & $(0.60, 1.88)$ & $(4.92^{+0.38}_{-0.33})\mathrm{e+}2$ & $0.15^{+0.01}_{-0.01}$ &  & $(5.83^{+1.63}_{-1.29})\mathrm{e-}9$ &  & $(6.10^{+0.56}_{-0.52})\mathrm{e-}9$ & $0.054^{+0.005}_{-0.005}$\\
104 (LG)& $70^{+20}_{-10}$ & $460^{+110}_{-90}$ & $970^{+80}_{-70}$ & $340^{+50}_{-40}$ & $(0.91, 1.10)$ & $(1.23^{+0.04}_{-0.04})\mathrm{e+}5$ & $49.01^{+1.71}_{-1.48}$ & $5.46^{+2.39}_{-1.66}$ & $(1.19^{+0.70}_{-0.44})\mathrm{e-}6$ & $(6.06^{+3.09}_{-2.06})\mathrm{e-}5$ & $(4.39^{+1.27}_{-0.97})\mathrm{e-}6$ & $36.57^{+10.55}_{-8.07}$\\
106 (LG)& $50^{+120}_{-20}$ &  & $1020^{+200}_{-370}$ &  & $(0.74, 1.40)$ & $(7.40^{+0.71}_{-4.39})\mathrm{e+}3$ & $4.35^{+0.42}_{-2.58}$ &  & $(6.39^{+23.69}_{-3.87})\mathrm{e-}8$ &  & $(2.97^{+1.14}_{-1.01})\mathrm{e-}7$ & $4.635^{+1.780}_{-1.586}$\\
118 (LG)& $30^{+10}_{-10}$ & $260^{+120}_{-80}$ & $1370^{+190}_{-170}$ & $420^{+90}_{-80}$ & $(0.53, 1.72)$ & $(4.12^{+0.34}_{-0.26})\mathrm{e+}4$ & $16.48^{+1.37}_{-1.04}$ & $5.46^{+2.39}_{-1.66}$ & $(5.58^{+6.24}_{-2.95})\mathrm{e-}8$ & $(6.61^{+6.99}_{-3.42})\mathrm{e-}6$ & $(6.27^{+2.78}_{-1.94})\mathrm{e-}7$ & $7.13^{+3.16}_{-2.20}$\\
119 & $70^{+60}_{-30}$ &  & $960^{+220}_{-200}$ &  & $(0.60, 1.92)$ & $(3.93^{+1.45}_{-1.19})\mathrm{e+}2$ & $0.79^{+0.29}_{-0.24}$ &  & $(2.50^{+3.52}_{-1.38})\mathrm{e-}9$ &  & $(9.80^{+2.90}_{-2.14})\mathrm{e-}9$ & $0.350^{+0.104}_{-0.076}$\\
121 & $130^{+150}_{-50}$ &  & $930^{+180}_{-240}$ &  & $(0.80, 1.29)$ & $(1.04^{+0.09}_{-0.54})\mathrm{e+}3$ & $0.74^{+0.07}_{-0.38}$ &  & $(7.92^{+9.90}_{-4.73})\mathrm{e-}9$ &  & $(1.79^{+0.13}_{-0.62})\mathrm{e-}8$ & $0.319^{+0.022}_{-0.110}$\\
125 (SG)& $490^{+170}_{-100}$ &  & $570^{+50}_{-60}$ &  & $(0.65, 1.68)$ & $(9.78^{+0.32}_{-1.11})\mathrm{e+}2$ & $0.61148^{+0.02005}_{-0.06943}$ &  & $(1.00^{+0.61}_{-0.35})\mathrm{e-}7$ &  & $(3.54^{+2.15}_{-1.22})\mathrm{e-}9$ & $0.033^{+0.020}_{-0.011}$\\
137 (SG)& $660^{+240}_{-220}$ &  & $630^{+110}_{-70}$ &  & $(0.85, 1.18)$ & $(2.02^{+0.37}_{-0.35})\mathrm{e+}2$ & $0.03704^{+0.00686}_{-0.00652}$ &  & $(1.20^{+0.64}_{-0.57})\mathrm{e-}8$ &  & $(9.21^{+4.90}_{-4.36})\mathrm{e-}10$ & $0.009^{+0.005}_{-0.004}$\\
140 (SG)& $1170^{+310}_{-340}$ &  & $570^{+80}_{-50}$ &  & $(0.89, 1.14)$ & $(6.40^{+0.02}_{-0.19})\mathrm{e+}1$ & $0.00640^{+0.00002}_{-0.00019}$ &  & $(6.44^{+3.84}_{-3.30})\mathrm{e-}9$ &  & $(8.11^{+4.83}_{-4.15})\mathrm{e-}10$ & $0.010^{+0.006}_{-0.005}$\\
124-22 & $190^{+80}_{-60}$ &  & $900^{+140}_{-110}$ &  & $(0.39, 2.43)$ & $(1.03^{+0.10}_{-0.20})\mathrm{e+}1$ & $0.00412^{+0.00049}_{-0.00071}$ &  & $(9.67^{+6.49}_{-4.60})\mathrm{e-}11$ &  & $(1.49^{+0.26}_{-0.35})\mathrm{e-}10$ & $0.0017^{+0.0003}_{-0.0004}$\\
\enddata
\tablecomments{Best-fit \dustem~SED models provide the radius of component 1, $r_1$, the dust temperature of component 1, T$_{d1}$, the total IR luminosity of emitting dust, L$_{IR}$, the fraction of stellar luminosity re-radiated in the IR by dust, L$_{IR}$/L$_{*}$, the dust mass in component 1, M$_{d1}$, the dust production rate, $\dot{M}_d$, and the mass fraction of carbon in the WC wind condensed into dust, $\chi_C$. For the 2-component dust models, the radius ($r_2$), temperature (T$_{d2}$), dust mass (M$_{d2}$), and luminosity ratio L$_{IR,1}$/L$_{IR,2}$ for component 2 are fit. $T_d$ values correspond to the dust temperature of the smallest grain component in the grain size distribution: $a=0.01$ $\mu$m and $0.1$ $\mu$m for the small and large distributions, respectively. $\pm1\sigma$ uncertainties from the model fitting are shown for each derived parameter. $\pm1\sigma$ uncertainties in the adopted distance are provided as multiplicative factors in the ``$d_\mathrm{unc}$ Factor" column which applies to L$_*$, L$_{IR}$/L$_{*}$, M$_{d1}$, M$_{d2}$, $\dot{M}_d$, and $\chi_C$. Note that $r_1$, $r_2$, T$_{d1}$, and T$_{d2}$ are independent of the adopted dustar distance. The WR names with ``(LG)" or ``(SG)'' indicate models where the large or small grain size distribution were favored based on spatial size constraints. All other dustars SED models used the small grain size distribution.}
\label{tab:Results}
\end{deluxetable}
\end{turnpage}
\clearpage


\begin{deluxetable}{lllllll}
\tablecaption{BPASS Constant SF Dust Models (at $t = 1.0$ Gyr)}
\tablewidth{0pt}
\tablehead{ & Low-$Z$ & LMC-like & Solar \\
& Z = 0.001 & Z = 0.008 & Z = 0.020 }
\startdata
	$\chi_{d/g}$ & 0.0008 & 0.0018 & 0.007 \\ 
	$m_g$ (M$_\odot$) & 3088 & 1947 & 1518 \\ 
N$_{AGB}$ & 1739 (100\%) & 5465 (100\%) & 6613 (100\%) \\
N$_{RSG}$ & 2332 (65\%) & 6070 (30\%)& 6900 (28\%) \\
N$_{WC}$ & 3 (100\%) & 27 (100\%) & 47 (56\%)\\
R$_{SN}$ (yr$^{-1}$) & $2.6 \times10^{-2}$ & $1.7 \times10^{-2}$ & $1.2 \times10^{-2}$ \\
$\dot{M}_{d,AGB}$ (M$_\odot$ yr$^{-1}$) & $4.4\times10^{-7}$ [$2.5\times10^{-10}$] & $5.7\times10^{-6}$ [$1.1\times10^{-9}$]& $1.9\times10^{-5}$ [$2.9\times10^{-9}$]\\
$\dot{M}_{d,RSG}$ (M$_\odot$ yr$^{-1}$) & $3.1\times10^{-7}$ [$2.0\times10^{-10}$] & $3.0\times10^{-6}$ [$1.7\times10^{-9}$]& $7.4\times10^{-6}$ [$3.8\times10^{-9}$]\\
$\dot{M}_{d,WC}$ (M$_\odot$ yr$^{-1}$) & $1.7\times10^{-10}$ [$6.5\times10^{-11}$] & $1.7\times10^{-6}$ [$6.2\times10^{-8}$]& $2.6\times10^{-5}$ [$9.9\times10^{-7}$]\\
$\dot{M}_{d,SN}$ (M$_\odot$ yr$^{-1}$) & $3.1\times10^{-3}$ [$0.12$ M$_\odot$] & $1.8\times10^{-3}$ [$0.10$ M$_\odot$] & $1.1\times10^{-3}$ [$0.10$ M$_\odot$] \\
$\dot{M}_{SN,dest}$ (M$_\odot$ yr$^{-1}$) & $-6.5\times10^{-2}$ [$-2.5$ M$_\odot$] & $-5.9\times10^{-2}$ [$-3.5$ M$_\odot$] & $-8.2\times10^{-2}$ [$-10.6$ M$_\odot$] \\
Log$(t_{0,AGB})$ & 7.9 & 7.9 & 7.8 \\
Log$(t_{0,RSG})$ & 6.3 & 6.4 & 6.5 \\
Log$(t_{0,WC})$ & 6.3 & 6.3 & 6.4 \\
Log$(t_{0,SN})$ & 6.5 & 6.6 & 6.7 \\
\enddata
\tablecomments{Dust source populations from the constant SFR BPASS dust model for AGBs, RSGs, and WC binaries, the SN rate (R$_{SN}$), the total and average (in brackets) DPR for each source at time $t = 1.0$ Gyr. $\chi_{d/g}$ is the adopted dust-to-gas mass ratio and $m_g$ is the total ISM mass swept up by each SN as determined by Eq.~\ref{eq:mg} at each metallicity. $N$ shows the total numbers of sources present at this time and the percentage indicates the fraction contribute to the dust input and do not end their lives as SN. The AGB, RSG, and WC binary DPRs are determined from the population that do not end their lives as SN. The SN DPR value in brackets is the average mass of dust formed/destroyed per SN. The time, $t_0$ (yr), corresponds to the onset time of each dust source.}
\label{tab:BPASSTab}
\end{deluxetable}
\clearpage


\begin{deluxetable}{lllllll}
\tablecaption{BPASS $10^6$ M$_\odot$ Stellar Population Dust Models (at $t = 1.0$ Gyr)}
\tablewidth{0pt}
\tablehead{ & Low-$Z$ & LMC-like & Solar \\ 
& Z = 0.001 & Z = 0.008 & Z = 0.020}
\startdata
$M_{d,AGB}$ (M$_\odot$) & $1.0\times10^{1}$ & $1.0\times10^{2}$ & $2.7\times10^{2}$ \\
$M_{d,RSG}$ (M$_\odot$) & $2.9\times10^{-1}$ & $2.1\times10^{0}$ & $4.3\times10^{0}$ \\
$M_{d,WC}$ (M$_\odot$) & $1.9\times10^{-5}$ & $2.3\times10^{-1}$ & $3.7\times10^{0}$ \\
$M_{d,SN}$ (M$_\odot$) & $1.4\times10^{3}$  & $9.1\times10^{2}$  & $5.4\times10^{2}$ \\
$\Delta t_{AGB}$ (yr) & $1.5\times10^{9}$ & $1.9 \times10^{9}$ & $2.4 \times10^{9}$ \\
$\Delta t_{RSG}$ (yr)& $2.0\times10^{8}$ & $1.6 \times10^{8}$ & $2.5 \times10^{8}$ \\
$\Delta t_{WC}$ (yr)& $5.2\times10^{5} $& $4.3 \times10^{6}$ & $2.5 \times10^{6}$ \\
$\Delta t_{SN}$ (yr)& $9.7\times10^{7} $&$ 7.5 \times10^{7}$ & $5.8 \times10^{7}$ \\
\enddata
\tablecomments{Model results show the cumulative dust mass, $M_d$, produced by each source and the timescale, $\Delta t$ that each source is actively forming dust. These are results show the population of RSG and WC stars that do not end their lives as SNe}
\label{tab:BPASSTab106}
\end{deluxetable}

\clearpage


\begin{deluxetable}{lll}
\tablecaption{BPASS LMC Dust Model Table}
\tablewidth{10pt}
\tablehead{ & BPASS ($t = 12.6$ Gyr) & From Obs. }
\startdata
N$_{AGB}$ & 20000 (100\%) & 20040 \\
N$_{RSG}$ & 11887 (30\%) & 3589 \\
N$_{WC}$ & 52 (100\%) & 2 \\
R$_{SN}$ (yr$^{-1}$) & $3.3 \times10^{-2}$ & $2.6\times10^{-3}$ \\
$\dot{M}_{d,AGB}$ (M$_\odot$ yr$^{-1}$) & $1.4\times10^{-5}$ [$7.1\times10^{-10}$] & $1.4\times10^{-5}$ [$6.9\times10^{-10}$]\\
$\dot{M}_{d,RSG}$ (M$_\odot$ yr$^{-1}$) & $5.9\times10^{-6}$ [$1.6\times10^{-9}$] & $1.4\times10^{-6}$ [$4.0\times10^{-10}$] \\
$\dot{M}_{d,WC}$ (M$_\odot$ yr$^{-1}$) & $3.3\times10^{-6}$ [$6.2\times10^{-8}$] & $\gtrsim3\times10^{-8}$\\
$\dot{M}_{d,SN}$ (M$_\odot$ yr$^{-1}$) & $3.4\times10^{-3}$ [$0.10$ M$_\odot$] & $1.7\times10^{-3}$ [$0.65$ M$_\odot$]* \\
$\dot{M}_{SN,dest}$ (M$_\odot$ yr$^{-1}$) & $-1.1\times10^{-1}$ [$-3.5$ M$_\odot$] & $\sim-3\times10^{-2}$ [$\sim-8$ M$_\odot$] \\
Log$(t_{0,AGB})$ & 7.9 &- \\
Log$(t_{0,RSG})$ & 6.4 & - \\
Log$(t_{0,WC})$ & 6.3 & - \\
Log$(t_{0,SN})$ & 6.6 & - \\
\enddata
\tablecomments{Results from the LMC BPASS dust models at time $t = 12.6$ Gyr compared to observationally derived values \citep{Temim2015,Srinivasan2016, Williams2019}. Values in this table are presented in the same format as Tab.~\ref{tab:BPASSTab}. Note that the observed value of $N_{WC}$ corresponds to the number of known dust forming WC systems in the LMC. For the SN dust destruction in the BPASS model, it is assumed that the dust-to-gas mass ratio in the LMC is $\chi_{d/g}=0.0018$ and that the SN shock clears dust from a total surrounding ISM mass of $m_g = 1947$ M$_\odot$. The dust destruction rates in the right column are from \citet{Temim2015}, who derive $m_g \sim 2000$ M$_\odot$.\\ *\citet{Temim2015} estimate the SN dust production rate in the LMC assuming $100\%$ dust condensation efficiency and present these as absolute upper limits on the injected dust mass.}
\label{tab:BPASSLMCTab}
\end{deluxetable}
\clearpage

\end{document}